\documentclass[fleqn,usenatbib]{mnras}

\usepackage[T1]{fontenc}
\usepackage{ae,aecompl}

\usepackage{amsmath}	% Advanced maths commands
\usepackage{amssymb}	% Extra maths symbols
\usepackage{xcolor}     % added by PM to add comments
\usepackage{soul}       % added by PM, allows highlights and strikethroughs
\usepackage{url}        % added by MRH, for hyperlinks in urls
\usepackage{pdflscape}  % added by Ilja, for landscape Figures and Tables

\usepackage[dvipdfmx]{graphicx}
\usepackage{bmpsize}

\usepackage{newtxtext,newtxmath}

\usepackage{subfig}
\usepackage{multirow}

\usepackage{todonotes}
\newcommand{\td}{\text{d}}                  % differential in math mode
\newcommand{\glref}[1]{Eq.~(\ref{#1})}     % produced eq. ()
\newcommand{\tbref}[1]{Table~\ref{#1}}      % produces Table x
\newcommand{\figref}[1]{Fig.~\ref{#1}}     % produces Figure x
\newcommand{\ein}[1]{\text{ #1}}            % creates the correct unit font in math modes

\defcitealias{2013Sci...342E...1I}{Aartsen et al. 2013}
\title[Neutrino emission from SMBBH and BBH Mergers]{Extragalactic neutrino emission induced by Supermassive and Stellar Mass Black Hole mergers}

\author[Ilja Jaroschewski et al.]{
Ilja Jaroschewski,$^{1,2}$\thanks{E-mail: ilja.jaroschewski@ruhr-uni-bochum.de}
Julia Becker Tjus$^{1,2}$, Peter L.~Biermann$^{3,4}$
\\
$^{1}$Ruhr-Universit\"at Bochum, Theoretische Physik IV: Plasma-Astroteilchenphysik, Universit\"atsstrasse 150, 44801 Bochum, Germany\\
$^{2}$Ruhr Astroparticle and Plasma Physics Center (RAPP Center), Ruhr-Universit\"at Bochum, 44780 Bochum, Germany\\
$^{3}$MPI for Radioastronomy, 53121 Bonn, Germany\\
$^{4}$Department of Physics \& Astronomy, University of Alabama, Tuscaloosa, AL 35487, USA\\
}

\date{Accepted XXX. Received YYY; in original form ZZZ}

\pubyear{2022}

\begin{document}
\label{firstpage}
\pagerange{\pageref{firstpage}--\pageref{lastpage}}
\maketitle

\begin{abstract}
The recent detections of binary stellar mass black hole mergers by the LIGO and Virgo Collaborations suggest that such mergers are common occurrences. 
Galaxy mergers further indicate that supermassive black holes in centers of galaxies also merge and are typically expected to have had at least one merger in their lifetime, possibly many. 
In the presence of a jet, these mergers are almost always accompanied by a change of the jet direction and a connected jet precession motion, leading to interactions of the jet with ambient matter and producing very high-energy particles, and consequently high-energy gamma-rays and neutrinos.

In this work, we investigate the possibility  under which conditions such mergers
could be the sources of the diffuse astrophysical neutrino flux measured by the IceCube Neutrino Observatory. The main free parameters in the calculation concern the frequency of the mergers and the fraction of energy that is transferred from the gravitationally released energy to neutrinos. 
We show that the merger rate for SMBBHs must lie between $\sim 10^{-7}$ and $10^{-5}$~Gpc$^{-3}$~yr$^{-1}$. The ratio of energy going to neutrinos {during such mergers lies then} between {$\sim 10^{-6} - 3\cdot 10^{-4}$}. For stellar mass BBH mergers, the rate needs to be $\sim 10-100$~Gpc$^{-3}$~yr$^{-1}$ and the expected ratio of neutrino to gravitational wave energy lies in a comparable range as for SMBBHs, {$\sim 2 \cdot 10^{-5} - 10^{-3}$}. These values lie in a {reasonable} parameter range, so that the production of neutrinos at the level of the detected neutrino flux is a realistic possibility.

\end{abstract}

\begin{keywords}
astroparticle physics -- black hole mergers -- quasars: supermassive black holes -- galaxies: starburst -- neutrinos -- gravitational waves 
%-- methods: analytical
\end{keywords}

%%%%%%%%%%%%%%%%%%%%%%%%%%%%%%%%%%%%%%%%%%%%%%%%%%

%%%%%%%%%%%%%%%%% BODY OF PAPER %%%%%%%%%%%%%%%%%%

\numberwithin{equation}{section}

\section{Introduction}
\label{sec:introduction}
Nearly a decade after the discovery of a diffuse, astrophysical flux of neutrinos by IceCube, its origin is still largely unknown \citep{2013Sci...342E...1I, 2014PhRvL.113j1101A}. 
The detection of a high-energy neutrino from the direction of the blazar TXS~0506+056 in combination with a coincident gamma-ray flare in September 2017 \citep{2018Sci...361.1378I} opened up the possibility that these sources are the origin of extragalactic neutrinos (e.g. \citealt{2011NuPhS.217..284B, 2018Sci...361..147I}).
A detailed analysis of this blazar indicates that it could harvest an ongoing merger of a supermassive binary black hole (SMBBH) at its core \citep{halzen2019cosmic, 2019A&A...630A.103B, 2020ApJ...905L..13D}.
This is reinforced by the detection of a third high-energy neutrino from the direction of TXS~0506+056 on September 18, 2022 by IceCube \citep{IC220918A_gcn1}.
With three, possibly more, blazars identified as high-energy neutrino sources, the hint gets strengthened that they are accelerators of high-energy neutrinos (\citealt*{2017MNRAS.466L..34K}, \citealt*{2019MNRAS.483L..42K}, \citealt{2021ApJ...911L..18K}).

Furthermore, starburst galaxies are believed to be sources of cosmic rays \citep*{2006JCAP...05..003L, 2007ApJ...654..219T} and high-energy neutrinos \citep*{2003ApJ...586L..33R, 2015ApJ...805...95C} as well.
This was discussed with a specific stellar mass binary black hole (BBH) merger being an UHECR accelerator in the starburst galaxy M82 \citep{2018AdSpR..62.2773B}, identified by a compact radio structure \citep*{1985ApJ...291..693K}.  A hint in the point source analysis of IceCube data for a 3-sigma signal from NGC~1068 points to the possibility that Seyfert-Starburst composite galaxies could be an interesting source class \citep{2020PhRvL.124e1103A}. It has been shown that a pure starburst contribution {due to supernova remnants} cannot explain the high-energy gamma-ray emission \citep{2014ApJ...780..137Y, 2016ApJ...821...87E}, but that the emission from the core could explain the multimessenger signatures, including a possible neutrino signal \citep{2020PhRvL.125a1101M, 2022icrc.confE1006E, 2021ApJ...922...45K}. The scenario of stellar mass BBH mergers, however, has not been tested yet.

Since the first detection of gravitational waves (GWs) from a stellar mass BBH merger in 2015 \citep{PhysRevLett.116.061102}, the number of detected mergers rose continuously, making these events common occurrences and leading to the creation of the Gravitational-Wave Transient Catalogues GWTC-1, GWTC-2, GWTC-2.1 and \mbox{GWTC-3} \citep{2019PhRvX...9c1040A, 2021PhRvX..11b1053A, 2021arXiv210801045T, 2021arXiv211103606T}.

In this paper, the hypothesis is investigated whether mergers of supermassive BBHs and/or stellar mass BBHs could be  dominant sources of the diffuse astrophysical neutrino flux measured by the IceCube neutrino observatory.
This work is motivated by the ultra-high energy cosmic ray (UHECR) anisotropy study by the Pierre Auger Collaboration, which found in their analysis that about 10 per cent of arriving UHECRs could be clustered around nearby starburst galaxies, M82 being one of them, with the rest contributed by other sources, here assumed to be active galactic nuclei (AGN) \citep{2018ApJ...853L..29A}.
Based on the model of \cite{2017PhRvD..96h3003D}, an approach that can be applied to both source classes is used in this paper to constrain the total emitted energy in neutrinos during a merging event with the emitted GW energy.

\subsection{Merging mechanism of SMBBHs}
\label{subsec:Merging_SMBBH}

A necessary requirement for SMBBH mergers is an associated merging of their host galaxies, each containing an AGN or a former, now dormant AGN with a supermassive black hole (SMBH) at its core. 
Latest observations and investigations suggest, that nearly all massive galaxies with a redshift near 0 underwent at least one merger \citep{1974ApJ...187..425P, 1980ApJ...236..351D, 1997ApJ...490..577D, 2003AJ....126.1183C, 2012ApJ...753..140B, 2016ApJ...826...91A}, with galaxies in centers of clusters undergoing more mergers than those at its edges \citep{2019ApJ...875..141P}.
After the initial galaxy merger, the merging mechanism of SMBBHs can be described in 4 stages: 

(1) In the \textbf{dynamical friction stage}, the binary SMBH loses orbital energy by gravitationally interacting with stars and gas clouds until it reaches a binary separation of a few parsec \citep{2019Galax...7...63G}. Due to the lack of necessary star masses in the central region at these separations, dynamical friction becomes highly inefficient in driving the binary together. 
Orbital energy loss via gravitational radiation alone on the other hand is too weak at pc separations, requiring time scales above the Hubble time for a merger of the binary (see e.g.~\citealt*{2003ApJ...582..559V, 2003ApJ...583..616J} and \citealt{2019Galax...7...63G}).
This lead to the assumption that the binary was stalled at such separations, the so-called 'final parsec problem', with several proposed solutions as discussed in step (2).
For the time scale of the dynamical friction stage, \citet{1980Natur.287..307B} derived the following relation, which was adjusted by \cite{2002MNRAS.331..935Y} as:
\begin{equation}
	t_{dyn} = 
	\frac{4 \cdot 10^6 \ein{yr}}{\log N_*}
	\cdot \frac{\sigma_c}{200\ein{km}\ein{s}^{-1}}
	\cdot \left( \frac{r_c}{100\ein{pc}} \right)^2 
	\cdot \frac{10^8 \ein{M}_{\sun}}{m_2} \,,
	\label{eq:t_dyn}
\end{equation}
with the mass of the smaller SMBH $m_2$, the number of stars in the central galactic region $N_*$, the velocity dispersion in that region $\sigma_c$ and the radius of the center region $r_c$.
A detailed discussion on the derivation and validity of the relation can be found in \cite{2002MNRAS.331..935Y} and \cite{2003ApJ...583..616J}.

(2) \textbf{The final parsec stage:} As a solution to the final parsec of the merging, the migration of both SMBHs as well as molecular clouds via gas dynamics into the center region of the merged galaxy, similar to the theory of planet migration, is suggested \citep*{1980Natur.287..307B, 2003ApJ...583..616J}.
Such a molecular flow was indeed observed in several parsec distance from the center of the Milky Way \citep*{1993A&A...269..169V}.
In addition, \citet*{2001A&A...377...23Z} showed that a torus-shaped stellar cluster with a mass in the order of the SMBBH is sufficient for overcoming the separation of the final parsec in time scales of $10^7 \ein{yr}$ by ejecting stars out of the inner region.
Besides that, simulations show that 3-body interactions of the SMBHs with molecular clouds in the inner galactic region lead to a separation reduction to a sub-parsec range in $t_{\rm fin} = (1 - 100) \ein{Myr}$ \citep{2016IAUFM..29B.285V}. 
Another suggested and prominent solution to this problem requires a third SMBH in the central galaxy region, as a consequence of a previous galaxy merger of one of the considered galaxies. 
By the presence of another massive object in a pc scale separation from the other two, it comes to a 'gravitational slingshot' \citep*{2003ApJ...582..559V}:
The lightest SMBH is thrown out of the system, while the heavy binary gets more tightly bound at sub-parsec separations. 
However, as this mechanism requires the existence of a third, not yet merged SMBH from a previous galaxy merger, the probability of such a constellation is consequently low.

There are also considerations that this 'final parsec problem' is non-existing and observations are biased by the fact that sub-parsec separations of SMBBHs are difficult to resolve, with the smallest directly measured separation being $7.3 \ein{pc}$ \citep{2006ApJ...646...49R}.
The event horizon telescope can resolve even smaller separations in the low pc or even sub-pc range (e.g. \citealt{2019ApJ...875L...1E}), so that an identification of smaller separated SMBBHs is possible in the future.
A loss of orbital energy by interacting magnetic winds of the binary is one possibility, how they could migrate to below pc separations.
One of these suggestions or a combination of them solves the 'final parsec problem'.
Thus we will refer to this stage as the 'stage for overcoming the final parsec' or just the final parsec stage.

(3) \textbf{Inspiral Stage:} The end of the previous stage is marked by gravitational radiation becoming the leading dissipative effect \citep{2009ApJ...697.1621G}. 
This is the most important stage for neutrino production and is referred to as the inspiral stage.
As a result of the turbulence caused by the merging galaxies at sub-parsec separations, the gravitationally bound accretion disks around each SMBH is fed by enough matter so that a powerful jet can be produced, either directly from the accretion disk \citep{1982MNRAS.199..883B} or via accretion onto the SMBH by angular momentum transport \citep{1977MNRAS.179..433B}.
During this stage, the unaligned, probably nearly maximal spins \citep{2019ApJ...886...37D} of the SMBHs start to precess and realign themselves with the direction of the orbital angular momentum (spin), causing a spin-flip of the jets \citep{2009ApJ...697.1621G}.
As a consequence, the jets sweep through the surrounding matter, leading to proton-proton interactions and producing neutrinos \citep{2018Univ....4...24K}.
It is therefore assumed that neutrinos are mainly produced in the inspiral stage, but not solely, as new material like molecular clouds can enter the jet at any time during his lifetime.
If a jet points at Earth during the spin-flip, neutrinos could be detectable with IceCube.
However, the jet has to drill a new path through the dusty torus or molecular and gaseous environment in the newly formed galaxy  \citep{2003MNRAS.342..399G}.
That is why its light is absorbed or reflected, while neutrinos only slightly interact with matter, leading to the chance of a neutrino detection even before a possible gamma-ray detection of the jet.

The time scale for this inspiral stage 
and thus neutrino production 
ranges from around $10^5 \ein{yr}$ to over $10^9 \ein{yr}$, depending on the total SMBBH mass and mass ratio, and will be discussed in more detail in Sec.~\ref{subsec:merging_time_scales}.

(4) \textbf{Actual Merger Stage:} By the end of the inspiral stage, the binary reaches the innermost
stable orbit (ISCO) at a few Schwarzschild radii separation from each other \citep{2017AnP...52900209A} and enters the merger stage.
Due to their aligned spin after the spin-flip, the spin and jet orientation stay the same hereafter. 
However, for highly eccentric orbits, the SMBBHs could merge before a successful spin-flip of the jets \citep*{2021PhRvD.103h4025K}, causing a different spin and jet orientation of the newly merged SMBH. 
The gravitational wave frequency reaches a maximum during the merger \citep{PhysRevLett.116.061102}.
The time scale for the actual merger after the ISCO is around years, depending on the binaries total mass, and will be addressed in Sec.~\ref{subsec:merging_time_scales}.

As the merged SMBH receives most of the angular and linear momentum of the two merging SMBHs, it can fly out of the galactic center \citep{2007PhRvL..98w1102C}.
This is called \textit{gravitational recoil} and explains the observations of different locations of the jet basis and SMBH in several galaxies \citep{2020ApJ...888...36R}.
Depending on the linear momentum, the SMBH can then begin precessing around the galactic center, or, in rare cases, could escape the galaxy.

\subsection{Merging mechanism of stellar mass BBHs}
\label{subsec:Merging_BBH}

The merging mechanism of stellar mass BBHs (just BBHs hereafter) differs from the merging of SMBBHs in several aspects.
In this work, we focus specifically on BBHs in starburst galaxies, gas rich galaxies with an extraordinary high star formation rate, see e.g.~\citep{bartos2015spectral} and references therein, as more massive stars compared to other galaxies can grow in them.
The most massive stars are born preferably in multiple star systems \citep{2012MNRAS.424.1925C, 2013EAS....64..155C}, 
consisting of one or more binary star systems (or a binary system with an orbiting third star), and develop into a super giant star with a subsequent supernova in scales of million years \citep{2018AdSpR..62.2764B}. 
While most of the stellar mass is pushed away via the supernova, the black hole left going into a merger has a mass ranging from $5 \ein{M}_{\sun}$ as the GWTC-1 to GWTC-3 data indicate \citep{2019PhRvX...9c1040A, 2021PhRvX..11b1053A, 2021arXiv210801045T, 2021arXiv211103606T}.

We are discussing explicitly the scenario, where all stars in the multiple star system have similar masses and thus co-evolve into BHs \citep{2012MNRAS.424.1925C}, which get fed by the surrounding matter and produce relativistic jets, see e.g.~ \citet{2018AdSpR..62.2773B} for a review. 
In the following, we describe the evolution of a quadruple star or rather BH system, consisting of two binary systems with orbital periods typically around years (see e.g. \citealt{2019A&A...630A.128Z}), with the orbital periods of the close binary in the range of days (see e.g. \citealt{2014ApJS..211...10S}).
A triple star system can evolve analogously.
The spins of the stellar mass BHs in each sub-binary system are parallel before the BBH merger, as they have enough time to align through tidal torques during their lifetime as stars, so that a spin-flip of the jets does not occur. 
This is in comparison with the aligned spins of the BHs prior to the merger in the GWTC-2 to GWTC-3 data. 
The merged BHs of the sub-binary systems then build a binary BH system and merge after some time as well. 
However, as the time before their inspiral stage is usually not sufficient to realign their spins, a spin-flip of the jets during the second BBH merger, similar to the merging event of two SMBHs, occurs.
In the process, the jets sweep through the gas and dust rich environment, leading to proton-proton interactions and resulting in neutrino productions.
During this spin-flip of the jet, a double cone free of cold, dense matter is left due to the precession of the unaligned jets.
Since the merged BH, just like with SMBHs, receives most of the angular and linear momentum of the two merging BHs, it moves away from the center of the cone due to gravitational recoil.
There is an observational hint to this scenario described here with a slightly displaced radio source in an emission free cone: 
the compact source 41.9+58 in the starburst galaxy M82 {\citep*{1985ApJ...291..693K, 2018AdSpR..62.2773B}}. 
This scenario can describe the different measured spins in the detection of BBH mergers in GWTC-2 to GWTC-3 \citep{2021PhRvX..11b1053A, 2021arXiv210801045T, 2021arXiv211103606T} with two generations of stellar mass BBH mergers.\\

This paper is organized as follows:
In Sec.~\ref{sec:method} we describe our analytical derivation of the connection between the diffuse astrophysical neutrino flux and the gravitational wave energy emitted by SMBBH as well as stellar mass BBH mergers, assuming the scenarios described in here to be the primary neutrino production mechanisms.
The main derivations of the diffuse neutrino flux are covered in Sec.~\ref{subsec:Source_model}, while
the cosmological evolution of the sources is discussed in Sec.~\ref{subsec:Cosmological_evolution}.
A detailed discussion on the relevant time scales of SMBBH mergers and the connection between neutrinos and the expected merging rates of SMBBHs in the inspiral stage is carried out in Sec.~\ref{sec:time_scales}.
In Sec.~\ref{sec:SMBBH_Merger_Rate}, the number of total SMBBHs mergers is estimated and used to determine the total merging time for one SMBH and the associated SMBBH merger rates in comparison to the expected neutrino flux.
For stellar mass BBH mergers, the connection between neutrinos and GWs is shown in Sec.~\ref{sec:BBH_Merger_Rates}.
Our results are compared in Sec.~\ref{subsec:comparison}, with Sec.~\ref{subsec:constraints} serving as an attempt to constrain the SMBH mass distribution. 
A discussion of the results is performed in Sec. \ref{sec:Discussion}, while Sec.~\ref{sec:Conclusions} summarizes the main results.
Finally, an outlook is given in Sec.~\ref{sec:Outlook}.

\section{Modeling the diffuse neutrino flux from (SM)BBHs}
\label{sec:method}

This chapter builds up a connection between the diffuse astrophysical neutrino flux and the gravitational wave energy emitted by SMBBH as well as stellar mass BBH mergers.

In this chapter, the general ansatz of deriving the diffuse neutrino flux will be explained together with the specific assumptions on the source model, luminosity function and redshift distribution.
It is assumed that neutrinos are produced at any point in time during the inspiral stage, so that a connection to GWs can be established.
That is why the time scales during the merger event are discussed in the following section in detail as well.

The total diffuse neutrino flux at Earth $\Phi $ can be represented as a double integral over the redshift $z$ and luminosity $L$ of the spectrum of a single source $d\Phi_\nu/dE_\nu$ \citep[e.g.][for a review]{2008PhR...458..173B}:
\begin{equation}
	\Phi(E_{\nu}^{0}) 
	= \int_{z} \int_{L} {\td} z {\td} L \frac{{\td} \Phi_\nu}{{\td} E_{\nu}}\left(E_{\nu}^{0}, L, z\right) 
	\cdot \frac{{\td} n}{{\td}V {\td} L}(L, z) 
	\cdot \frac{{\td} V}{{\td} z} 
	\cdot \frac{1}{4 \pi d_{L}(z)^{2}} \,.
	\label{eq:diffuse_neutrino_flux}
\end{equation}
Due to the expansion of the Universe, neutrino energies are shifted towards lower values at the detector $E^0_\nu$ compared to their energy at the source $E_\nu$ according to $E^0_\nu = E_\nu/(1 + z)$.
Here, the number of sources at redshift $z$ per volume and luminosity is given by $\td n/(\td V \td L)$. 
The expression $\td V/\td z$ captures the comoving volume, while the factor $1/(4\pi d_L^2)$ accounts for the decrease of the flux with the luminosity distance $d_L$.
This connection is only sufficient for time averages of high-energy neutrino and gamma-ray emissions, not for single emission episodes.

In the following derivations, the hypothesis is tested whether the diffuse neutrino flux detected by IceCube can be explained by our model and what parameter space would be allowed. We therefore compare the results we achieve from our calculation applied to Eq.~(\ref{eq:diffuse_neutrino_flux}) with current IceCube results of the flux weighted with a factor $0< \lambda <1$ that can account for the fact that only a fraction of the flux is made up by either of the source classes investigated. 

With this approach in mind, the details on the source model and evolution will be discussed in the following paragraphs. 

\subsection{Source model}
\label{subsec:Source_model}
Following the approach by \cite{2017PhRvD..96h3003D}, a parameter $f^\nu_\text{BBH}$ is introduced at this point.
This parameter describes the energy fraction that the total produced neutrinos receive, compared to the total gravitational potential loss of the system during the binary black hole merger.
The substantial majority of the potential loss is emitted in form of gravitational waves, so that the connection
\begin{equation}
	E^\text{total}_\nu = f^\nu_\text{BBH} \cdot E_\text{GW} \,,
	\label{eq:parameter_f}
\end{equation}
with the total emitted gravitational wave energy $E_\text{GW}$ can be established.
For the sake of simplicity, we do not distinguish at this stage between a stellar mass binary black hole merger and a supermassive binary black hole merger.
The formalism is the same for both, while the absolute and relative numbers differ.
The cosmological conditioned shift of the energy to lower values affects neutrinos as well as gravitational radiation, so that their relation stays the same.
That is why a distinction between these energies at the source or at the detector is not performed in \glref{eq:parameter_f}.\\

Splitting the single source spectrum per time and energy into the derivative of neutrino number per energy and time results in:
\begin{equation}
    \frac{{\td} \Phi_\nu}{{\td} E_{\nu}}  
    = \frac{{\td} \dot{N}_\nu}{{\td}E_\nu}
    = \frac{{\td}^2 N_\nu}{{\td}t \ {\td}E_\nu}
    = \frac{\td}{\td t} \left( \frac{{\td}N_\nu}{{\td}E_\nu} \right) \,,
    \label{eq:source_spectrum_E_GW_1}
\end{equation}
By integrating over the neutrino number per energy, the total neutrino energy of a point source is obtained.
\begin{equation}
	E^\text{total}_\nu = \int E_\nu \ \frac{\td N_\nu}{\td E_\nu} \ \td E_\nu \,.
	\label{eq:E_tot}
\end{equation}
Taking a neutrino flux of the form $E^{-p}$ with the spectral slope $p$, the neutrino number per energy of a single point source is expressed with a constant $A$ according to:
\begin{equation}
	\frac{{\td}N_\nu}{{\td}E_\nu} = A \cdot E_\nu^{-p} \,.
	\label{eq:dN_dE}
\end{equation}
Combining \glref{eq:E_tot} with \glref{eq:dN_dE} and integrating using \glref{eq:parameter_f} yields:
\begin{equation}
	A = \kappa_p \cdot f^\nu_\text{BBH} \cdot E_\text{GW} \,,
	\label{eq:constant_A}
\end{equation}
with
\begin{equation}
    \kappa_p 
    = \begin{cases}
            \frac{ (2- p)}{ 
            \left( E_{\max}^{ (2- p)} - E_{\min}^{ (2-p)} \right) } \ein{GeV}^{- (2- p)} 
            & \text{for} \quad p \neq 2 \,, \\
            \frac{1}{ \ln \left( \frac{E_{ \max}}{E_{\min}} \right) } 
            & \text{for} \quad p = 2 \,.
      \end{cases}
    \label{eq:B_p}
\end{equation}
The minimum and maximum neutrino energies are $E_{\min}$ and $E_{\max}$, respectively.
The resulting single source spectrum is then
\begin{equation}
	\frac{{\td} \Phi_\nu}{{\td} E_{\nu}}  
	= \frac{\td}{\td t} \left( \frac{{\td}N_\nu}{{\td}E_\nu} \right)
	= E_\nu^{-p} 
	\cdot \kappa_p \cdot f^\nu_\text{BBH}
	\cdot \frac{{\td}E_\text{GW}}{{\td}t} \,,
	\label{eq:source_spectrum_E_GW}
\end{equation}
as only the radiated GW energy has a time dependency in here.\\

\subsection{Cosmological evolution}
\label{subsec:Cosmological_evolution}

As the comoving volume is strongly dependent of cosmological parameters, it can be described via the luminosity distance $d_L$ in a flat Universe as \citep{1999astro.ph..5116H}
\begin{equation}
	\frac{{\td}V}{\td z}
	= \frac{4\pi \cdot d_L^2(z)}{(1 + z)^2} \cdot \frac{c \cdot t_H}{E(z)}
	= 4\pi \cdot d_H \frac{r(z)^2}{E(z)} \,.
	\label{eq:comoving_volume}
\end{equation}
The inverse of the Hubble constant $H_0$ is the Hubble time $t_H$ and its product with the speed of light the Hubble distance $d_H = c \cdot t_H$, while $r(z)$ is the comoving distance defined as $r(z) = d_H \int_0^z \frac{\td z'}{E(z')}$.
The function $E(z) \equiv \sqrt{\Omega_{\mathrm{M}}(1+z)^{3}+\Omega_{k}(1+z)^{2}+\Omega_{\Lambda}}$ contains the dependency on the cosmological density parameters. 
They are dimensionless and describe the ratio of energy or matter to the total matter and energy in the Universe \citep[see][]{1999astro.ph..5116H}. 
The baryonic and dark matter density parameter is $\Omega_{\mathrm{M}}$ and $\Omega_\Lambda$ the dark energy density parameter.
The parameter $\Omega_\text{k}$ on the other hand measures the curvature of space.
A value of $0$ means the space is flat and therefore Euclidean, while a positive value means the space is elliptic and a negative value that it is hyperbolic.
{Per definition, the sum of these three density parameters is $1$.}
As seen by the latest observations of the Planck collaboration in \tbref{tab:Planck}, the Universe is dark energy dominated and most likely flat due to the small error in the measurement of $\Omega_\text{k}$.
Thus, a $\Lambda$-CDM Universe (Lambda cold dark matter Universe) is assumed because it is the simplest cosmological model in full agreement with the Planck data. 
Here, $h$ is the dimensionless Hubble constant: $H_0/ (100 \ein{km}\ein{s}^{-1}\ein{Mpc}^{-1})$.
In the following, values from Table~\ref{tab:Planck} are used for the calculations.

In analogy to \cite{2001MNRAS.322..536W} and \cite{2014PhRvD..89l3005B}, the luminosity function is divided into a luminosity and a redshift dependent part: 
\begin{equation}
	\frac{\td n}{\td V \td L} = g(L) \cdot f(z) \,.
	\label{eq:luminosity_function}
\end{equation}
By inserting \glref{eq:source_spectrum_E_GW} along with \glref{eq:comoving_volume} and \glref{eq:luminosity_function} into the diffuse neutrino flux at Earth, \glref{eq:diffuse_neutrino_flux} is reshaped as:
\begin{align}
	E_\nu^{p } \Phi(E_{\nu}) 
	& =
	\underbrace{ \kappa_p \cdot f^\nu_\text{BBH} \cdot c \cdot t_\text{H} }_{\zeta_c}
	\cdot \underbrace{\int_{z} \ \frac{f(z)}{(1 + z)^2 \cdot E(z)} \ {\td}z}_{\xi_z}
	\notag \\
	& \ \cdot \underbrace{\int_{L} \ \frac{{\td}E_\text{GW}}{{\td}t} \cdot g(L) \ {\td}L}_{\zeta_L} \,,
	\label{eq:E3_diffuse_neutrino_flux}
\end{align}
with the constant term $\zeta_c$, the redshift dependent activity-integral $\xi_z$ and the luminosity dependent integral $\zeta_L$.  
In the following, both integrals are considered in more detail.

\subsubsection{Redshift dependency}
\label{subsubsec:redshift}

The redshift dependent part $f(z)$ of the luminosity function in \glref{eq:luminosity_function} is the source evolution density: it describes, how the sources evolved with the age of the Universe.
In general, stellar mass BBH mergers are expected to follow the star formation rate (SFR), as their evolution is primarily linked to that of stars, being the remnants of the most massive stars (see Sec.~\ref{subsec:Merging_BBH}).
We note that some stellar mass binary black hole merging models allow a decoupling of their evolution from the SFR, due to an elongated binary stage and delayed merger in isolated binaries \citep[e.g.][for a review]{2022ApJ...931...17V}.
However, as we focus in this work on stellar mass BBH mergers in starburst galaxies, which evolve from multiple star systems and are non-isolated as described in Sec.~\ref{subsec:Merging_BBH}, we apply the SFR for their evolution, following \citet*{2009PhRvD..79h3009A}:
\begin{equation}
	f_\text{SFR}(z) 
	= \left\{   
	\begin{array}{ll}
    	{(1+z)^{3.4}} 
    	& {z<1} \,, \\
    	{2^{3.7}} \cdot (1+z)^{-0.3} 
    	& {1<z\leq 4 \,,} \\
    	{2^{3.7} \cdot 5^{3.2} \cdot (1+z)^{-3.5}} 
    	& {z\geq 4 \,.}
	\end{array} \right.
\end{equation}
As the evolution of SMBBH mergers only slightly follows the SFR, their source evolution density will be approximated at this point.
However, an approximation with AGN source evolution densities from the literature contains several biases:
(i) They  only consider certain wavelengths of the AGN and not all of their spectrum. 
(ii) The AGN source evolution densities only focus on active AGN, the dormant ones are not included.
Dormant AGN usually still have a relativistic jet \citep{1984A&A...130L..13P, 1989A&A...221L...3C}, which is not very powerful ($
\leq 10^{40} \ein{erg}\ein{s}^{-1}$) \citep{2000ApJ...542..186N, 2002A&A...392...53N}.
It is assumed that AGN are about 1 per cent of their lifetime active and dormant for the rest (e.g. \citealt{1993A&A...269..169V}). 
Thus the approximation is done using estimates for the highest and lowest possible values for $f(z)$.
The lower limit, $f_{\rm low}(z)$, was chosen as:
\begin{equation}
	f_{\rm low}(z) =
	\left\{
	\begin{array}{ll}{(1+z)^{2.5}} 
    	& {z < 2 \,,} \\
    	{3^{6.5} \cdot (1+z)^{-4}} 
    	& {z \geq 2 \,,}
	\end{array}\right.
\end{equation}
and the upper limit $f_{\rm up}(z)$:
\begin{equation}
	f_{\rm up}(z) =
	\left\{
    	\begin{array}{ll}{(1+z)^{4}} 
    	& {z < 3 \,,} \\
    	{4^{6} \cdot (1+z)^{-2}} 
    	& {z \geq 3 \,.}
	\end{array}\right.
\end{equation}

\begin{figure}
	\centering
	\includegraphics[width=0.49\textwidth]{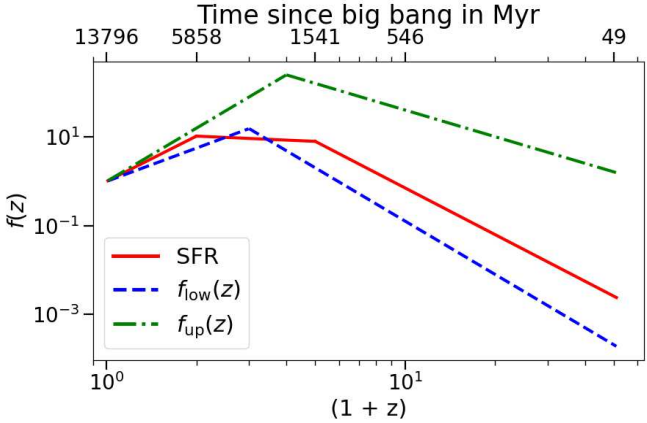}
	\caption{ Plot of the source evolution densities considered. The bottom x-axis shows the redshift in the form $(1 + z)$ for a better visualization. The x-axis on top shows the time since the beginning of the Universe as a reference. The SFR is illustrated in red, $f_{\rm low}(z)$ in blue and $f_{\rm up}(z)$ in green. }
	\label{fig:f_z}
\end{figure}
\begin{figure}
	\centering
	\includegraphics[width=0.49\textwidth]{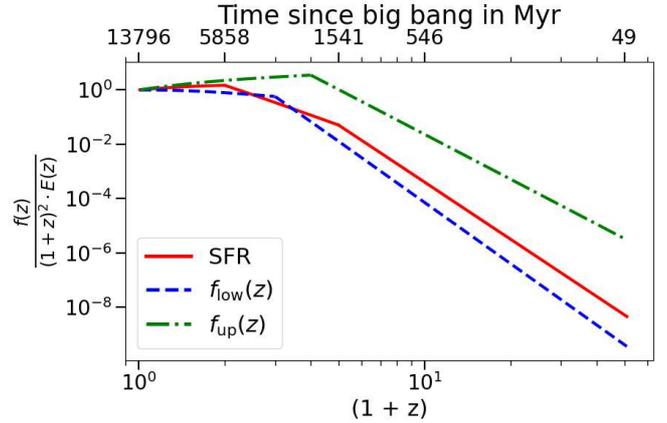}
	\caption{ Plot of the expression in the activity-integral $\xi_z$. The bottom x-axis shows the redshift in the form $(1 + z)$ for a better visualization. The x-axis on top shows the time since the beginning of the Universe as a reference. }
	\label{fig:xi_z}
\end{figure}

The red solid curve in Fig.~\ref{fig:f_z} shows the course of $f_\text{SFR}(z)$ in its dependence of the redshift $z$.
The x-axis on the bottom shows the redshift in form $(1 + z)$, while the top x-axis shows the time since the big bang in Myr. 
Stars and objects that evolved similar to stars like stellar mass BHs or neutron stars reached a density maximum between a redshift of $ 2 $ and $ 3 $ (corresponding to $(1+z) = 3$ and $4$), {as the mass range of stars evolving into BHs increases with lower metallicity \citep[see e.g.][]{2011A&A...528A.149M, 2014ARA&A..52..415M}}. 
This is why the peak of the lower estimate $f_{\rm low}(z)$ was taken as $z = 2$ and the peak at $z = 3$ for the upper estimate $f_{\rm up}(z)$.
The lower estimate is represented as the blue dashed line, while the upper estimate is seen as the green dashed and dotted line.
Additionally, the plot of the expression in the activity-integral $\xi_z$ is seen in \figref{fig:xi_z}.
It has the same x-axes and descriptions of the graphs as \figref{fig:f_z}.
Standing out in this figure is that with higher redshift, the contribution to the activity-integral gets lower.

Calculating the activity-integral $\xi_z$ with the SFR over all redshifts yields $\xi_z \approx 2.4$ for stellar mass BBH mergers.
Integrating the lower and upper limits delivers for SMBBHs:
\begin{equation}
	\xi_z = 2.4^{+8.3}_{-0.5} \,.
	\label{eq:xi_z_error}
\end{equation}

\subsubsection{The Luminosity Function}
\label{subsubsec:luminosity}

The luminosity dependent integral in $\zeta_L$ requires a separate processing of BBHs and SMBBHs due to the different physical origin of the two source classes.

\paragraph{Application to Supermassive Binary Black Holes \label{paragraph:SMBBH} \qquad}
For our work, it is assumed that all observed SMBHs with masses greater than $ 3 \cdot 10^6  \ein{M}_{\sun}$ are formed in SMBBH mergers emitting GWs as described in Sec.~\ref{subsec:Merging_SMBBH}. 
The mass range of the observed SMBHs was specifically chosen to extend from $ \sim 3 \cdot 10^6 \ein{M}_{\sun}$, in order to include the mass of the SMBH in the Milky Way, Sgr A$^*$, up to the mass of the SMBH in the galaxy M87 of $\sim 7 \cdot 10^9  \ein{M}_{\sun}$.
The poor statistics on SMBHs with lower masses \citep{2007ApJ...670...92G} indicates a lower-cut of the mass distribution \citep{2010A&A...521A..55C}, while masses higher than the M87 SMBH are scarce and contribute only negligibly to the distribution. 
We do not consider intermediate mass black holes (IMBHs) in centers of small galaxies with masses between stellar mass BHs and SMBHs at this point due to the uncertainty, if they merge at all. Their small masses in comparison with SMBHs could indicate that they did not merge and thus maintained their jet direction, if they have a jet at all.

\begin{figure}
	\centering
	\includegraphics[width=0.49\textwidth]{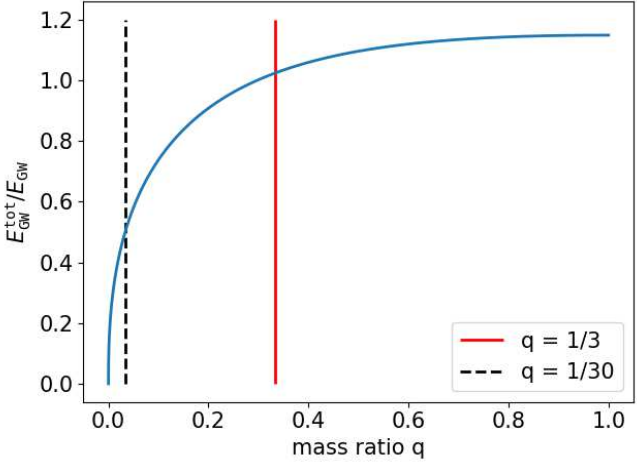}
	\caption{ Representation of the ratio of $E_\text{GW}^\text{tot}$ to $E_\text{GW}$ at different mass ratios. }
	\label{fig:E_GW_tot_E_GW}
\end{figure}

The GW energy radiated by a single SMBBH merger, until the end of the inspiral stage, can be approximated by the difference of the orbital energy at an infinitely large separation and at the separation of the ISCO before the merger \citep{2017AnP...52900209A}.
The former is designated as $E_\text{orb}^i$ and the latter as $E_\text{orb}^\text{ISCO}$.
The orbital energy of a binary object is defined as $E_\text{orb} = - \frac{G m_1 m_2}{2a}$, so that follows:
\begin{equation}
	E_\text{GW}^{\rm insp} = E_\text{orb}^i - E_\text{orb}^\text{ISCO} = 0 - \left(- \frac{G m_1 m_2}{2 \, a}\right)
	=  \frac{G m_1 m_2}{2 \, a} \,.
	\label{eq:E_GW_0}
\end{equation}
The gravitational constant is $G$ and the masses of the SMBHs are $m_1$ and $m_2$ with the mass ratio $q = m_2/m_1 \leq 1$ and $m_1 \geq m_2$, while the separation is parameterized as $a$.
For an equal mass system, so $q = 1$, the separation at the ISCO is approximated by \cite{PhysRevLett.116.241102} as:
\begin{equation}
	a \approx \frac{5 \, G M}{c^2} = 2.5 \cdot r_S \,,
	\label{eq:a_merge_Q105}
\end{equation}
with $M = m_1 + m_2 = m_1 \cdot (1 + q)$ and the Schwarzschild radius $r_S$.
Inserting \glref{eq:a_merge_Q105} into \glref{eq:E_GW_0} delivers
\begin{equation}
	E_\text{GW}^{\rm insp} = \frac{1}{10} \, \mu \, c^2 = \frac{1}{10} \cdot \frac{q}{(1+q)^2} \cdot M \cdot c^2 \,.
\end{equation}
After the ISCO, the binary enters the merger stage, during which additional gravitational waves are emitted.
A detailed general relativistic treatment reveals that approximately the same amount of energy is emitted during this stage as in the previous stages \citep{PhysRevLett.116.241102}, leading to $E_\text{GW}^\text{tot} \approx 2 \cdot E_\text{GW}^{\rm insp}$.
This expression of the GW energy can be approximated by 10 per cent of the chirp mass $\mathcal{M} = (\mu^3 M^2 )^{\frac{1}{5}} = 10 \, k(q) \cdot M$ with $k(q) = 0.1 \cdot \sqrt[5]{\frac{q^3}{(1 + q)^6}} $.
Thus, the GW energy is expressed as:
\begin{equation}
	E_\text{GW} \approx 0.1 \cdot \mathcal{M} \cdot c^2 
	= k(q) \cdot M \cdot c^2 \,.
	\label{eq:E_GW_1}
\end{equation}
The relationship between $E_\text{GW}^\text{tot}$ and $E_\text{GW}$ is visualized in \figref{fig:E_GW_tot_E_GW}.
According to \cite{2009ApJ...697.1621G}, the most common mass ratios for SMBBH mergers lie between $q= 1/30$ and $q= 1/3$. 
These values are drawn in \figref{fig:E_GW_tot_E_GW} as vertical lines.
The red solid line represents the mass ratio $q = 1/3$, which results in $E_\text{GW}^\text{tot}/E_\text{GW} \sim 1$, while the black dashed line represents the mass ratio $q =1/30$, resulting in $E_\text{GW}^\text{tot}/E_\text{GW} \sim 0.5$.

As $E_\text{GW}^\text{tot}$ considers an equal mass system ($q = 1$), it is inaccurate for smaller mass ratios.
That is why the GW energy $E_\text{GW}$ is taken according to \glref{eq:E_GW_1} hereafter.\\

\begin{figure}
	\centering
	\includegraphics[width=0.45\textwidth]{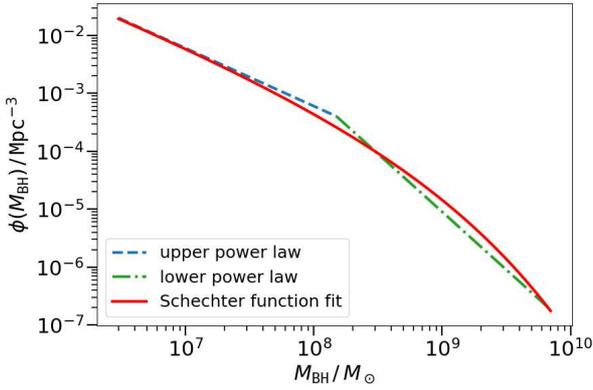}
	\caption{ Integrated mass function of SMBH per $\text{Mpc}^{3}$ up to a distance of $100 \ein{Mpc}$ with a Schechter function fit \citep{2010A&A...521A..55C}. } 
	\label{fig:Schechter_fit}
\end{figure}

Under the assumption that the now detected mass of the merged SMBH, $M_{\rm post}$, is equal to the total mass of the merging SMBHs $M$ minus the energy emitted in gravitational waves, it follows:
\begin{equation}
	M_{\rm post} = M - E_\text{GW} \cdot c^{-2} = M - k(q) \cdot M = (1 - k(q)) \cdot M \,,
	\label{eq:M_now}
\end{equation}
and therefore $M = M_{\rm post} \cdot (1 - k(q))^{-1} $.
This assumption is justified, as GWs, neutrinos and gamma-rays from the merger reach the Earth with a small delay after the merger, so that the observed SMBH would not have much time to grow through accretion.
Inserted into \glref{eq:E_GW_1}, the expression
\begin{equation}
	E_\text{GW} = \frac{k(q)}{(1 - k(q))} \cdot M_{\rm post} \cdot c^2 = h(q) \cdot M_{\rm post} \cdot c^2
	\label{eq:E_GW_2}
\end{equation}
is obtained.
Here, $h(q)$ is the percentage of the detected SMBH mass that was emitted into GWs dependent on the mass ratio $q$.
With this, the luminosity dependent integral $\zeta_L$ in \glref{eq:E3_diffuse_neutrino_flux} delivers 
\begin{equation}
	\zeta_L = h(q) \cdot \int_{L} \ \frac{\td M_{\rm post} \cdot c^2}{\td t} \cdot g(L) \ \td L \,.
\end{equation}
In the following, instead of $M_{\rm post}$, the expression $M$ is used for the SMBH mass after the merger, unless stated otherwise.\\
Luminosity, in general, is defined by the time derivative of the emitted matter or energy: $L = \td E/\td t = \td(M \cdot c^2)/\td t$.
With the assumption that the luminosity of each SMBH inside an AGN is a certain percentage $l_{\rm lum}$ of the Eddington luminosity with $L_\text{Edd} \cong 3.2 \cdot 10^4 \left( {M}/{\text{M}_{\sun}}\right) \ein{L}_{\sun}$, the former can be expressed as
\begin{equation}
	L = l_{\rm lum} \cdot L_\text{Edd} 
	\cong l_{\rm lum} \cdot 3.2 \cdot 10^4 \left( \frac{M}{\text{M}_{\sun}}\right) \ein{L}_{\sun} 
	= a \cdot M \,,
\end{equation}
with the constant $a$.
Describing $g(L)$ as a Schechter function $n(L)$ divided by the luminosity $L$, the integral over the luminosity can then be converted into an integral over the observed SMBH mass:
\begin{align}
	\int_{L} \ \frac{\td(M \cdot c^2)}{\td t} \cdot g(L) \ \td L 
		= \int_{M} \ \frac{\td(M \cdot c^2)}{\td t} \cdot g(M) \ {\td}M \,,
\end{align}
with the Schechter function in the form \citep[see e.g.][]{1976ApJ...203..297S, 2004MNRAS.354.1020S}
\begin{equation}
	n(M) =  \rho_0 
	\cdot \left( \frac{M}{M_\star}\right)^{-\alpha} 
	\cdot \exp\left[ - \left( \frac{M}{M_\star} \right)^{\beta} \right] \,,
	\label{eq:Schechter_M}
\end{equation}
and $g(M) = \frac{n(M)}{M}$, so that the normalization constant for the SMBH distribution $\rho_0$ has the unit of density.
It follows that the luminosity dependent integral $\zeta_L$ in \glref{eq:E3_diffuse_neutrino_flux} can be rewritten as purely mass dependent:
\begin{equation}
	\zeta_L = \zeta_M = h(q) \cdot \int_{M} \ \frac{\td M_{\rm post} \cdot c^2}{\td t} \cdot g(M) \ \td M \,.
	\label{eq:zeta_L_M}
\end{equation}
This integral can be determined using the mass function of SMBHs, which is described by \cite{2010A&A...521A..55C}:
\begin{equation}
	\phi_{\rm low} = 6 \cdot 10^{-3 \pm 0.4}\left(\frac{M}{10^{7} \ein{M}_{{\sun}}}\right)^{-1.0} \ein{Mpc}^{-3} 
	\label{eq:unMasse}
\end{equation}
for masses between $3 \cdot 10^6  \ein{M}_{\sun}$ and $1.5 \cdot 10^8 \ein{M}_{\sun} $ and
\begin{equation}
	\phi_{\rm upp} = 9 \cdot 10^{-4 \pm 0.4}\left(\frac{M }{10^{8} \ein{M}_{{\sun}}}\right)^{-2.0} \ein{Mpc}^{-3} 
	\label{eq:obMasse}
\end{equation}
for masses larger than $1.5 \cdot 10^8 \ein{M}_{\sun}$.
A Schechter function fit (red line) is carried out on these data, which can be seen in \figref{fig:Schechter_fit}.
The parameters for the fit are presented in \tbref{tab:Schechter_parameter}.
The fit underestimates the SMBH density at about $10^8 \ein{M}_{\sun}$ and overestimates it in the range of $10^9 \ein{M}_{\sun}$.
Assuming these over- and underestimates cancel each other mostly out, the fit is evaluated as suitable to describe the course of the SMBH density.
The mass integral of the Schechter function results in a SMBH mass density of $\int_{M} n(M) \ {\td}M = 2.69 \cdot 10^{5 \pm 0.4} \ein{M}_{\sun} \ein{Mpc}^{-3}$ in the mass range considered.
\begin{table}
	\centering
	\caption{ Parameter for the Schechter function \glref{eq:Schechter_M} in \figref{fig:Schechter_fit}, which was fitted to \glref{eq:unMasse} and \glref{eq:obMasse}. }
	\label{tab:Schechter_parameter}
	\begin{tabular}{ c|c}
	    \hline
		parameter& value\\
		\hline
		$\rho_0$& $ 7.00 \cdot 10^{-4\pm0.4} \ \text{Mpc}^{-3}$\\
		$M_\star$& $ 1.50 \cdot 10^8 \ein{M}_{\sun}$ \\
		$\alpha$& $ 0.90 $\\
		$\beta$& $0.41 $\\
		\hline
	\end{tabular}
\end{table}

\paragraph{Application to stellar mass Binary Black Holes \qquad}
\label{paragraph:BBH}
Stellar mass BBHs require a different approach than SMBBHs, due to their distinct formation history and position inside a galaxy.
As we focus in this work on BBH mergers in starburst galaxies, all the interactions addressed in Sec.~\ref{subsec:Merging_BBH} along with their probabilities as well as formation rates and star multiplicities have to be considered and averaged over several starburst galaxies, in order to get a numerically meaningful stellar BH mass function.

However, in contrast to SMBBH mergers, BBH mergers were already detected by the LIGO and Virgo Collaborations, so that at this point, the rates inferred can be used.
But first, the physical interpretation and the unit of the luminosity dependent integral $\zeta_L$ has to be discussed.
The units of the luminosity dependent part $g(L)$ are chosen to be the same as the units of $\td n/(\td V \, \td L)^{-1}$: $\text{s} \ein{Mpc}^{-3} \ein{eV}^{-1}$ with $\text{eV}$ the unit for energy and $\text{Mpc}^{-3}$ the unit for density.
Consequently the redshift dependent part $f(z)$ has no physical dimensions. 
The integral over the luminosity provides the unit $\text{erg} \ein{s}^{-1}$ and can be converted into $\text{eV} \ein{s}^{-1}$, while the time derivation of the gravitational wave energy has the unit $\text{eV} \ein{s}^{-1}$ as an example.
This means that the units of the integral are: $\text{eV} \ein{Mpc}^{-3} \ein{s}^{-1}$.
To put it in other words, the luminosity dependent integral describes the radiated gravitational wave energy per time element of the considered sources in a certain volume.
Therefore, it can be rewritten as a product of the mean radiated gravitational wave energy $\langle E_\text{GW}\rangle$ and the rate $R$ of these events:
\begin{equation}
	\zeta_L 
	= \int_{L} \ \frac{{\td}E_\text{GW}}{{\td}t} \cdot g(L) \ {\td}L
	= \langle E_\text{GW}\rangle \cdot R \,. 
	\label{eq:zeta_L}
\end{equation}
The mean radiated gravitational wave energy during a stellar mass binary black hole merger can be calculated from the gravitational wave transient catalogues GWTC-1 to GWTC-3, while their rates are specified in the analysis of the observations (\citealt{2021ApJ...913L...7A}, consistent with recent findings in \citealt{2021arXiv211103634T}), so that this integral $\zeta_L$ can be determined from detections alone.

\subsection{Comparison with the measured diffuse neutrino flux}
\label{subsec:Compasison_diffuse_flux}
Applying the results from Sections \ref{subsubsec:redshift} and \ref{subsubsec:luminosity}, \glref{eq:E3_diffuse_neutrino_flux} can be rearranged and expressed as
\begin{equation}
	f^\nu_\text{BBH} 
	= \frac{E^{p} \Phi(E_{\nu})}{ \kappa_p \cdot c \cdot t_\text{H} \cdot \xi_z} 
	\cdot \frac{ 1 }{\zeta_L} \,.
	\label{eq:f_BBH_comp}
\end{equation}
We now assume that SMBBHs and BBHs make up a fraction $0< \lambda <1$ of the diffuse neutrino flux, such that 
\begin{equation}
    \left. E_\nu^p \Phi(E_{\nu})\right|_{\rm SMBBH} = \lambda \cdot \left. E_\nu^p \Phi(E_\nu)\right|_{\rm obs } \,,  
    \label{eq:lambda_SMBBH}
\end{equation}
and 
\begin{equation}
    \left. E_\nu^p \Phi(E_{\nu})\right|_{\rm BBH} = (1 - \lambda) \cdot \left. E_\nu^p \Phi(E_\nu)\right|_{\rm obs } \,,  
    \label{eq:lambda_BBH}
\end{equation}
with $\Phi_{\rm obs}$ as the diffuse neutrino flux as measured with IceCube.
Here, we compute that the entire diffuse neutrino flux originates from these two sources, with SMBBH mergers making up a fraction $\lambda$ of the flux and stellar mass BBH mergers the rest $(1 - \lambda)$.
A potential re-scaling, under the assumption that additional source types make up some part of the astrophysical diffuse neutrino flux, is still possible by inserting only a part of the flux into the equations above.
However, in this work, only these two source classes are considered. 

The following results are obtained using the diffuse astrophysical starting tracks neutrino flux measured in $10.3$ years by IceCube \citep{TeVPA22}:
\begin{align}
    \left. \Phi(E_\nu)\right|_{\rm obs }
    = \left( 1.68_{-0.22}^{+0.19} \right) 
    \left( \frac{E_\nu}{100 \ein{TeV}} \right)^{ -2.57_{-0.09}^{+0.09} } 
    \notag \\
    \qquad \qquad \qquad \qquad 
    \cdot 3 \cdot 10^{-18} \ein{GeV}^{-1} \mathrm{~cm}^{-2} \mathrm{~s}^{-1} \mathrm{sr}^{-1}
    \,.
    \label{eq:neutrino_flux}
\end{align}
We chose to use the measurement of the muon neutrino flux in combination with the starting tracks analysis method as compared to the previous analysis method for the measured muon neutrino flux in 9.5 years \citep{2019ICRC...36.1017S} because the starting tracks analysis reduces the atmospheric muon background more efficiently, resulting in a more precise measurement of the diffuse astrophysical neutrino flux.
Compared to the high-energy staring event (HESE) \citep{2021PhRvD.104b2002A} and cascade \citep{2020PhRvL.125l1104A} results, the muon neutrino sample is also based on a large-statistic sample and contains data with the largest time span so far (10.3 years).
It should be noted that the different detection channels result in somewhat different combinations of normalization and spectral index \citep{TeVPA22, 2021PhRvD.104b2002A}.
Here, we chose the all-flavour neutrino flux, thus the factor $3$ appears in \glref{eq:neutrino_flux}.
The high-energy neutrino fluxes measured with 9.5 years of muon tracks data, the HESE data and cascade data are applied to the comparison of the combined results and shown in the Appendix.

It is assumed that the spectrum - measured in the range of $1 \ein{TeV}$ to $10 \ein{PeV}$ -  can be extended to the energy range from $E_{\min} = 100 \ein{GeV}$ to $E_{\max} = 100 \ein{PeV}$, corresponding the approximate lower energy threshold for proton-proton interactions and the maximum energy of the accelerator, respectively.

\subsubsection{Application to Supermassive Binary Black Holes}
\label{subsubsec:SMBBH}
For supermassive binary black hole mergers, \glref{eq:zeta_L_M} is inserted in the expression in \glref{eq:f_BBH_comp}, which can be rearranged to describe the ratio of energy going into neutrino production from gravitational waves in SMBBH mergers: 
\newpage
\begin{equation}
	f^\nu_\text{SMBBH} 
	= \lambda \cdot \frac{\left. E_\nu^p \Phi(E_\nu)\right|_{\rm obs }}{ \kappa_p \cdot c \cdot t_\text{H} \cdot h(q) \cdot \xi_z} 
	\cdot \frac{ 1 }{\int_{M} \ \frac{\td(M \cdot c^2)}{\td t} \cdot g(M) \ \td M}\,.
	\label{eq:f_SMBBH}
\end{equation}
The parameter $\lambda$ takes into account that only a fraction of the total diffuse neutrino flux is expected to be produced in SMBBH mergers.
In order to determine $f^\nu_\text{SMBBH}$, a closer view of the time scales is necessary, see Sec.~\ref{sec:time_scales}.

\subsubsection{Application to stellar mass Binary Black Holes}
\label{subsubsec:BBH}

For stellar mass binary black hole mergers, \glref{eq:zeta_L} can be inserted into \glref{eq:f_BBH_comp} to receive the fraction of gravitational wave energy going into neutrino production for these sources in dependence of their detected merging rates:
\begin{equation}
	f^\nu_\text{BBH} = (1 - \lambda) \cdot \frac{\left. E_\nu^p \Phi(E_\nu)\right|_{\rm obs }}{ \kappa_p \cdot c \cdot t_\text{H} \cdot \xi_z} 
	\cdot \frac{ 1 }{\langle E_\text{GW}\rangle} \cdot \frac{1}{R} \,,
	\label{eq:f_BBH}
\end{equation}
with the same parameter $\lambda$ as in \glref{eq:f_SMBBH}.

\section{Inspiral time merger rates}
\label{sec:time_scales}

The term for the fraction $f^\nu_\text{SMBBH}$ of gravitational wave energy that goes into neutrino production during each merger of supermassive binary black holes in \glref{eq:f_SMBBH} contains the integral $\int d(M \cdot c^2)/dt\cdot g(M)\,dM$ (see \glref{eq:zeta_L_M}).
The time derivation and thus this integral are only valid during the inspiral stage, since this is the time during which GWs are radiated away while neutrinos are produced.
By introducing the SMBBH merger rate after the binary entered the inspiral stage, $R_{insp}$, this integral can then be expressed as 
\begin{equation}
	\zeta_M /h(q) =  
	\int_{M} \ \frac{\td(M \cdot c^2)}{\td t} \cdot g(M) \ \text{dM} 
	= {M}_0 \cdot c^2 \cdot R_{insp} \,,
	\label{eq:zeta_M}
\end{equation}
with the constant mass parameter ${M}_0$.
We define this rate, $R_{insp}$, as proportional to the density of SMBHs in the volume observed, $n_\text{SMBH}$, divided by the mean time the binary remains in the inspiral stage $ \bar{t}_{insp} $:
\begin{equation}
	R_{insp} \propto \frac{n_\text{SMBH}}{\bar{t}_{insp}} \,.
	\label{eq:R_neutrino}
\end{equation}
The mean inspiral time, $ \bar{t}_{insp} $, is determined according to:
\begin{equation}
	\bar{t}_{insp} = \frac{\int_M (M \cdot c^2) \cdot g(M) \ {\td}M}{\int_M \frac{M \cdot c^2 \cdot g(M)}{t_{insp}(M)} \ {\td}M} \,,
	\label{eq:t_mean_insp}
\end{equation}
with $g(M) = n(M) \, M^{-1} $ (see \glref{eq:Schechter_M}).
The SMBH density is derived from the SMBH mass distribution presented by \cite{2010A&A...521A..55C} in Sec.~\ref{paragraph:SMBBH}, which includes $2.4 \cdot 10^{4\pm0.4}$ SMBHs in a radius of $100 \ein{Mpc}$.
This radius corresponds to a volume of $V = 4.19 \cdot 10^{6} \ein{Mpc}^3$, thus the density is $n_\text{SMBH} = 2.4 \cdot 10^{4\pm0.4}/(4.19 \cdot 10^{6}) \ein{Mpc}^{-3} = 5.73 \cdot 10^{-3\pm0.4} \ein{Mpc}^{-3}$ in this volume or $5.73 \cdot 10^{6\pm0.4} \ein{Gpc}^{-3}$ extrapolated to a cubic Gpc.
This value for the SMBH density is taken for each calculation hereafter. \\
Following the assumptions made in Sec.~\ref{subsubsec:redshift} that observed AGN make up only about 1 per cent of all AGN (and thus SMBHs), with the rest being dormant, we assume further that AGN are active due to an ongoing SMBBH merger event in the inspiral stage accompanied by a change of jet direction at its core.
This means that only 1 per cent of SMBHs are currently in the inspiral stage, resulting in:
\begin{equation}
	R_{insp} = \frac{0.01 \cdot n_\text{SMBH}}{\bar{t}_{insp}} \,.
	\label{eq:R_insp}
\end{equation}
These assumptions include that the time the AGN remain active after the prior SMBBH merger is short compared to the time they remain in the inspiral stage.

The connection in \glref{eq:zeta_M} is used to calculate the presented rate $R_{insp}$ and adjust \glref{eq:f_SMBBH} into 
\begin{equation}
	f^\nu_\text{SMBBH} 
	= \lambda \cdot \frac{\left. E_\nu^p \Phi(E_\nu)\right|_{\rm obs }}{ \kappa_p \cdot c \cdot t_\text{H} \cdot h(q) \cdot \xi_z} 
	\cdot \frac{ 1 }{{M}_0 \cdot c^2} \cdot \frac{1}{R_{insp}} \,.
	\label{eq:f_SMBBH_2}
\end{equation}
Based on the assumption made that most neutrinos detected as the diffuse astrophysical flux by IceCube are produced during the inspiral stage of SMBBH mergers, this SMBH inspiral merger rate reflects the energy fraction, which neutrinos get from the gravitational wave energy.

\subsection{Inspiral stage time scales}
\label{subsec:merging_time_scales}

\begin{figure}
	\centering
	\includegraphics[width=0.49\textwidth]{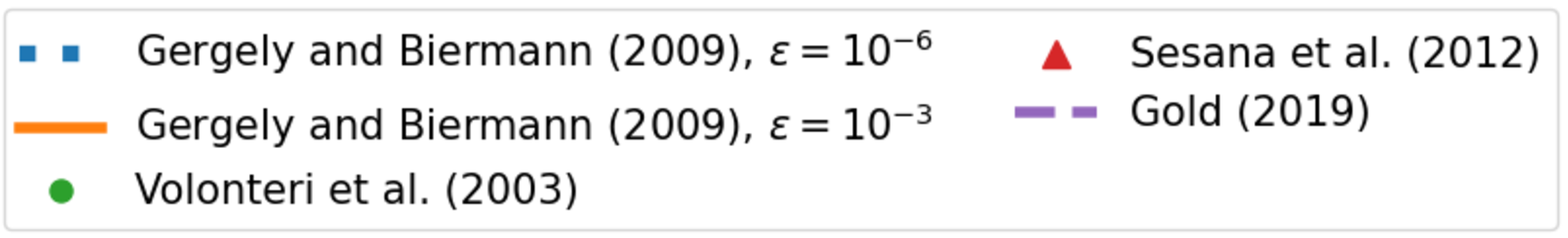}
	\vspace*{-7pt}
	\subfloat[$ q = 1/3$]{%
		\includegraphics[width=0.45\textwidth]{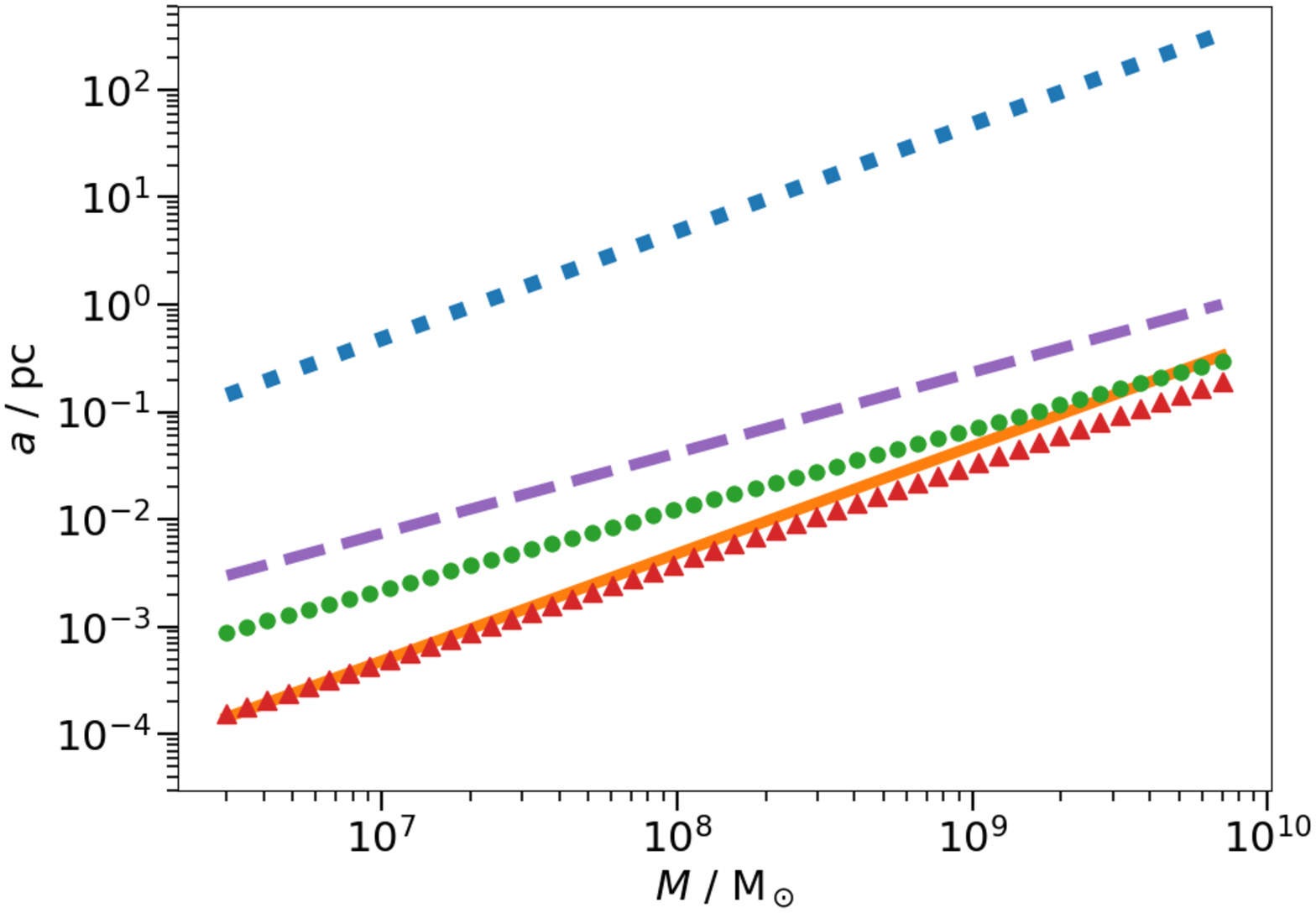}
		\label{fig:a_fin_parsec_a}}
	\hfill
	\subfloat[$ q = 1/30$]{%
		\includegraphics[width=0.45\textwidth]{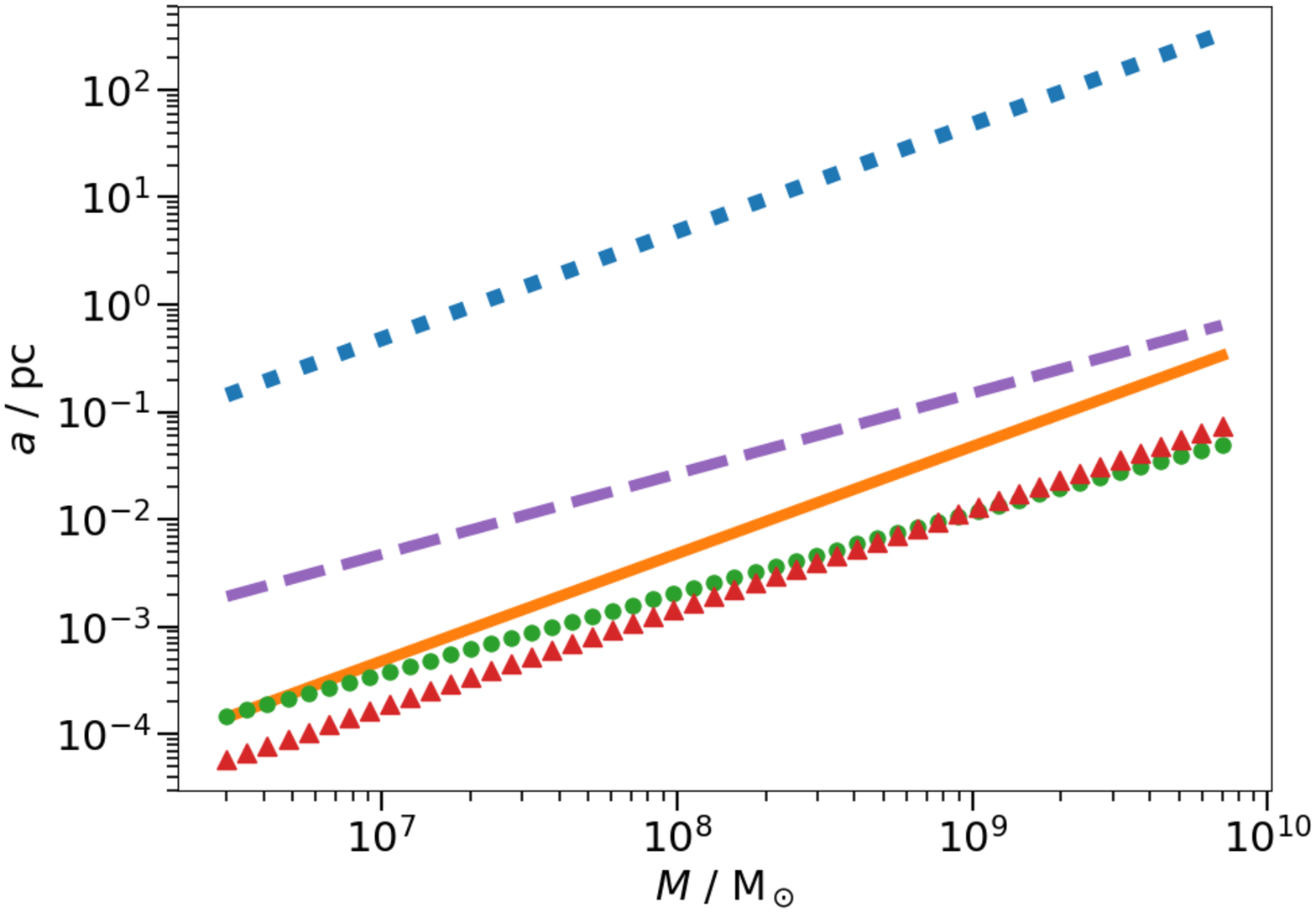}
		\label{fig:a_fin_parsec_b}}
	\caption{ Different models for the mass dependent separation of the binary SMBH at the end of the final parsec stage. The separation at the beginning of this stage in blue squares is listed as a comparison.	}
	\label{fig:a_fin_parsec}
\end{figure}

The SMBBH merger phase can be divided into two main stages: the merging of the host galaxies with its time scale $t_{gal}$ and the merging of the SMBHs after the prior galaxy merger $t_\text{SMBH}$. 
The former time scale is highly uncertain, as it depends on the exact location of the galaxy in the cluster, and will be addressed in Sec.~\ref{subsec:estimation_SMBBH_mergers}.
The latter, $t_\text{SMBH}$, is of particular interest for this work, as it contains the relevant time scale for neutrino emission, the inspiral stage with the redirection of the rotation axis due to the merger and, connected to it, the rearrangement of jets in the system.
As addressed in Sec.~\ref{subsec:Merging_SMBBH}, the merging mechanism of SMBBHs after the initial host galaxy merger can be described in 4 stages: (1) the dynamical friction stage, (2) the final parsec stage, (3) the inspiral stage, in which neutrinos are produced and (4) the actual merger stage.
The time scales for overcoming the stages (1), (2) and (4) are given in Sec.~\ref{subsec:Merging_SMBBH}, while several models for (3), the inspiral stage, are investigated in more detail in this section.\\

Since neutrinos are mainly produced during the inspiral stage, the inspiral time $t_{insp}$ is of great importance, because it defines how long and thus how many neutrinos are produced.
For that, the starting separation of this stage as well as its end, the innermost stable orbit before the merger, are estimated as accurately as possible.
Figure~\ref{fig:a_fin_parsec} shows the separation of the binary at the beginning of the final parsec stage in comparison to the separation at the beginning of the inspiral stage.
The blue squares describe the distance of the SMBHs at the start of the final parsec stage.
They were determined using 
\begin{equation}
	a = \frac{G M}{c^2 \varepsilon} \,,
	\label{eq:Q9_G_B}
\end{equation}
given by \cite{2009ApJ...697.1621G}.
The total mass of the binary is $M = m_1 + m_2$, while the expression $\varepsilon$ describes the post-Newtonian (PN) parameter and is defined as $\varepsilon \approx \upsilon^2/c^2$ with $\upsilon$ being the orbital velocity of the reduced mass of the SMBHs.
The PN parameter gets larger, the closer $\upsilon$ gets to the speed of light.
It approximates the relativistic behavior via a Newtonian approach, hence post-Newtonian.
A value of $\varepsilon \approx 10^{-6}$ was derived for the separation at the end of the dynamical friction stage by \cite{2009ApJ...697.1621G} (blue squares).

The value $\varepsilon \approx 10^{-3}$ marks the end of the final parsec stage according to \cite{2009ApJ...697.1621G} (orange, solid line).
As \glref{eq:Q9_G_B} is independent of the mass ratio $q$, the two curves remain unchanged in Figures \ref{fig:a_fin_parsec_a} and \ref{fig:a_fin_parsec_b}, which represent the two mass ratios $q = 1/3$ and $q=1/30$. 
Here, 3 additional approaches were investigated to confirm the derived separations $a_{insp}$. \\
The model by \citet*{2003ApJ...582..559V} (Green dots) marks the beginning of the inspiral stage, if the SMBBH reduces its separation during the final parsec stage by ejecting stars out of the inner orbit of the binary:
\begin{equation}
	a_{insp} 
	= 0.014 \cdot \frac{q}{(1 + q)^2} \cdot \left( \frac{M^3}{10^{21.3} \ein{M}_{\sun}^3}  \right)^{1/4} \ein{pc} \,.
\end{equation}
For $q=1/3$ at smaller SMBH masses, this function is about an order of magnitude larger compared to what is expected at the end of the final parsec stage (orange, solid line) in the \cite{2009ApJ...697.1621G} behavior, but approaches and crosses it at high masses.
With the mass ratio of $q=1/30$, the green dotted curve crosses the orange curve at small SMBH masses and passes below it at higher masses.
The red triangles refer to predictions from \cite{10.1111/j.1365-2966.2011.20097.x}, in which interactions of the binary with both accretion discs leads to the separation at the beginning of the inspiral stage of:
\begin{align}
	& a_{insp}  = 
	0.31 \cdot \alpha_{0.3}^{-4/25} 
	\cdot \left( \frac{\dot{m}_{0.3}}{\epsilon_{0.1}} \right)^{-1/5}
	\cdot \left( \frac{M}{10^8 \ein{M}_{\sun}} \right)^{-2/25}
	\cdot \delta^{7/25}\notag \\
	& \qquad
	\cdot \left( \frac{4 q}{(1 + q^2)} \right)^{11/25}
	\cdot \left( (1-e^2)^{-7/2} \cdot f(e) \right)^{8/25}
	\cdot 10^3 \ r_S \,.
\end{align}
The parameter $\alpha_{0.3} = \alpha/{0.3}$ is the viscosity parameter of the thin circumbinary disk around the BHs, while $\dot{m}_{0.3} = \dot{m}/{0.3}$ and $\epsilon_{0.1} = \epsilon/{0.1}$. 
The Schwarzschild radius is $r_S = \frac{2 G M}{c^2}$ and $\delta$ is dependent on their semi-major axis and has a value of 2.
The factor $\dot{m}$ describes the accretion rate normalized to the accretion in the Eddington limit of the SMBHs: $\dot{m} = \dot{M}/\dot{M}_\text{Edd}$.
The expression $\dot{M}_\text{Edd}$ is linked to the Eddington luminosity $L_\text{Edd}$ via: $\dot{M}_\text{Edd} = L_\text{Edd}/(\epsilon c^2)$.
The parameter $\epsilon$ is the efficiency of the radiation conversion, for which we adopt a reference value of $0.1$.
As all these parameters have to be identified by observations, the values in each of their subscripts are taken as an estimate in this work.
The function $f(e)$ describes the dependency of the orbital eccentricity $e$ and is defined here as: 
\begin{equation}
	f(e) = 
    \left(1+\frac{73}{24} e^{2}+\frac{37}{96} e^{4}\right) \,.
	\label{eq:eccentricity}
\end{equation}
The eccentricity is assumed to shrink faster than the binary separation and to be around 0 by the time of the merger. 
This was observed with the detected gravitational waves from the binary neutron star merger GW170817 and the neutron star with black hole or another neutron star merger GW190425.
Their eccentricities at the time of the merger were measured to be $e \leq 0.024$ and $e \leq 0.048$, respectively \citep*{2020MNRAS.497.1966L}.
For the sake of simplicity, the eccentricity is assumed to be $e = 0$, so that the function takes the value of $1$.

For the above characterized values and $q = 1/3$, this curve is nearly identical with the one by \cite{2009ApJ...697.1621G} (orange, solid line) with $\varepsilon \approx 10^{-3}$.
Only at large masses do the two curves diverge.
However, because it was plotted with uncertain values for the parameters $\alpha$ and $\dot{m}$, the error of this estimate is huge.
For a mass ratio of $1/30$ in \figref{fig:a_fin_parsec_b}, these red triangles lie nearly parallel to the orange curve and mark the lowest separations at the end of the final parsec stage.\\
Lastly, the dashed violet line follows the separation by \cite{2019Galax...7...63G}:
\begin{equation}
	a_{insp} =
	2 \cdot 10^{-3} 
	\cdot f(e)^{ - 7 / 8} 
	\cdot \frac{q^{1 / 4}}{\sqrt{1+q}}
	\cdot \left(\frac{M}{10^{6} \ein{M}_{{\sun}}}\right)^{3 / 4} \ein{pc} \,.
\end{equation}
The eccentricity is also assumed to be $0$ at this point, so that $f(0) = 1$.
As can be seen in \figref{fig:a_fin_parsec_a}, this graph overestimates the separation at the innermost stable orbit when compared to the end of the final parsec stage in the \cite{2009ApJ...697.1621G} approach (orange, solid line) by a mean factor of $10$.
However, with a lower mass ratio as illustrated in \figref{fig:a_fin_parsec_b}, the curve approaches the reference by \cite{2009ApJ...697.1621G} at high masses.
As a result, the separation according to \cite{2009ApJ...697.1621G} is evaluated as a good estimate for the beginning of the inspiral stage, as it is for the most part in agreement with the other models and stays constant with varying mass ratios. 
Only the dashed violet curve expected by \cite{2019Galax...7...63G} probably overestimates the separation.

\begin{figure}
	\centering
	\includegraphics[width=0.49\textwidth]{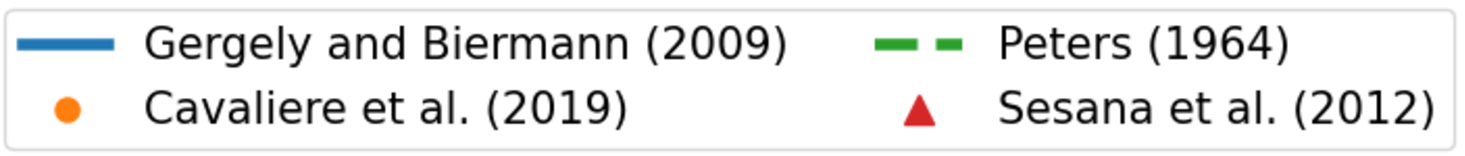}
	\vspace*{-8pt}
	\subfloat[$q = 1/3$]{%
	    \centering
		\includegraphics[width=0.4\textwidth]{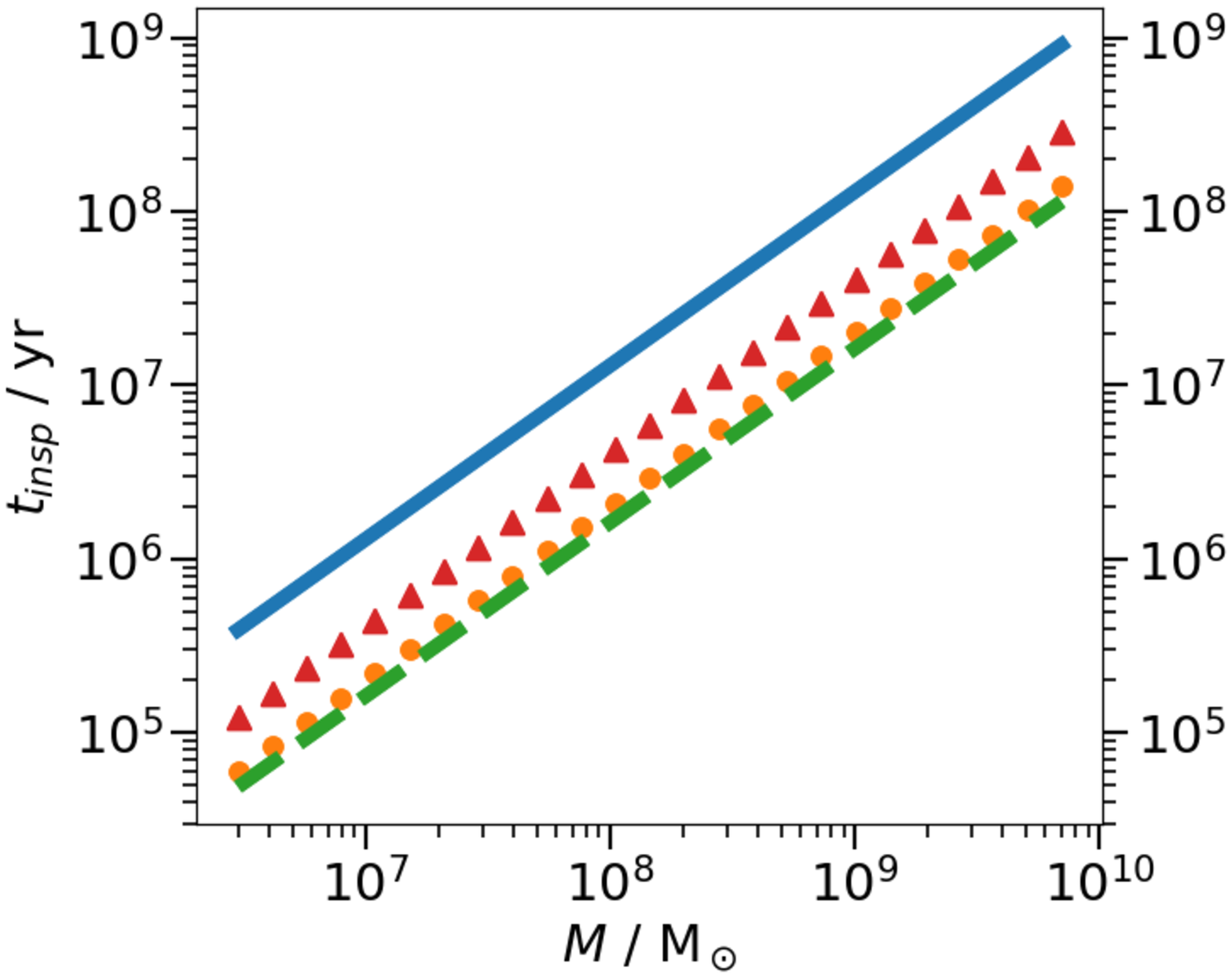}
		\label{fig:t_insp_a}}
	\hfill
	\subfloat[$q = 1/30$]{%
	    \centering
		\includegraphics[width=0.4\textwidth]{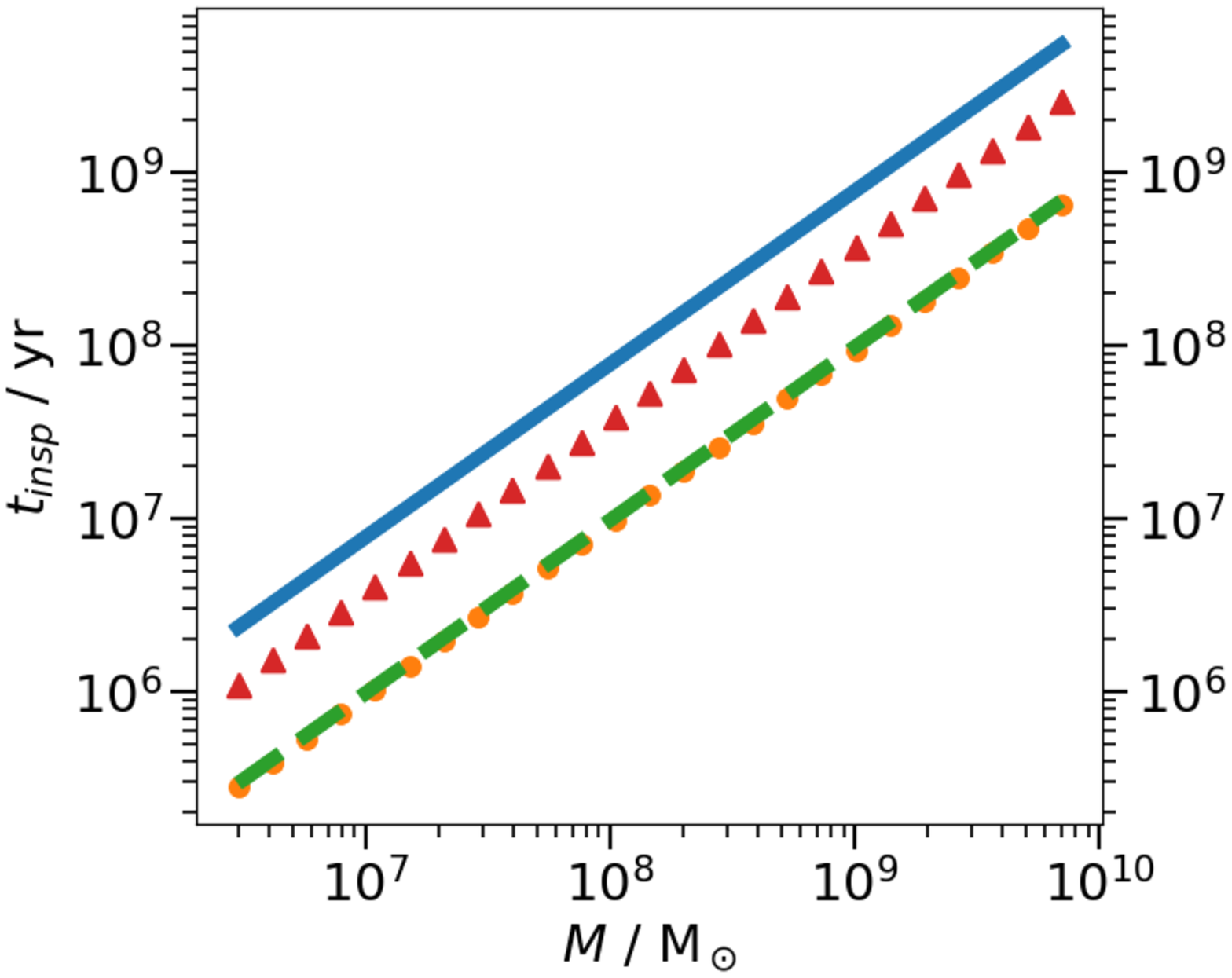}
		\label{fig:t_insp_b}}
	\caption{ Plot of different inspiral times against the total SMBBH mass.}
	\label{fig:t_insp}
\end{figure}

By deriving the separation at the start of the inspiral stage, the inspiral time for the binary system can now be estimated.
For that, 4 models are used to get a high diversity of possible time scales.
The first one, by \cite{2009ApJ...697.1621G}, has the following form:\\
\begin{equation}
	t_{insp, \rm GB}
	\approx (3.9 \cdot 10^7 \ein{yr}) 
	\cdot \left( \frac{M}{3 \cdot 10^8 \ein{M}_{\sun}} \right)
	\cdot \left( \frac{10^{-3}}{\varepsilon} \right)^{4}
	\cdot \left( \frac{(1 + q)^2}{5.33 \cdot q} \right) \,.
	\label{eq:t_Q9}
\end{equation}
It is represented by the blue solid line in \figref{fig:t_insp}.
The model by \cite{2019ApJ...875L..22C} is represented by orange dots:
\begin{align}
	t_{insp, \rm C} =
	(6.0 \cdot 10^6 \ein{yr})
	\cdot \left( \frac{M}{3 \cdot 10^8 \ein{M}_{\sun}} \right)
	\cdot \left( \frac{10^{-3}}{\varepsilon} \right)^{4}
	\cdot \left( \frac{(1 + q)^3}{7.11 \cdot q} \right) \,.
	\label{eq:t_Q73}
\end{align}
The red triangles are derived by \cite{10.1111/j.1365-2966.2011.20097.x} as:
\begin{equation}
	t_{insp, \rm S} = 
	\left( 1.2 \cdot 10^7 \ein{yr} \right) 
	\cdot \left( \frac{M}{3 \cdot 10^8 \ein{M}_{\sun}} \right)
	\cdot \left( \frac{10^{-3}}{\varepsilon} \right)^{4}
	\cdot \left( \frac{1 + q^2}{4 \cdot 0.833 \cdot q} \right) \,. 
	\label{eq:t_Q115}
\end{equation}
Finally, \cite{PhysRev.136.B1224} derived the following connection (dashed green line):
\begin{equation}
	t_{insp, \rm P} 
	\approx (4.9 \cdot 10^6 \ein{yr}) 
	\cdot \left( \frac{M}{3 \cdot 10^8 \ein{M}_{\sun}} \right)
	\cdot \left( \frac{10^{-3}}{\varepsilon} \right)^{4}
	\cdot \left( \frac{(1 + q)^2}{5.33 \cdot q} \right) \,,
	\label{eq:t_Q95}
\end{equation}
which is about a factor 8 smaller than $t_{insp, \rm GB}$.

All inspiral times presented here only depend on the total SMBBH mass $M$, the mass ratio $q$ and the PN parameter $\varepsilon$ (as the separation dependency was transformed into a PN parameter dependency according to \glref{eq:Q9_G_B}).
Furthermore, the mass dependency is linear.
{The value that stands in the same brackets as the mass ratio $q$ was chosen especially so that inserting $q=1/3$ will result in the value $1$ for the brackets.}
The mass ratio $q = 1/3$ is shown in \figref{fig:t_insp_a} and $q = 1/30$ in \figref{fig:t_insp_b}. The PN-parameter was taken as $\varepsilon = 10^{-3}$, as it marks the beginning of the inspiral stage.

The models by \cite{2019ApJ...875L..22C} and \cite{PhysRev.136.B1224} lie close together with $q = 1/3$ and are even nearly identical with the smaller mass ratio of $1/30$.
They predict the lowest inspiral times among the 4 models.
On the other hand, the model by \cite{2009ApJ...697.1621G} predicts the largest inspiral times with both mass ratios used.
The model by \cite{10.1111/j.1365-2966.2011.20097.x} lies in between the other three models.
The inspiral time scales of all models rise about a factor of 5 to 10 by changing the mass ratio from $1/3$ to $1/30$.
All numbers still appear reasonable, as even the inspiral time of the heaviest binary at the lowest mass ratio is with $\sim 6 \cdot 10^9 \ein{yr}$ smaller than the Hubble time (see \figref{fig:t_insp_b}). \\

For a full picture of the necessary time scales, the merger stage and times are investigated in more detail at this point as well: 
As the inspiral stage ends with the ISCO of the two BHs, the estimation for the ISCO in \glref{eq:a_merge_Q105}
is used for their separation.
This equals $\varepsilon = 0.2$ in the PN approach.
By inserting this PN parameter in \glref{eq:t_Q9} to \glref{eq:t_Q95}, the merging time scales are estimated.
Compared with the inspiral time in \figref{fig:t_insp}, the merging time is about a factor $(10^{-3}/0.2)^4 \approx 10^{-9.2}$ smaller than the appropriate inspiral times for the different models.
The heaviest SMBHs with $q=1/30$ take up to 4 years to merge, with lighter ones or smaller mass ratios taking less time.
The merging time scale $t_{merge}$ is thus neglected at this point, as it is by several orders in magnitude lower than the other time scales. \\

\begin{figure}
	\centering
	\includegraphics[width=0.49\textwidth]{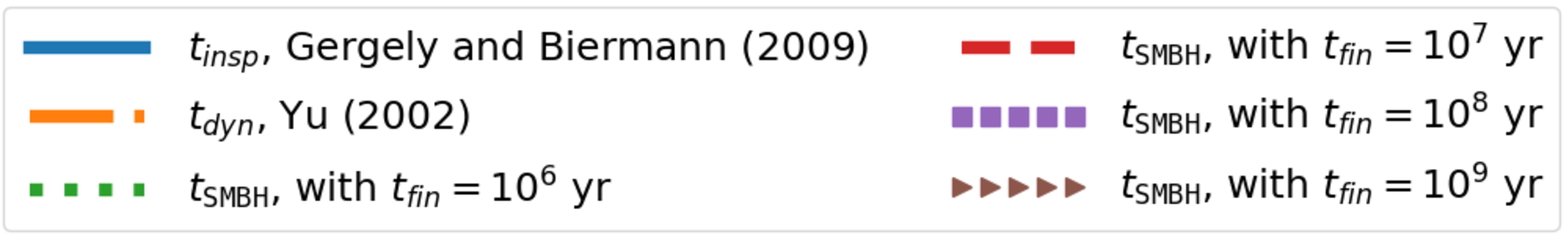}
	\vspace*{-18pt}
	\flushleft
	\subfloat[$q = 1/3$]{%
		\includegraphics[width=0.48\textwidth]{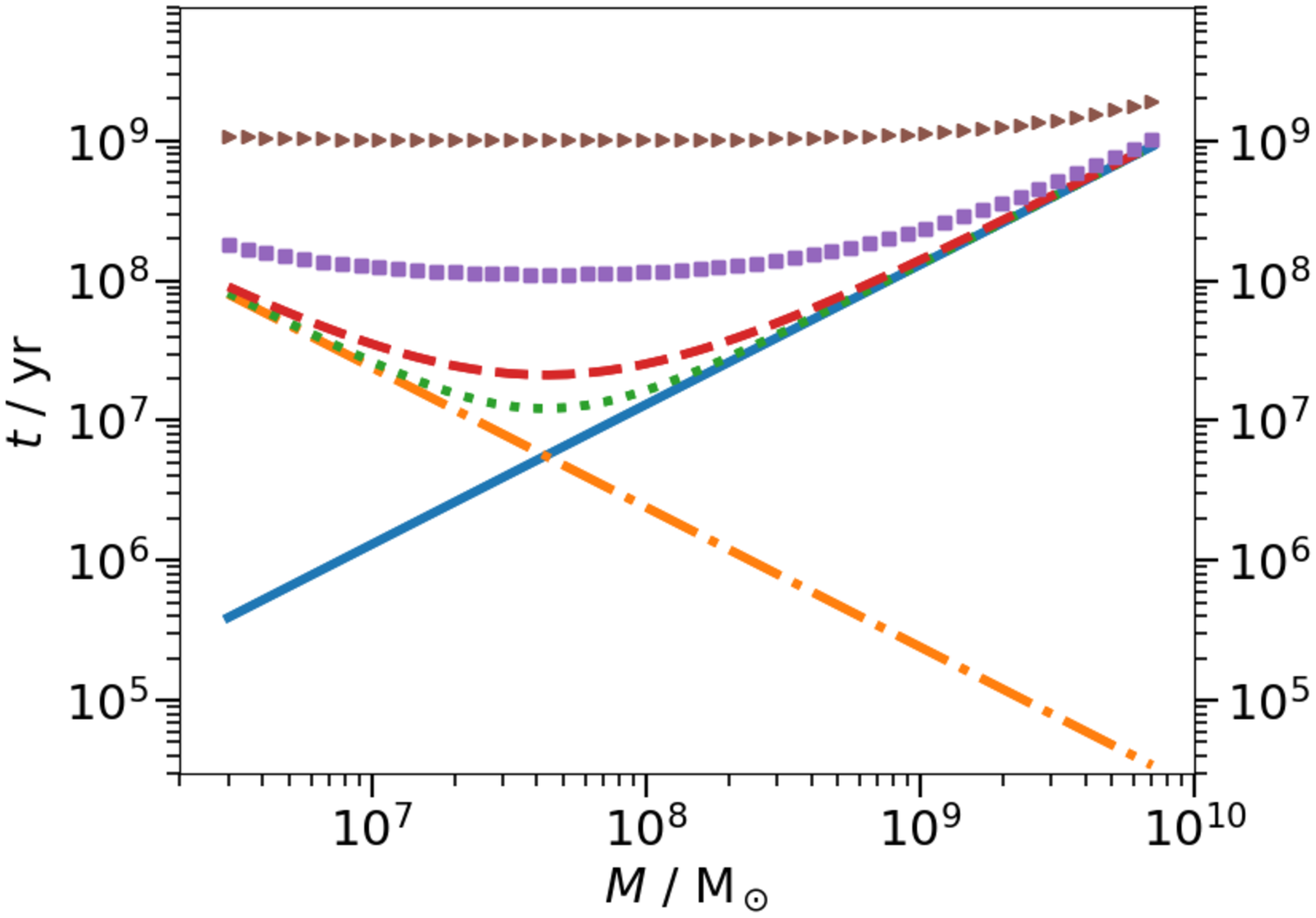}
		\label{fig:t_total_a}}
	\hfill
	\subfloat[$q = 1/30$]{%
		\includegraphics[width=0.48\textwidth]{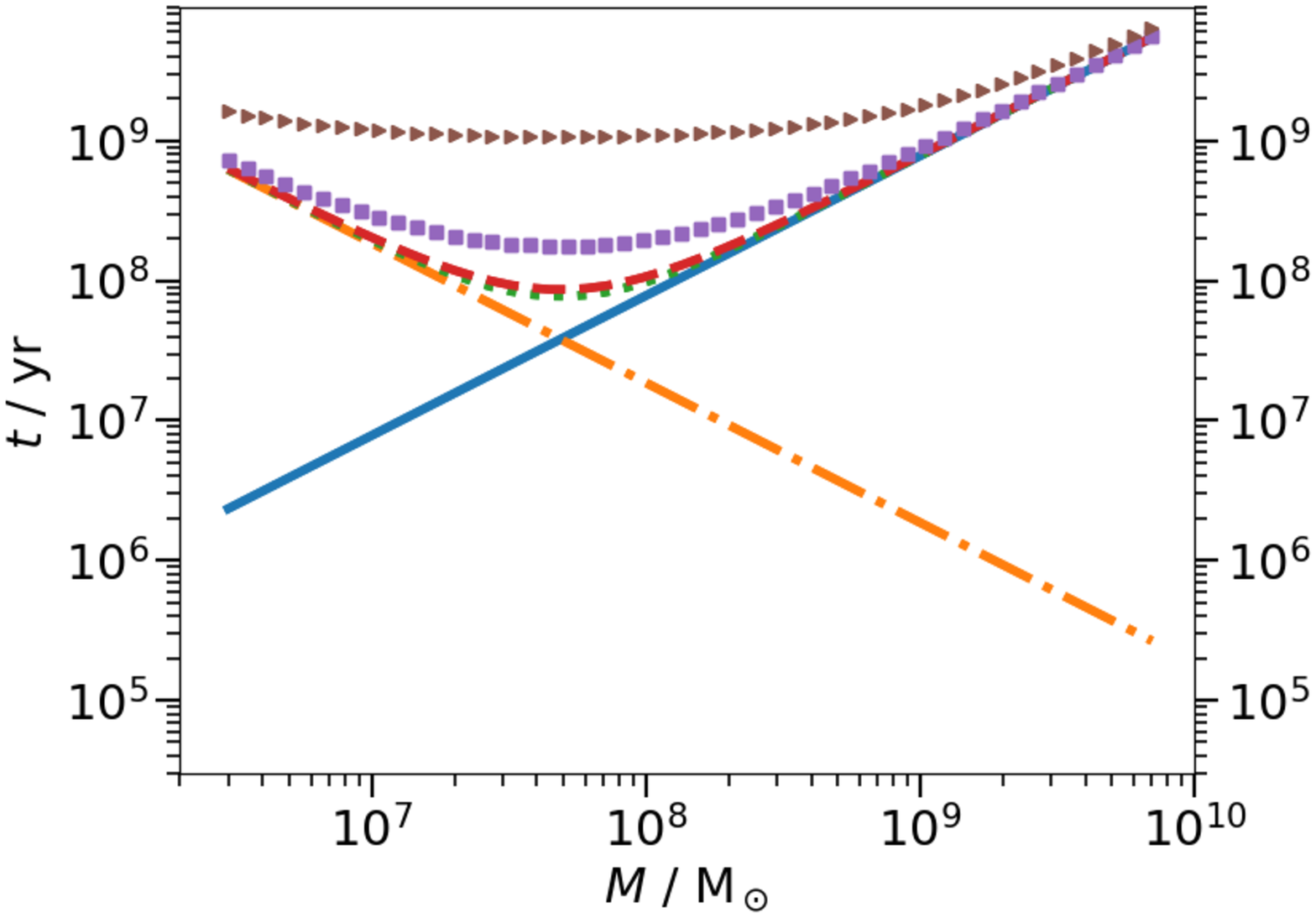}
		\label{fig:t_total_b}}
	\caption{ Estimates for the total merging time of SMBBHs after a prior galaxy merger. The dynamical friction and inspiral times are fixed, while the final parsec time scale is varied.}
	\label{fig:t_total}
\end{figure}

\figref{fig:t_total} combines the time scales discussed.
The blue, solid curve marks the longest inspiral time by \cite{2009ApJ...697.1621G}, while the orange dashed and dotted line shows the estimated dynamical friction time by \cite{2002MNRAS.331..935Y} in \glref{eq:t_dyn} with {values that describe galaxies like M87:} $r_c = 100 \ein{pc}$, $N_* \approx 10^{10}$ and $\sigma_c = 300 \ein{km}\ein{s}^{-1}$ \citep{1980Natur.287..307B}.
The other curves describe the total SMBBH merging time excluding a prior galaxy merger $t_\text{SMBH}$ by considering the time scales mentioned above with different values for the time scale of the final parsec stage of $10^6 \ein{yr}$ (green, dotted line), $10^7 \ein{yr}$ (red, dashed line) and $10^8 \ein{yr}$ (violet squares).
A value of $10^9 \ein{yr}$ was additionally taken for this time (brown triangles), to get a conservative upper estimate. 
In \figref{fig:t_total_a}, the mass ratio $q= 1/3$ was taken and $q=1/30$ was used in \figref{fig:t_total_b}.

The total merging time of SMBBHs of each total mass in the observed mass spectrum lies beneath the Hubble time of $t_H \approx 10^{10} \ein{yr}$. 
Even an overestimate of the final parsec time scale to $1 \ein{Gyr}$ confirms that, theoretically, such merging events can happen in the lifetime of the Universe. 
However, the time scales for galaxy merging are not considered in this representation, due to high uncertainties of it.
That is why these time scales can only be assumed valid after the galaxy merger event.

\subsection{Neutrinos during the Inspiral Stage}
\label{subsec:neutrinos_insp_stage}
Following the analysis from \cite{2018ApJ...853L..29A} that about 10 per cent of arriving UHECRs could be clustered around nearby starburst galaxies, it is assumed in the following that
also 10 per cent of the neutrino producing CRs have their origin in starburst galaxies.
We thus assume that 10 per cent of the diffuse, astrophysical neutrino flux comes also from starburst galaxies, or more precisely from BBH mergers in them, with the rest contributed by other sources, in our case SMBBH mergers. 
This ensures that only neutrinos from these two source classes are considered (see Sec.~\ref{subsec:Compasison_diffuse_flux}). \\

In order to determine the rate $R_{insp}$, the mean inspiral time according to \glref{eq:t_mean_insp} is required. 
It is specified in \tbref{tab:mean_time}, along with the values for the integral of $\zeta_M$ according to \glref{eq:zeta_M}.
All values were determined using the mass ratios $1/3$ and $1/30$.
\begin{table*}
    \renewcommand{\arraystretch}{2}
	\centering
	\caption{Values for the integral of the mass dependent part $\zeta_M$, the mean inspiral time $\bar{t}_{insp}$, the rates $R_{insp}$ and the resulting neutrino energy fraction of GW energy $f^\nu_\text{SMBBH}$ for the used inspiral time models.
	Values for the mass ratios $1/3$ and $1/30$ are presented separately. }
	\label{tab:mean_time}
	\begin{tabular}{ p{2.7cm}|cc|cc|cc|cc} 
		\centering model
		& \multicolumn{2}{p{3.3cm} |}{\centering $\zeta_M/h(q) / 10^{\pm 0.4}$ in $\text{M}_{\sun} \ein{Mpc}^{-3} \ein{yr}^{-1}$}
		& \multicolumn{2}{c}{$\bar{t}_{insp}$ in $\text{yr}$} 
		& \multicolumn{2}{p{2.3cm} |}{\centering $R_{insp}/10^{-2 \pm 0.4}$ in $\text{Gpc}^{-3}
            \ein{yr}^{-1}$} 
		& \multicolumn{2}{c}{$f^\nu_\text{SMBBH}$}
		\\
		\hline
		& $q = 1/3$ 
		& $q = 1/30$ 
		& $q = 1/3$ 
		& $q = 1/30$ 
		& $q = 1/3$ 
		& $q = 1/30$
		& $q = 1/3$ 
		& $q = 1/30$
		\\
		\cline{2-3}
		\cline{4-5}
		\cline{6-7}
		\cline{8-9}
		Gergely and Biermann (2009) 
		& \multirow{2}{*}{ $ 1.44 \cdot 10^{-3} $ }
		& \multirow{2}{*}{ $ 2.40 \cdot 10^{-4} $ }
		& \multirow{2}{*}{ $ 1.86 \cdot 10^6 $ }
		& \multirow{2}{*}{ $1.12 \cdot 10^7$ }
		& \multirow{2}{*}{ $3.08 $ }
		& \multirow{2}{*}{ $ 0.51 $ }
		& \multirow{2}{*}{ $ 1.88_{ - 1.79}^{ + 8.80} \cdot 10^{-6} $ }
		& \multirow{2}{*}{ $ 3.40_{ - 3.23}^{ + 15.89} \cdot 10^{-5} $ }
		\\
		Sesana et al.\ (2012)
		& $ 4.60 \cdot 10^{-3} $
		& $ 5.11 \cdot 10^{-4} $ 
		& $ 5.84 \cdot 10^5 $
		& $5.26 \cdot 10^6$ 
		& $ 9.80$
		& $ 1.09 $
		& $ 5.90_{ - 5.61}^{ + 27.62} \cdot 10^{-7} $
		& $ 1.60_{ - 1.52}^{ + 7.48 } \cdot 10^{-5} $
		\\
		Cavaliere et al.\ (2019)
		& $ 9.34 \cdot 10^{-3} $
		& $ 2.01 \cdot 10^{-3} $ 
		& $ 2.88 \cdot 10^5 $
		& $1.34 \cdot 10^6$
		& $ 19.90$
		& $ 4.27 $
		& $ 2.91_{ - 2.77}^{ + 13.62} \cdot 10^{-7} $
		& $ 4.07_{ - 3.87}^{ + 19.06} \cdot 10^{-6} $
		\\
		Peters (1964)
		& $ 1.16 \cdot 10^{-2}$
		& $ 1.92 \cdot 10^{-3} $
		& $ 2.33 \cdot 10^5  $
		& $1.40 \cdot 10^6$
		& $ 24.61$
		& $ 4.10 $
		& $ 2.35_{ - 2.23}^{ + 11.01} \cdot 10^{-7} $
		& $ 4.24_{ - 4.03}^{ + 19.87} \cdot 10^{-6}  $
	\end{tabular}
\end{table*}
Using these values, the mass parameter $M_0$ is calculated to a value of ${M_0} = 4.70 \cdot 10^7 \ein{M}_{\sun}$ for both mass ratios considered. 
This mass parameter corresponds to the mass, at which these mean times can be read off from the graphs in \figref{fig:t_insp}.

\begin{figure}
	\centering
		\includegraphics[width=0.48\textwidth]{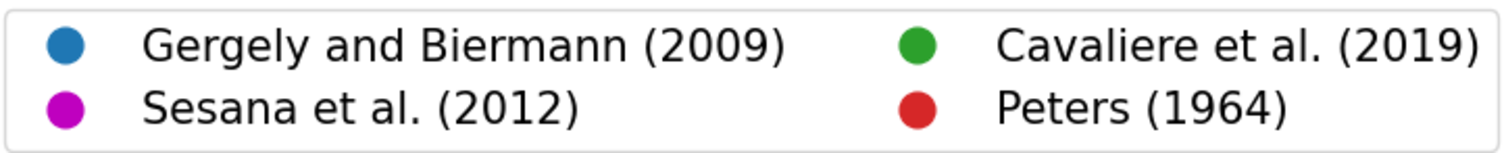}
	\subfloat[$q = 1/3$]{%
		\includegraphics[width=0.45\textwidth]{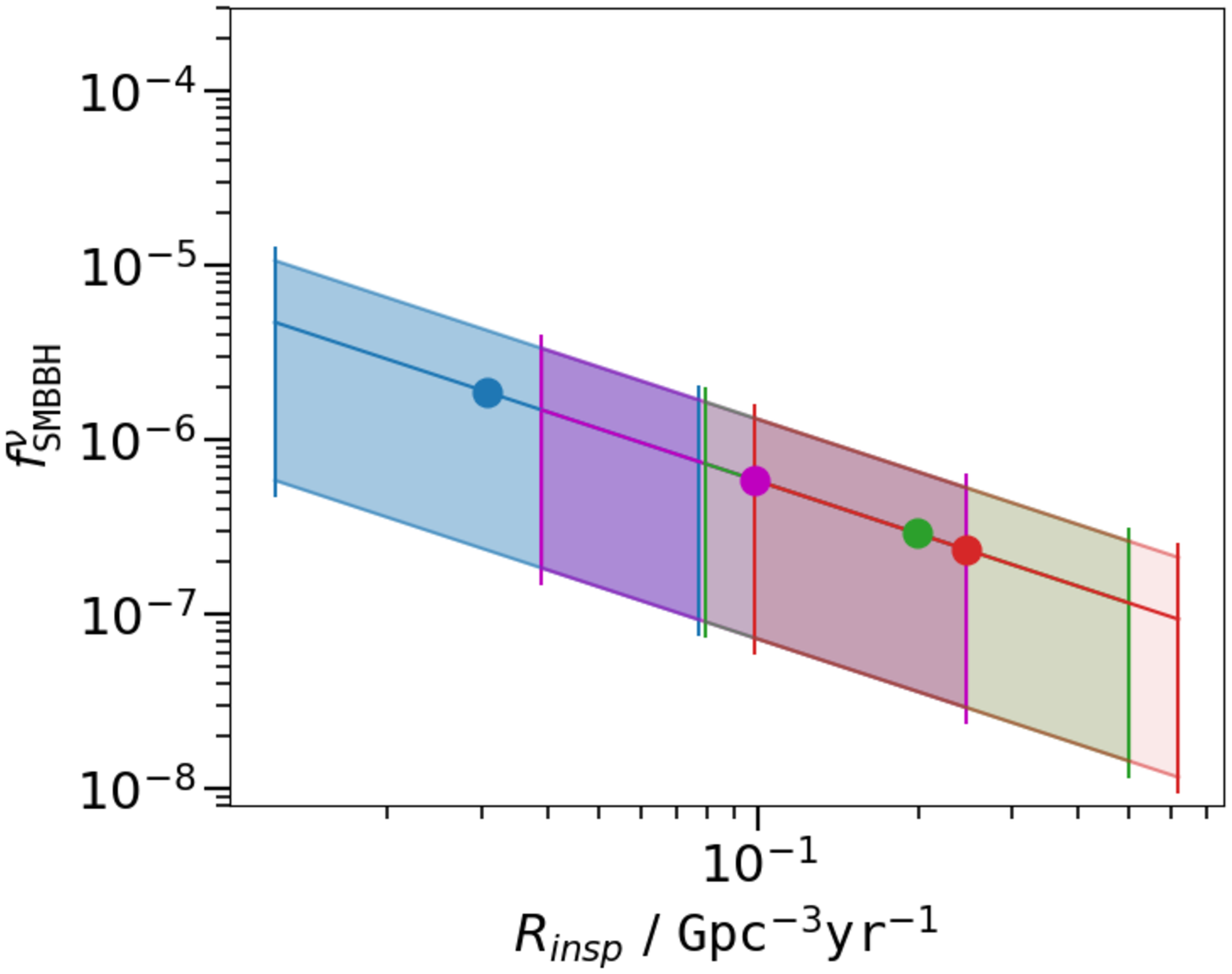}
		\label{fig:f_SMBBH_q_1_3a}}
	\hfill
	\subfloat[$q = 1/30$]{%
		\includegraphics[width=0.45\textwidth]{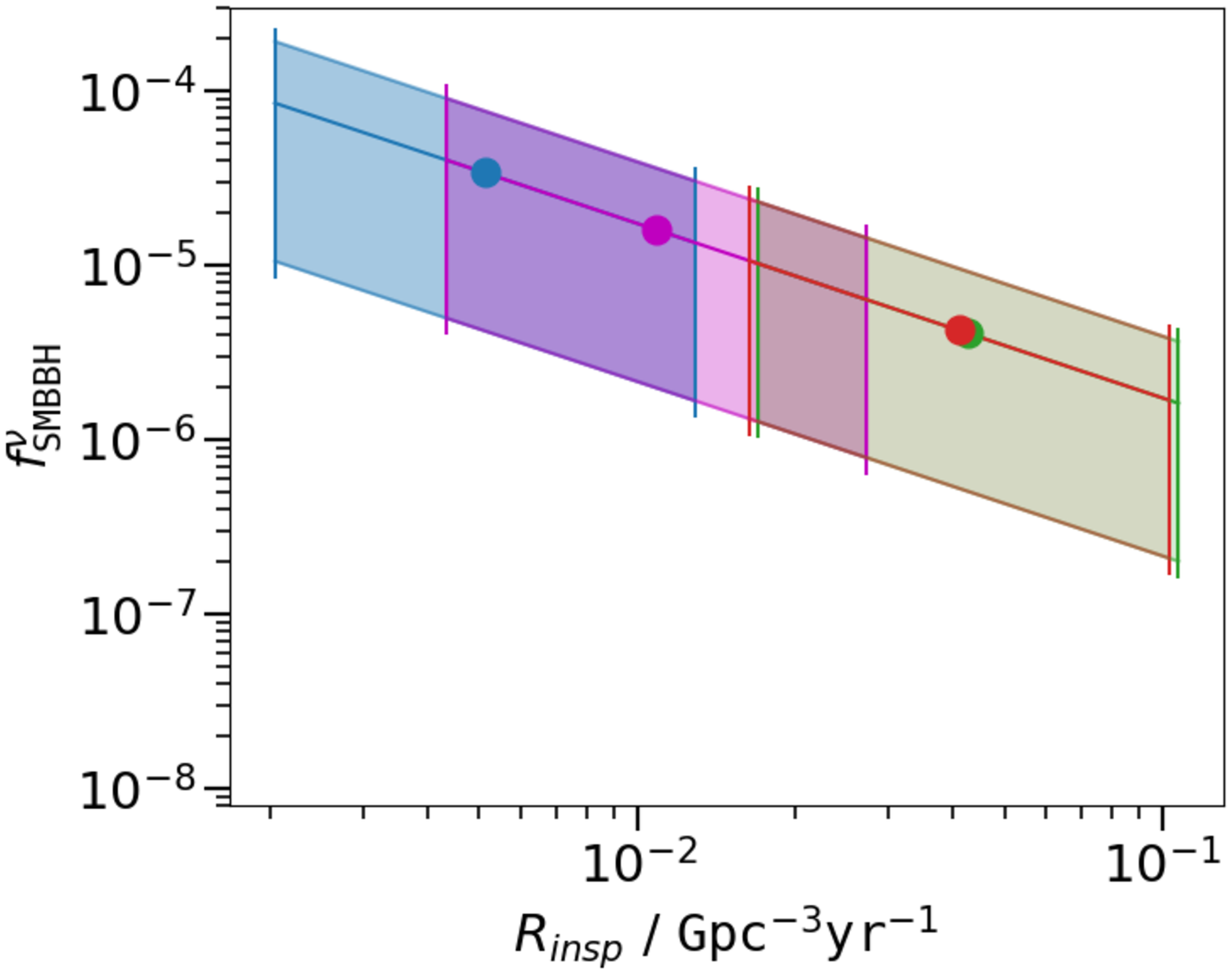}
.		\label{fig:f_SMBBH_q_1_30}}
	\caption{Representation of the fraction of GW energy that goes into neutrinos with an energy in the range between $E_{\min} = 100 \ein{GeV}$ and $E_{\max} = 100 \ein{PeV}$ during each SMBBH merger $f^\nu_\text{SMBBH}$ in respect to the merger rate after the binary entered the inspiral stage $R_{insp}$ for the different inspiral time models. $R_{insp}$ was determined using the mean inspiral times.}
	\label{fig:f_SMBBH_q_1_3}
\end{figure}

The factor $f^\nu_\text{SMBBH}$ determined for each of the models considered is listed in \tbref{tab:mean_time} as well and can be seen in \figref{fig:f_SMBBH_q_1_3}.
In order to understand the figure better, the model for the inspiral time by Gergely and Biermann in \figref{fig:f_SMBBH_q_1_3a} with the mass ratio $q = 1/3$ is discussed in more detail at this point:
The blue point shows the rate $R_{insp} = 3.08 \cdot 10^{-2} \ein{Gpc}^{-3} \ein{yr}^{-1}$ with the diffuse neutrino flux defined in \glref{eq:neutrino_flux} of $\left. E^{2.57} \Phi(E)\right|_{\text {obs }} = 3\cdot 1.68 \cdot (100 \ein{TeV})^{2.57} \cdot 10^{-18} \ein{GeV}^{-1} \ein{cm}^{-2} \ein{s}^{-1} \ein{sr}^{-1}$, a $p$-value of $2.57$ in \glref{eq:B_p}, the redshift dependent parameter $\xi_z = 2.4$ and a corresponding $f^\nu_\text{SMBBH} = 1.88 \cdot 10^{-6}$.
The uncertainty in the rate of $ + \Delta R_{insp} = 10^{+0.4} \ein{Gpc}^{-3} \ein{yr}^{-1}$or $ - \Delta R_{insp} = 10^{-0.4} \ein{Gpc}^{-3} \ein{yr}^{-1}$ results from the error in the SMBH density and thus the number of SMBHs in the volume under consideration.
It is displayed by the solid blue line on which the point is located.
A rate of $R_{insp} = 3.08 \cdot 10^{- 2.4} \ein{Gpc}^{-3} \ein{yr}^{-1} = 1.23 \cdot 10^{-2} \ein{Gpc}^{-3} \ein{yr}^{-1}$ requires a higher value for the parameter $f^\nu_\text{SMBBH}$ due to the anti-proportional relationship. 
The redshift dependent factor $\xi_z$ stays the same for this blue line displayed.
The upper blue line is constructed taking $R_{insp}$ with its error bars, the lower error in $\xi_z$ from \glref{eq:xi_z_error} and the upper limit on the diffuse neutrino flux of $\left. E^{2.66} \Phi(E)\right|_{\text {obs }} = 3\cdot 1.87 \cdot (100 \ein{TeV})^{2.66} \cdot 10^{-18} \ein{GeV}^{-1} \ein{cm}^{-2} \ein{s}^{-1} \ein{sr}^{-1}$ with $p = 2.66$ because of the linear dependence of the parameter $f^\nu_\text{SMBBH}$ on the neutrino flux and its anti-proportional dependence on $\xi_z$. 
Accordingly the lower blue line is constructed taking the lower limit on the diffuse neutrino flux and the upper error of $\xi_z$.
As the error bars of the models overlap, their upper and lower rates are marked by vertical lines in their respective color. 

The model for the inspiral time by Gergely and Biermann yields the longest mean inspiral time for each mass ratio, so that the determined rate with a value of $R_{insp} = 3.08 \cdot 10^{-2 \pm 0.4} \ein{Gpc}^{-3} \ein{yr}^{-1}$ for $q = 1/3$ is the lowest among these models.
The neutrino fraction is with $f^\nu_\text{SMBBH} = 1.88_{ - 1.79}^{ + 8.80} \cdot 10^{-6} $ thus the greatest, as can be seen in \figref{fig:f_SMBBH_q_1_3a}.
These errors are determined using the Gaussian error propagation.
Using a mass ratio of $q=1/3$, the model by Sesana et al.\ delivers an inspiral rate of $R_{insp} = 9.80 \cdot 10^{-2\pm 0.4} \ein{Gpc}^{-3} \ein{yr}^{-1}$ along with an associated $f^\nu_\text{SMBBH} = 5.90_{ - 5.61}^{ + 27.62} \cdot 10^{-7}$ (pink point).
For the same mass ratio, the inspiral rates by Cavaliere et al.\ lie close to the ones of Peters with a value of $R_{insp} = 19.90 \cdot 10^{-2 \pm 0.4} \ein{Gpc}^{-3} \ein{yr}^{-1}$ and $R_{insp} = 24.61 \cdot 10^{-2 \pm 0.4} \ein{Gpc}^{-3} \ein{yr}^{-1}$, respectively.
Due to the asymmetric error in the rate, the rates calculated with the Peters model have the highest value range among the models. 
The corresponding values for $f^\nu_\text{SMBBH}$ are $ 2.91_{ - 2.77}^{ + 13.62} \cdot 10^{-7}$ for the model by Cavaliere et al.\ and $ 2.35_{ - 2.23}^{ + 11.01} \cdot 10^{-7}$ for the model by Peters.
Thus it can be seen, that the $f^\nu_\text{SMBBH}$-value extends over one order of magnitude for these models. 

Switching to the mass ratio $q = 1/30$ in \figref{fig:f_SMBBH_q_1_30}, the mean inspiral times get longer, resulting in lower inspiral rates $R_{insp}$ than with the mass ratio $1/3$.
Comparing \figref{fig:f_SMBBH_q_1_3a} with \ref{fig:f_SMBBH_q_1_30}, the corresponding values for the parameter $f^\nu_\text{SMBBH}$ are about one order of magnitude larger for $q=1/30$ than for $q=1/3$.
This can be seen by comparing the values for $f^\nu_\text{SMBBH}$ at different mass ratios in \tbref{tab:mean_time}.
The largest value for $f^\nu_\text{SMBBH}$ among all models and mass ratios is $3.40_{ - 3.23}^{ + 15.89} \cdot 10^{-5} $ and is obtained for the Gergely and Biermann model using a mass ratio of $1/30$.
This means that with the most optimistic {inspiral time} model {and the mass ratio $1/30$}, the diffuse neutrino flux can be explained via SMBBH merger events {after entering the inspiral stage}, if only a fraction of $10^{-5}$ of the total emitted GW energy is going into neutrino production.
For higher production rates, an even smaller energy fraction is required to go into neutrino production.
With an upper mass ratio of $1/3$, the fraction in the most optimistic model decreases to $10^{-6}$.

We want to validate the order of magnitude of the fraction determined that neutrinos receive from the gravitational wave energy.
For that, we first roughly estimate that about $10^{-2} \text{ to } 10^{-3}$ of GW energy goes into CRs and about 5 per cent of their energies into neutrinos due to charged pion decay (see e.g. \citealt{2008PhR...458..173B} and \citealt{2018PrPNP.102...73A}).
This means that neutrinos detected at Earth receive about $10^{-4} \text{ to } 10^{-5}$ of the GW energy. 
At first glance, this seems that the determined fraction of GW energy that neutrinos receive during each SMBBH merger $f^\nu_\text{SMBBH}$ is about a factor $10^{-1} \text{ to } 10^{-2}$ (considering both mass ratios) smaller than the estimated value. 
However, in our calculations, we made the simplified assumption that 1 per cent of SMBH in the distribution undergo a current merger. 
If this fraction is overestimated and only about 0.1 per cent of them merges currently, the density of detected merging SMBBHs decreases and, due to the anti-proportional relationship of the density to $f^\nu_\text{SMBBH}$, the values for $f^\nu_\text{SMBBH}$ increases by a factor of 10, making them, in general, consistent with the simple estimation above.
On top of that, the possible fraction of gravitational wave energy that goes into CRs is unknown and only roughly estimated here.

Nevertheless, we can estimate the fraction of gravitational wave energy that can go into gamma-rays by investigating the binary neutron star merger GW170817, since this is the only binary merger so far that was detected in both gravitational waves as well as coincident $\gamma$-rays, the gamma-ray Burst GRB 170817A \citep{2017PhRvL.119p1101A}.
Its radiated gravitational wave energy was measured as at least $0.04 \ein{M}_\odot c^2$ \citep{2019PhRvX...9c1040A} {or $7.15 \cdot 10^{52} \ein{erg}$}, while the isotropic energy release in gamma-rays was determined as $E_{\text{iso}, \gamma} = (5.3 \pm 1.0) \cdot 10^{46} \ein{erg}$ for a detailed best fit \citep{2017ApJ...848L..13A}.
For this source, this means that about $(7.42 \pm 1.40) \cdot 10^{-7}$ of the GW energy was converted into gamma-rays during the merger of the binary. 
These gamma-rays can be produced in neutral pion decay due to $p\gamma$ and $pp$ interactions of accelerated CRs in a jet \citep{2018PrPNP.102...73A}.
Consequently, charged pions could also be produced in these interactions, leading to neutrino production.
These neutrino have, at most, half the energy of the gamma-rays. 
Keeping in mind that this estimation is only valid for the binary neutron star merger GW170817 in the case that neutrinos would indeed be produced and in a similar jet as the one from black holes, they would receive about $(3.71 \pm 0.70) \cdot 10^{-7}$ of the GW energy.
This is about a factor of $10^{-2}$ less than what resulted from the rough estimate above, but is partly consistent with the findings in \tbref{tab:mean_time} for $q= 1/3$.

\section{SMBBH merger rates in the Universe}
\label{sec:SMBBH_Merger_Rate}

The inspiral time merger rate $R_{insp}$, determined in Sec.~\ref{sec:time_scales}, describes only the SMBBH merger rate once the binary enters the inspiral stage. 
However, this rate is not specifiable from current or future GW detections from SMBBH mergers, as a possible determinable merger rate includes all time scales from a binaries life, not just the time scale in the inspiral stage. 
That is why we introduce the SMBBH merger detection rate at this point, which is the same as the rate of the total SMBBH merger including all time scales between two binary mergers, as:
\begin{equation}
	R_\text{tot} = 
	\frac{\frac{\bar{t}_{insp}}{\bar{t}_{total}} \cdot n_\text{SMBH}}{\bar{t}_{total}} \,,
	\label{eq:R_SMBH}
\end{equation}
with the total mean time for a SMBBH merger including a prior galaxy merging event $\bar{t}_{total}$. 
Unlike in \glref{eq:R_insp}, this rate is not proportional to 1 per cent of the SMBH density, as, using the inspiral stage models introduced in Sec.~\ref{subsec:merging_time_scales} and an estimation of the total merging time $\bar{t}_{total}$, which will be performed in Sec.~\ref{subsec:estimation_SMBBH_mergers}, we can determine the actual fraction of SMBHs that are currently undergoing a merging event in the inspiral stage and thus are located inside an active AGN, possibly producing neutrinos.
This fraction is captured by ${\bar{t}_{insp}}/{\bar{t}_{total}}$ and will be determined for each of the used inspiral stage models.

The possible SMBBH merger detection rate $R_{\rm tot}$ captures all relevant time scales during a SMBBH merging process and not only the time scale of gravitational wave radiation and neutrino production
That is why a correction is necessary in order to put it in relation to the energy fraction, which neutrinos receive from the gravitational wave energy during each merger.
Since neutrinos and GWs are only emitted in the inspiral stage, the time in the denominator in \glref{eq:R_SMBH} has to be adjusted.
For that, the mean total merging time is multiplied by ${\bar{t}_{insp}}/{\bar{t}_{total}}$, so that the mean inspiral time $\bar{t}_{insp}$ remains in the denominator, changing \glref{eq:R_SMBH} into:
\begin{equation}
	R_\text{tot} \cdot \frac{1}{\frac{\bar{t}_{insp}}{\bar{t}_{total}}}
	= 
	\frac{\frac{\bar{t}_{insp}}{\bar{t}_{total}} \cdot n_\text{SMBH}}{\frac{\bar{t}_{insp}}{\bar{t}_{total}} \cdot \bar{t}_{total}} 
	= 
	\frac{\frac{\bar{t}_{insp}}{\bar{t}_{total}} \cdot n_\text{SMBH}}{\bar{t}_{insp}} 
	= \frac{n_\text{SMBH}}{\bar{t}_{total}} \,.
	\label{eq:R_tot}
\end{equation}
As can be seen, a reducing of the right term leads to the disappearing of the mean inspiral time, although this equation describes the adjusted SMBBH merger rate during the inspiral stage. 
Analogous to \glref{eq:zeta_M}, it follows for the integral in \glref{eq:f_SMBBH}:
\begin{equation}
    \zeta_M /h(q) =  
	\int_{M} \ \frac{\td(M \cdot c^2)}{\td t}  \cdot g(M) \ \td M 
	= {M}_0 \cdot c^2 \cdot \frac{n_\text{SMBH}}{\bar{t}_{total}} \,.
	\label{eq:zeta_M_2}
\end{equation}
The time derivation of the mass on the left side is performed with the inspiral time, just as in \glref{eq:zeta_M}.
The parameter $M_0$ is the same as in \glref{eq:zeta_M}.

As \glref{eq:zeta_M_2} shows, $\zeta_M /h(q)$ is independent of the model for the inspiral stage, but only dependent on the SMBH density and the mean total merging time.
Thus, \glref{eq:f_SMBBH} changes to
\begin{equation}
	f^\nu_\text{SMBBH,eff} 
	= \lambda \cdot \frac{\left. E_\nu^p \Phi(E_\nu)\right|_{\rm obs }}{ \kappa_p \cdot c \cdot t_\text{H} \cdot h(q) \cdot \xi_z} 
	\cdot \frac{ 1 }{{M}_0 \cdot c^2} \cdot \frac{\bar{t}_{total}}{n_\text{SMBH}} \,.
	\label{eq:f_SMBBH_4}
\end{equation}
This means that by estimating a total mean merging time for SMBBHs, the effective fraction of neutrino from GW energy $f^\nu_\text{SMBBH,eff}$ can be determined.

However, as we want to set the mean SMBBH merger detection rate in relation to the fraction, which neutrinos receive from the gravitational wave energy during each merger, we change the left term in \glref{eq:R_tot} for ${n_\text{SMBH}}/{\bar{t}_{total}}$, leading to:
\begin{equation}
	f^\nu_\text{SMBBH,eff} 
	= \lambda \cdot \frac{\left. E_\nu^p \Phi(E_\nu)\right|_{\rm obs }}{ \kappa_p \cdot c \cdot t_\text{H} \cdot h(q) \cdot \xi_z} 
	\cdot \frac{ 1 }{{M}_0 \cdot c^2} \cdot \frac{1}{R_\text{tot}}
	\cdot \frac{\bar{t}_{insp}}{\bar{t}_{total}} \,.
	\label{eq:f_SMBBH_3}
\end{equation}
Note that although at first glance a dependence on the inspiral stage model can be seen in \glref{eq:f_SMBBH_3}, the factor $f^\nu_\text{SMBBH,eff}$ is independent of it, as is seen in \glref{eq:f_SMBBH_4}. 
Only the mean SMBBH merger detection rate depends on the inspiral stage model (see \glref{eq:R_SMBH}).

\subsection{Estimation of the number of SMBBH mergers}
\label{subsec:estimation_SMBBH_mergers}

As the merger time scales of galaxies vary strongly because of their distributions in the cosmological structure, they cannot be specified easily. 
However, as mentioned in Sec.~\ref{sec:introduction}, observational data also indicates that nearly all massive galaxies with a redshift near $0$ have already merged at least once in their lifetime \citep[see e.g.][]{1997ApJ...490..577D, 2001PhDT.......173R, 2003AJ....126.1183C}.
This is why we assume that all observed SMBHs resulted from mergers.
Using this assumption, it can be estimated how many mergers a SMBH must have had in order to grow to the observed mass today.
This is done with an iteration, where the simplifying assumption of a constant mass ratio in each merger event is used.
It is further assumed that SMBHs gain a constant percent of their mass $\eta_{\rm acc}$ via accretion between each merger, i.e.~$M_{\rm acc} = \eta_{\rm acc} \cdot M$.
Additionally, we consider that a percentage of the mass $k(q)$ is emitted via GWs during each merger: $E_{\rm GW}= k(q) \cdot M \cdot c^2$ with $k(q) = 0.1 \cdot \sqrt[5]{\left(\frac{q^3}{(1 + q)^6}\right)}$, (see \glref{eq:E_GW_1}).
The SMBH post-merger mass detected now, $M_{\rm post}$, can then be expressed as follows:
\begin{align}
	M_{\rm post} 
	&= m_1 + m_2 - E_\text{GW}/c^2 + M_{\rm acc} \notag \\
	&= M - k(q) \cdot M + \eta_{\rm acc} \cdot M \notag \\
	&= (1 + \eta_{\rm acc} - k(q)) \cdot (1 + q) \cdot m_1 \,.
	\label{eq:M_now_ganz}
\end{align}
The total mass of the merging SMBHs is $M$ with $M = m_1 + m_2 = (1 + q) \cdot m_1$, with $m_1$ being the heavier SMBH mass and $q = m_2/m_1$ the mass ratio with the lighter SMBH mass $m_2$.
By taking an observed SMBH mass $M_{\rm post}$, the mass of the heavier SMBH can then be estimated according to this relation. 
Repeating this process once delivers the heavier SMBH mass $m^{(2)}_{1}$, which merged with a SMBH of the mass $m^{(2)}_{2}$ to form a SMBH with the mass $m^{(1)}_1 = m_1$.
Based on \glref{eq:M_now_ganz}, this merging process can be expressed as
\begin{equation}
	m^{(1)}_{1} = (1 + \eta_{\rm acc} - k(q)) \cdot (1 + q) \cdot m^{(2)}_1 \,,
\end{equation}
with the mass ratio $q = m^{(1)}_2/m^{(1)}_1 = m^{(2)}_{2}/m^{(2)}_{1}$ held constant.
Inserted in \glref{eq:M_now_ganz}
and repeating this process $i$-times, leads to:
\begin{equation}
	M_{\rm post} = \big[ (1 + \eta_{\rm acc} - k(q)) \cdot (1 + q) \big]^i \cdot m^{(i)}_1 \,.
	\label{eq:M_post_iteration}
\end{equation}
The number of steps is represented by $i$.
This process is repeated for a fixed mass $M_{\rm post}$ until $m^{(i)}_1$ is smaller than the lower mass limit of the mass spectrum considered.
The number of applied steps is then the estimate for the number of mergers required in order to achieve the observed SMBH mass, starting with a mass of nearly $3 \cdot 10^6 \ein{M}_{\sun}$.\\

\begin{figure}
	\centering
	\includegraphics[width=0.49\textwidth]{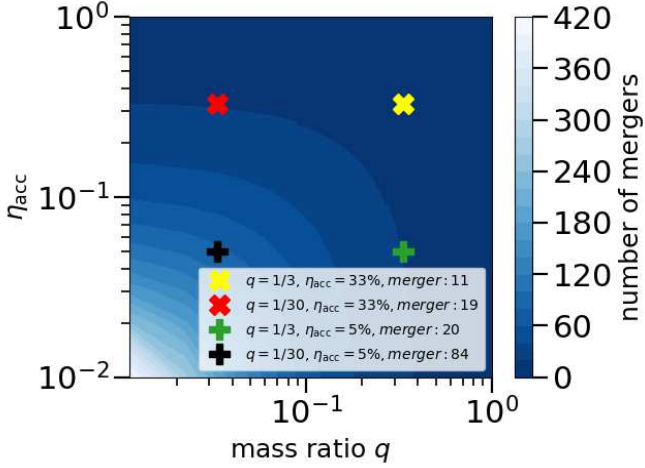}
	\caption{ Number of mergers required to receive a SMBH with a mass of $10^9 \ein{M}_{\sun}$.
	X-axis: constant mass ratio $q$ during each merger.
	Y-Axis: Constant percentage of mass, gained in the accretion between each merger $\eta_{\rm acc}$. The points indicate different combinations of $q$ and $\eta_{\rm acc}$.}
	\label{fig:1_mal_10_9_M}
\end{figure}

For an observed SMBH mass of $10^9 \ein{M}_{\sun}$, this iterative process is shown exemplary in \figref{fig:1_mal_10_9_M}.
The x-axis shows the mass ratio $q$, while the y-axis shows the relative amount of accreted mass between each merger $\eta_{\rm acc}$.
The color bar indicates the number of mergers required in order to reach the observed SMBH mass with varying mass ratios and accretion percentages.
The yellow cross displays the number of mergers with a mass ratio of $q = 1/3$ and an accretion percentage of $\eta_{\rm acc} = 33$ per cent.
The latter was chosen so that the mass increase via accretion, which is captured by the factor $(1 + \eta_{\rm acc})$, is the same as the mass increase via merger at the highest regarded mass ratio $1/3$, which is described by the total mass $M = (1 + q) \cdot m_1$.
With these parameters, 11 SMBBH mergers are required for the resulting SMBH to reach a mass of  $10^9 \ein{M}_{\sun}$, starting with $\sim 3 \cdot 10^6 \ein{M}_{\sun}$.
On the other hand, the red cross shows the value corresponding to a mass ratio of $q = 1/30$ and the same accretion percentage of $\eta_{\rm acc} = 33$ per cent.
A decrease of the mass ratio results in 19 required SMBBH mergers to form a SMBH of this mass.
Lowering the accretion percentage exemplarily to $\eta_{\rm acc} = 5$ per cent between each merger requires 20 SMBBH mergers with a mass ratio of $1/3$ (green plus) and 84 SMBBH mergers with $q=1/30$ (black plus).
The latter clearly overestimates the number of required mergers.
That is why in the following iterations, $\eta_{\rm acc}$ is kept as 33 per cent.\\

\begin{figure}
	\centering
	\includegraphics[width=0.49\textwidth]{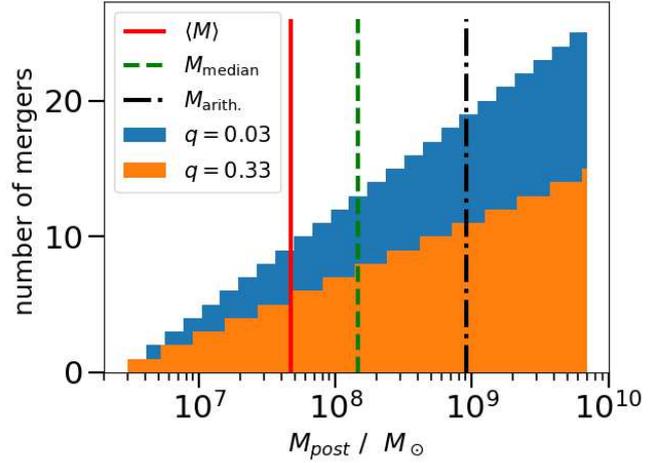}
	\caption{ Number of mergers required to receive the observed SMBH masses with a constant merger mass ratio $q = 1/30$ in blue and $q = 1/3$ in orange.
	The constant percentage of gained mass in the accretion between each merger is set to $\eta_{\rm acc} = 33$ per cent.}
	\label{fig:number_mergers}
\end{figure}

Using the mass ratios $q=1/3$ and $q=1/30$ and the accretion percentage $\eta_{\rm acc} = 33$ per cent, the necessary number of mergers to reach each mass in the mass range between $3 \cdot 10^6 \ein{M}_{\sun}$ and $7 \cdot 10^9 \ein{M}_{\sun}$ is determined. 
This is shown in \figref{fig:number_mergers} with increasing mass steps of $1 \cdot 10^6 \ein{M}_{\sun}$.
The x-axis shows the post-merger SMBH mass in a logarithmic scale, while the y-axis displays the number of mergers.
Starting with a mass of $3 \cdot 10^6 \ein{M}_{\sun}$, over 26 mergers are required for the heaviest SMBH masses of $7 \cdot 10^9 \ein{M}_{\sun}$, if a mass ratio of $1/30$ is taken and only 15 SMBBH mergers in the case of $q=1/3$.
In order to reach a certain SMBH mass, the number of necessary SMBBH mergers decreases with increasing mass ratio.
We introduce the mean SMBH mass of the mass distribution used in order to get a mean number of already happened SMBBH mergers. 
It is determined by dividing the mass density inferred from the Schechter function of the SMBH distribution in Sec.~\ref{paragraph:SMBBH}
by the SMBH density $n_\text{SMBH}$ (see Sec.~\ref{sec:time_scales}) and results in $\langle M\rangle \approx 4.70 \cdot 10^7 \ein{M}_{\sun}.$  
This mean SMBH mass is indicated as the solid red vertical line in \figref{fig:number_mergers} and corresponds to 6 SMBBH mergers with a mass ratio of $q=1/3$ and 9 SMBBH mergers with a mass ratio of $q=1/30$.
Due to the representation in this figure, the exact number of mergers cannot be read off easily.
However, a separate iteration similar to \figref{fig:1_mal_10_9_M} delivers these values.

In order to compare different numbers of SMBBH mergers, two particular masses are viewed at as well:
The median mass and the arithmetic mean mass.
The former takes all masses equally into account and thus slightly over-weights heavier masses, since they are not as common as lighter masses.
This mass is represented by the green dashed line and has a value of $M_{\rm median} \approx 1.45 \cdot 10^8 \ein{M}_{\sun}$.
Taking a mass ratio of $1/3$, the median mass equals 8 SMBBH mergers and taking $q=1/30$, 13 SMBBH mergers are required.
Additionally, the arithmetic mean SMBH mass clearly overestimates the weighting of heavier SMBH masses, as larger components have
a greater impact on its value. 
It is indicated as the black dotted line, has a value of $M_{\rm arith.} \approx 9.03 \cdot 10^8 \ein{M}_{\sun}$ and requires 11 ($q = 1/3$) or 19 ($q = 1/30$) SMBBH mergers in this iterative process.

At this point we are making the assumption that most mergers took place in galaxy groups and clusters, because of the relatively high galaxy densities in them, and happened between a redshift of $z = 1$ and $z = 3$. 
This time frame corresponds to where the estimated source evolution densities in \figref{fig:f_z} reach their maximum as well as where the Madau plot of the cosmic star formation history in \cite{2014ARA&A..52..415M} has its maximum.
This redshift range corresponds to a time span of $t_{1,3} = 3.71 \ein{Gyr}$.
However, since a distance $d(t)$ and the scale factor $a(t)$ scale with the redshift according to $d(t) \sim a(t) \sim (1 + z)^{-1}$ 
\citep{1993ppc..book.....P}, it follows with $v(t) \sim 1/a(t) \sim (1+z)$ \citep{2005pfc..book.....M}:
\begin{equation}
    t = \frac{d(t)}{v(t)} \sim (1+z)^{-2} \,.
    \label{eq:t_z_dependence}
\end{equation}
So the merger time scales get shorter with a redshift dependency of $(1+z)^{-2}$.
Note that \glref{eq:t_z_dependence} describes how a time at redshift $z_0 = 0$ behaves at a redshift $z > z_0$.
The other way around, time scales at $z > z_0$ are getting larger with $(1+z)^{2}$ at the redshift $z_0 = 0$. 
With 6 SMBBH mergers ($q = 1/3$) between $z = 1$ and $3$, the mean merger time for one SMBBH merger at the current redshift ($z=0$) is thus: 
\begin{equation}
	\bar{t}_{total} \big(\langle M\rangle, q = 1/3 \big) = \frac{3.71 \ein{Gyr}}{6} \cdot (1+1)^{2} \approx 2.47 \ein{Gyr} \,.
	\label{eq:mean_time_galaxy_1}
\end{equation}
This time contains all time scales discussed for the SMBBH after the preceding galaxy merger as well as the galaxy merger time scale itself:
\begin{equation}
	\bar{t}_{total} = \bar{t}_{gal} + \bar{t}_{dyn} + \bar{t}_{\rm fin} + \bar{t}_{insp} + \bar{t}_{merge} \,.
	\label{eq:t_total_SMBH}
\end{equation}
The other mean merger times are specified analogously and displayed in \tbref{tab:f_gal_1}.
Each of them are simultaneously the maximum mean merger times for SMBBHs assuming the corresponding merger numbers, as the mergers could have also happened in a fraction of the specified merger time.
In order to validate the total mean merger times, we plotted them along with the SMBBH merger rates with the relation in \cite{1979ApJ...229..242S} and obtained the same shape of the integrated mass function as can be seen in \figref{fig:Schechter_fit} with the Schechter function fit.
Comparing these times with the total merging times of SMBBHs after the galaxy merger in \figref{fig:t_total} makes two things clear:

(i) Considering a mass ratio of $q = 1/30$ for high SMBH masses, all 3 total mean merger time models are clearly shorter than the inspiral time by \cite{2009ApJ...697.1621G} alone, as is seen in \figref{fig:t_total_b}. 
Even the largest total mean merger time of $\bar{t}_{total}(\langle M\rangle, 1/30) = 1.65 \cdot 10^9 \ein{yr}$ with 9 mean SMBBH mergers ($q = 1/30$) is about 3 times shorter than the longest inspiral time in the \cite{2009ApJ...697.1621G} model for the highest SMBH masses.
This leads to the conclusion that, in general, larger mass ratios are required for the formation of the heaviest SMBHs, as otherwise the discrepancy $\bar{t}_{total} < t_{\rm SMBH}$ with $t_{\rm SMBH} = t_{dyn} + t_{\rm fin} + t_{insp} + t_{merge}$ arises.

(ii) By comparing the estimated total mean SMBBH merger times $\bar{t}_{total}$ and the SMBBH merger times without a galaxy merger $t_{\rm SMBH}$ in \figref{fig:t_total_a}, both with the mass ratio $q=1/3$,
it becomes apparent that even with a final parsec time scale of $1 \ein{Gyr}$ (brown triangles in \figref{fig:t_total_a}), the duration of $t_{\rm SMBH}$ is smaller than $\bar{t}_{total}$ for most masses. 
Even some $10^8-10^9 \ein{yr}$, depending on the SMBH mass and the estimated total mean SMBBH merger time $\bar{t}_{total}$, remain for the galaxy merger prior to the SMBBH merger process, which is not captured by $t_{\rm SMBH}$. 
Only for the most massive SMBH masses, the final parsec time scale has to be smaller (around $10^8 \ein{yr}$, violet squares) in order to have $\bar{t}_{total} = t_{\rm SMBH} + \bar{t}_{gal}$, provided the galaxy merger takes $10^8-10^9 \ein{yr}$ at most.

As can be seen, nearly all time scales discussed in Sec.~\ref{subsec:merging_time_scales} are in agreement with the total mean SMBBH merger times estimated in this section. Only the merger of the highest SMBH masses require restrictions like higher mass ratios or a smaller final parsec time scale in order to agree with the total mean SMBBH merger times estimated. 
The premise is, however, that the preceding galaxy mergers last only so long that \glref{eq:t_total_SMBH} is satisfied.\\

With the mean SMBBH merger times determined, the accretion rate can now be compared to the Eddington accretion of $\dot{M}_{\rm Edd} \approx 2.23 \cdot 10^{-8} M \ein{yr}^{-1}$ (compare to e.g. \cite{2009herb.book.....D} with an accretion efficiency of $\sim 10$ per cent). 
Taking the used accretion percentage between each merger of $\eta_{\rm acc} = 33$ per cent, the longest mean SMBBH merger time $\bar{t}_{total} \approx 2.47 \ein{Gyr}$ and taking into account that AGN are only active and produce a relativistic jet for about 1 per cent of their lifetime, we determine an accretion rate of $\dot{M} \approx 1.35 \cdot 10^{-8} M \ein{yr}^{-1} \approx 0.61 \, \dot{M}_{\rm Edd}$, so in agreement with accretion in the Eddington limit. 
If, however, AGN are active a fraction 
${\bar{t}_{insp}}/{\bar{t}_{total}}$
of their life time, as is assumed in Sec.~\ref{sec:SMBBH_Merger_Rate}, then the accretion rate changes to $\dot{M} \approx 8 \, \dot{M}_{\rm Edd}$ with the inspiral time model by \cite{2009ApJ...697.1621G} and each mean SMBBH merger total merger time received for $q = 1/3$.
It gets even higher for the other inspiral time models (see values for ${\bar{t}_{insp}}/{\bar{t}_{total}}$ in \tbref{tab:t_insp_t_total}).

This could indicate that the active stage of an AGN is longer than the inspiral stage alone, at least for high mass ratios.
As a result, the SMBBH merger rate, $R_{\rm tot}$, would get larger by the same value that the inspiral stage gets longer, as the merger rate is defined as proportional to the active stage of an AGN (see \glref{eq:R_SMBH}).
However, a short period of super-Eddington accretion, resulting in a higher accretion efficiency, as was suggested by \citet{1985A&A...153...99S} for neutron stars, could also explain a high accretion, but on smaller time scales than the inspiral stage time scale.
Whereas a smaller or variable accretion percentage between each SMBBH merger could lower the accretion rate below the Eddington limit, but would increase the number of SMBBH mergers, reducing the time for a SMBBH merger including a galaxy merger. 
Consequently, a combination of a slightly longer AGN active stage, short periods of super-Eddington accretion and smaller accretion percentage for high mass ratios can explain the high accretion rate determined in this section. 
That is why, for the mass ratio $q = 1/3$, the number of SMBBH mergers are viewed as a lower limit and the resulting mean SMBBH merger times determined as an upper limit, as this mass ratio is an upper limit for the mass ratio as well. 

Considering the mass ratio $q = 1/30$ and the respective values for the inspiral time and total mean time for one SMBBH merger with the same accretion percentage between each merger of $\eta_{\rm acc} = 33$ per cent, an accretion rate of $\dot{M} \approx 0.13 \, \dot{M}_{\rm Edd}$ is obtained. 
As the Eddington limit for accretion is not violated, this could indicate that mergers at highly unequal mass ratios are more realistic than mass ratios closer to unity.

\citet*{2003ApJ...582..559V} also derived a value for the mean mass accretion between each major SMBBH merger, which corresponds to $\Delta M_{\rm acc} \approx 0.1 M$ with large error bars, which we do not specify at this point. 
As the error extends over several orders of magnitude, we conclude that our value 
of $M_{\rm acc} = 0.33 M$ 
lies clearly in the range derived by others. 
For more details see \citet*{2003ApJ...582..559V}.

\subsection{Neutrinos from SMBBH Mergers}
\label{subsec:neutrinos_SMMBH}

\begin{table*}
	\renewcommand{\arraystretch}{1.5}
	\centering
	\caption{Overview over the estimated mean number of SMBBH mergers in the Hubble time. The mean merger times, resulting SMBBH merger rates as multiple of ${\bar{t}_{insp}}/{\bar{t}_{total}}$ and effective fraction of neutrino from GW energy during a SMBBH merger $f^\nu_\text{SMBBH,eff}$ are shown as well. 
	All values are separated in mass ratios $1/3$ and $1/30$.}
	\label{tab:f_gal_1}
	\begin{tabular}{ c | c | c | c | c |c}
		mass ratio $q$
		& \centering mass in ${\rm M}_\odot$
		& \centering number of mergers 
		& \centering $\bar{t}_{total}$ in yr
		& $R_\text{tot} \cdot \frac{\bar{t}_{total}}{\bar{t}_{insp}}$ in $\text{Gpc}^{-3} \ein{yr}^{-1}$
		& $f^\nu_\text{SMBBH,eff} $
		\\
		\hline
		\multirow{3}{*}{$1/3$} 
		& $4.7 \cdot 10^7 $
		& \centering $ 6 $
		& \centering $2.47 \cdot 10^9$ 
		& $ 2.32 \cdot 10^{-3 \pm 0.4}$ 
		& $ 2.50_{ - 2.37}^{ + 11.69} \cdot 10^{-5} $
		\\
		& $1.45 \cdot 10^8 $
		& \centering $ 8 $ 
		& \centering $1.85 \cdot 10^9$ 
		& $ 3.09 \cdot 10^{-3 \pm 0.4}$ 
		& $ 1.87_{ - 1.78}^{ + 8.77} \cdot 10^{-5} $
		\\
		& $9.03 \cdot 10^8 $
		& \centering $ 11 $ 
		& \centering $1.35 \cdot 10^9$ 
		& $ 4.25 \cdot 10^{-3 \pm 0.4}$ 
		& $ 1.36_{ - 1.30}^{ + 6.38} \cdot 10^{-5} $
		\\
		\hline
		\multirow{3}{*}{$1/30$}
		& $4.7 \cdot 10^7 $
		& \centering $ 9 $
		&\centering $1.65 \cdot 10^9$ 
		& $ 3.47 \cdot 10^{-3 \pm 0.4}$
		& $ 5.00_{ - 4.76}^{ + 23.43} \cdot 10^{-5} $
		\\
		& $1.45 \cdot 10^8 $
		& \centering $ 13 $
		&\centering $1.14 \cdot 10^9$ 
		& $ 5.02 \cdot 10^{-3 \pm 0.4}$
		& $ 3.46_{ - 3.29}^{ + 16.22} \cdot 10^{-5} $
		\\
		& $9.03 \cdot 10^8 $
		& \centering $ 19 $
		&\centering $7.81 \cdot 10^8$ 
		& $ 7.33 \cdot 10^{-3 \pm 0.4}$
		& $ 2.37_{ - 2.25}^{ + 11.10} \cdot 10^{-5} $
	\end{tabular}
\end{table*}
\renewcommand{\arraystretch}{1}

\begin{table*}
	\renewcommand{\arraystretch}{1.5}
	\centering
	\caption{Values for the fraction of the mean inspiral time from the mean SMBBH merger time for the four inspiral time models used, separated in mass ratios $1/3$ and $1/30$. Values for $\bar{t}_{total}$ are taken from \tbref{tab:f_gal_1}.}
	\label{tab:t_insp_t_total}
	\begin{tabular}{ p{2.5cm}|ccc|ccc} 
		\centering inspiral time model
		& \multicolumn{6}{c}{\centering $\frac{\bar{t}_{insp}}{\bar{t}_{total}}$}
		\\
		\hline
		& \multicolumn{3}{c}{$q = 1/3$ }
		& \multicolumn{3}{c}{$q = 1/30$ }
		\\
		\cline{2-7}
		& $\bar{t}_{total} = 2.47 \ein{Gyr}$
		& $\bar{t}_{total} = 1.85 \ein{Gyr}$
		& $\bar{t}_{total} = 1.35 \ein{Gyr}$
		& $\bar{t}_{total} = 1.65 \ein{Gyr}$
		& $\bar{t}_{total} = 1.14 \ein{Gyr}$
		& $\bar{t}_{total} = 0.78 \ein{Gyr}$
		\\
		\cline{2-7}
		Gergely and Biermann (2009) 
		& \multirow{2}{*}{ $ 7.53 \cdot 10^{-4} $ }
		& \multirow{2}{*}{ $ 1.00 \cdot 10^{-3} $ }
		& \multirow{2}{*}{ $ 1.38 \cdot 10^{-3} $ }
		& \multirow{2}{*}{ $ 6.79 \cdot 10^{-3}$ }
		& \multirow{2}{*}{ $ 9.82 \cdot 10^{-3} $ }
		& \multirow{2}{*}{ $ 1.43 \cdot 10^{-2} $ }
		\\
		Sesana et al.\ (2012)
		& $ 2.36 \cdot 10^{-4} $
		& $ 3.14 \cdot 10^{-4} $ 
		& $ 4.33 \cdot 10^{-4} $
		& $ 3.19 \cdot 10^{-3} $ 
		& $ 4.61 \cdot 10^{-3}$
		& $ 6.73 \cdot 10^{-3} $
		\\
		Cavaliere et al.\ (2019)
		& $ 1.17 \cdot 10^{-4} $
		& $ 1.55 \cdot 10^{-4} $ 
		& $ 2.13 \cdot 10^{-4} $
		& $ 8.12 \cdot 10^{-4} $ 
		& $ 1.18 \cdot 10^{-3}$
		& $ 1.72 \cdot 10^{-3} $
		\\
		Peters (1964)
		& $ 9.43 \cdot 10^{-5} $
		& $ 1.25 \cdot 10^{-4} $ 
		& $ 1.73 \cdot 10^{-4} $
		& $ 8.48 \cdot 10^{-4} $ 
		& $ 1.23 \cdot 10^{-3}$
		& $ 1.79 \cdot 10^{-3} $
	\end{tabular}
\end{table*}
\renewcommand{\arraystretch}{1}

After determining the total mean SMBBH merger times, the SMBBH merger or rather
detection rate can be calculated using \glref{eq:R_SMBH}.
Inserting these rates in \glref{eq:f_SMBBH_3}, the effective fraction of neutrino from GW energy during a SMBBH merger $f^\nu_\text{SMBBH,eff}$ can be specified.
Both these values as well as the derived mean total merger times above are presented in \tbref{tab:f_gal_1}.
Additionally, the fractions of mean inspiral time from the respective mean total merger times are shown in \tbref{tab:t_insp_t_total} for each inspiral time model used.

\begin{figure}
	\centering
	\includegraphics[width=0.4\textwidth]{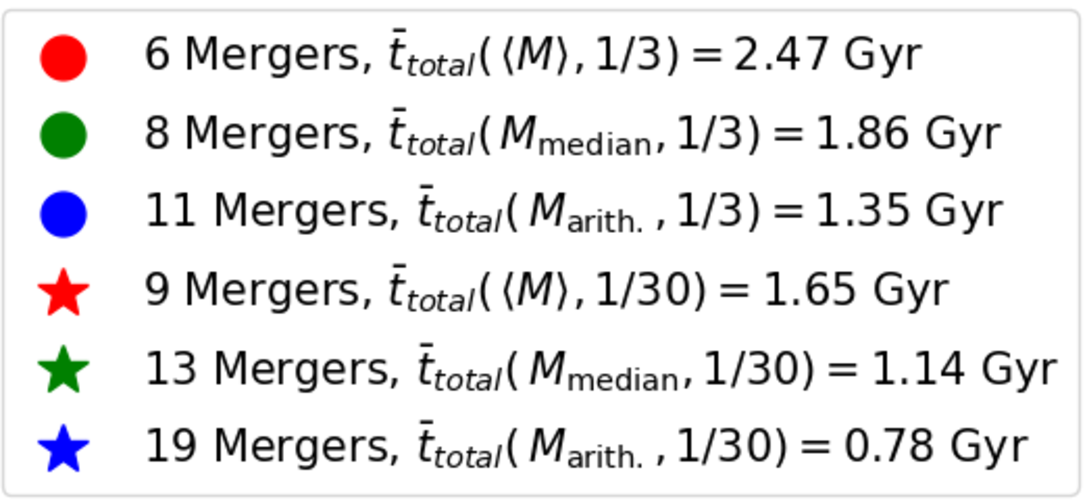}
	\vspace*{-15pt}
	\flushleft
	\includegraphics[width=0.49\textwidth]{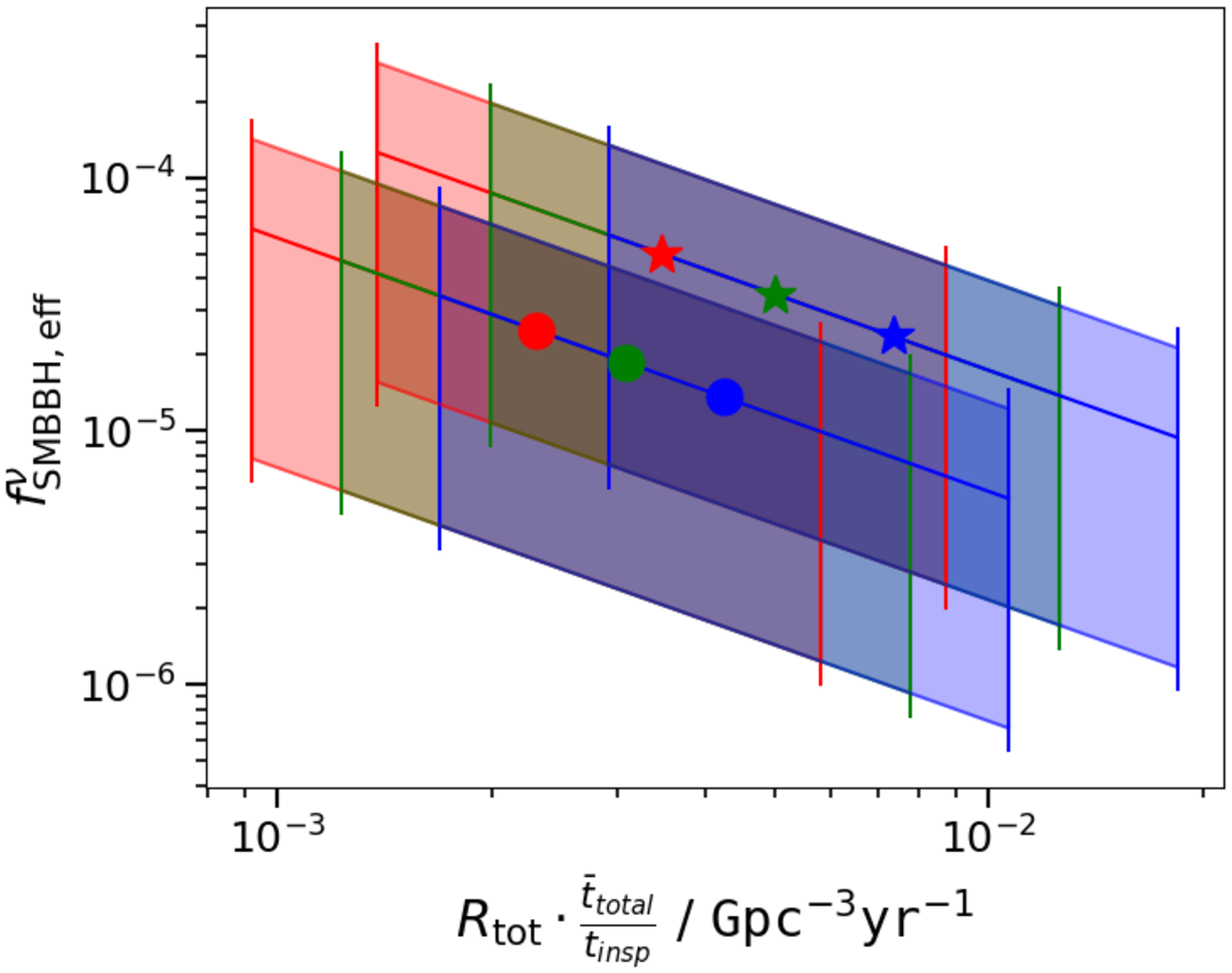}
	\caption{ $f^\nu_\text{SMBBH,eff}$ to total SMBBH merger rate $R_\text{tot}$ Plot. The x-axis and y-axis are in logarithmic scales. The points and stars with their respective error bars are marked for different iterated mean merger times using a mass ratio of $q=1/3$ and $q=1/30$.}
	\label{fig:f_SMBBH_gal}
\end{figure}

Figure \ref{fig:f_SMBBH_gal} shows $f^\nu_\text{SMBBH,eff}$ in dependency of the determined SMBH detection rates, which are in multiples of ${\bar{t}_{insp}}/{\bar{t}_{total}}$, as is seen in \tbref{tab:f_gal_1}.
The points represent the SMBBH merger rates with a mass ratio of $q=1/3$, which results in 6, 8 and 11 mean total SMBBH mergers, respectively.
The stars outline the SMBBH merger rates with 9, 13 and 19 mean SMBBH mergers, which are determined using a mass ratio of $1/30$.
The error bars are drawn analogously to \figref{fig:f_SMBBH_q_1_3}:
The upper and lower lines are created using the error bars of the diffuse neutrino flux and the parameter $\xi_z$, while the error bars in the rates result in a higher or lower parameter $f^\nu_\text{SMBBH,eff}$.
Due to the overlapping error bars of the points, the upper and lower limits of the rates are marked via vertical lines.

If the observed SMBHs are formed in 6 mergers on average, the total mean merger time is $\bar{t}_{total}(\langle M\rangle, 1/3) \approx 2.47 \ein{Gyr}$ (see \glref{eq:mean_time_galaxy_1}) and results in a total SMBBH merger rate of $R_{\rm tot} = 2.32 \cdot 10^{-3 \pm 0.4} \cdot \frac{\bar{t}_{insp}}{\bar{t}_{total}} \ein{Gpc}^{-3} \ein{yr}^{-1}$.
This is marked as the red point in \figref{fig:f_SMBBH_gal}. 
This rate delivers the greatest factor $f^\nu_\text{SMBBH,eff}$ for the mass ratio $1/3$ with a value of $ 2.50_{ - 2.37}^{ + 11.69} \cdot 10^{-5}$, which is independent of the inspiral time model used for determining $R_{\rm tot}$ (see \glref{eq:f_SMBBH_4}). 
Compared to the values for the true fraction of neutrino to GW energy $f^\nu_\text{SMBBH}$ at $q = 1/3$ in \tbref{tab:mean_time}, the value for the effective fraction of neutrino to GW energy $f^\nu_\text{SMBBH,eff}$ is by a factor of about $10-100$ larger, depending on the inspiral time model used for calculating $f^\nu_\text{SMBBH}${, and is consistent with the rough estimate performed in Sec.~\ref{subsec:neutrinos_insp_stage}}.

The largest value for $f^\nu_\text{SMBBH,eff}$ is $ 5.00_{ - 4.76}^{ + 23.43} \cdot 10^{-5}$ and is obtained using the total mean merger time of $\bar{t}_{total}(\langle M\rangle, 1/30) \approx 1.65 \ein{Gyr}$, which corresponds to 9 average mergers with a mass ratio of $1/30$.
Is is about a factor $1.5 - 10$ larger than the corresponding values for $f^\nu_\text{SMBBH}$ with the same mass ratio of $1/30$ in \tbref{tab:mean_time}. 
This total mean merger time equals a SMBBH merger rate of $R_{\rm tot} = 3.47 \cdot 10^{-3 \pm 0.4} \cdot \frac{\bar{t}_{insp}}{\bar{t}_{total}} \ein{Gpc}^{-3} \ein{yr}^{-1}$.
This value is indicated as the red star in \figref{fig:f_SMBBH_gal}.
As this comparison shows, a mass ratio of $1/30$ delivers greater values for $f^\nu_\text{SMBBH,eff}$ than the higher mass ratio $1/3$, although the rates $R_{\rm tot}$ of the former mass ratio are greater.

All values for different $R_{\rm tot}$ and $f^\nu_\text{SMBBH,eff}$ lie close together, because of the choice of the constant accretion percentage between each merger $\eta_{\rm acc}$.
A high accretion reduces the influence of the mass gain via mergers, as the factor $(1 + \eta_{\rm acc})$ gets larger than $(1 + q)$.
As a consequence, the choice of the mass ratio $q$ becomes less important, so that the values for $f^\nu_\text{SMBBH,eff}$ vary only slightly for all mass ratios.
A weaker accretion ensures that more SMBBH mergers are required in order to reach the observed SMBH masses and thus increases the influence of the mass ratio $q$.
In this case, a lower mass ratio causes significantly more SMBBH mergers than a higher one.
This is indicated in \figref{fig:1_mal_10_9_M} with an accretion percentage between each merger of $\eta_{\rm acc} = 5$ per cent, which leads to 84 SMBBH mergers for $q=1/30$ (black plus symbol) and only 20 mergers for $q=1/3$ (green plus symbol).
More mergers also mean that more energy in the form of GWs is emitted.
On the other hand, it implies that the total mean merger time decreases, so that the rate $R_\text{tot}$ increases and as a result, the values for $f^\nu_\text{SMBBH,eff}$ vary more.

\section{Neutrinos from stellar mass Binary Black Hole Mergers}
\label{sec:BBH_Merger_Rates}

This section covers  neutrinos produced in BBH mergers in starburst galaxies.
As more than 80 BBH mergers were already documented in the GWTC-1, GWTC-2, GWTC-2.1 and GWTC-3 catalogues \citep{2019PhRvX...9c1040A, 2021PhRvX..11b1053A, 2021arXiv210801045T, 2021arXiv211103606T}, the mean measured gravitational wave energy radiated in each merger is determined as $\langle E_\text{GW}\rangle \approx 2.79 \ein{M}_{\sun} \cdot c^2$ with a standard deviation of $\sigma \approx 1.61 \ein{M}_{\sun} \cdot c^2$. These values consider the radiated GW energies of 85 clear BH-BH mergers without the error bars in the measurements and determinations of the binaries masses and radiated GW energy. 

\begin{figure}
	\centering
	\includegraphics[width=0.49\textwidth]{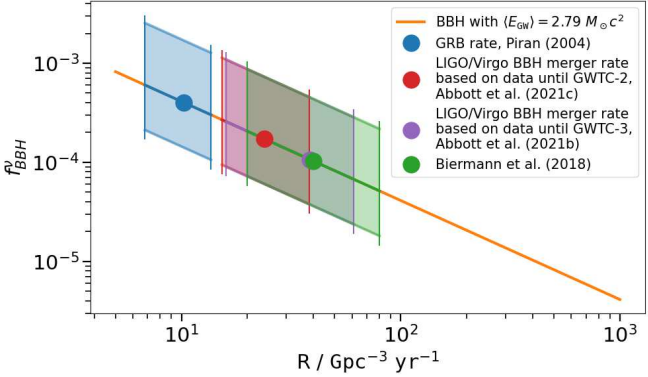}
	\caption{ Representation of the $f^\nu_\text{BBH}$ to BBH merger rates Plot. The x- and y-axis are in logarithmic scales. The points mark the different rates, which can be found in \tbref{tab:f_BBH}.}
	\label{fig:f_BBH}
\end{figure}

Using different rates, the values for $f^\nu_\text{BBH}$ are estimated with \glref{eq:f_BBH} and the diffuse neutrino flux presented in \glref{eq:neutrino_flux}.
The factor $\xi_z$ is taken as $2.4$, as BBHs follow the SFR (compare to Sec.~\ref{subsubsec:redshift}).
The BBH merger rates used as well as the resulting values for $f^\nu_\text{BBH}$ are listed in \tbref{tab:f_BBH} and displayed in \figref{fig:f_BBH}.
Here, the orange line represents the connection in \glref{eq:f_BBH}.

The blue point with its error bars is constructed using the predicted GRB rate estimated by \cite{2004RvMP...76.1143P}:
\begin{equation}
	R_\text{GRB} = (33 \pm 11) \ h^3 \ein{Gpc}^{-3}\ein{yr}^{-1} \,.
	\label{eq:R_GRB}
\end{equation}
It was taken in order to compare whether BBH mergers could make up at least a part of the observed GRBs or if the rates are not comparable.
Similar to the plots in \figref{fig:f_SMBBH_q_1_3}, the error bars were created using the uncertainties of the diffuse neutrino flux (\glref{eq:neutrino_flux}), as well as the uncertainties of the rates and the standard deviation of the observed radiated GW energy. 
The value $h$ was taken as $0.6766$ according to \tbref{tab:Planck}, so that the GRB rate becomes $R_\text{GRB} = (10.22 \pm 3.41) \ein{Gpc}^{-3}\ein{yr}^{-1}$.
This rate delivers a value of $f^\nu_\text{BBH} = 4.02^{ + 21.53}_{ - 2.96} \cdot 10^{-4}$. 
So a fraction of $10^{-4}$ of the gravitational potential is radiated in form of neutrinos in each BBH merger, if every GRB is produced by such a merger.

The green point follows the rate by \cite{2018AdSpR..62.2773B} of events similar to the compact source 41.9+58, which was suggested to be a BBH merger:
\begin{equation}
	R_{41.9 + 58} = 40 \ t_{3.4}^{-1} \ein{Gpc}^{-3}\ein{yr}^{-1} \,.
	\label{eq:R_Biermann}
\end{equation}
Since the parameter $ t_{3.4}^{-1} $ is highly uncertain and unspecified, it was set to the values $1/2$ and $2$ respectively, in order to get a large uncertainty in the rate.
In doing so, the rate becomes $40^{+40}_{-20} \ein{Gpc}^{-3}\ein{yr}^{-1}$ and is equivalent to $f^\nu_\text{BBH} = 1.03^{ + 7.68}_{ - 0.85} \cdot 10^{-4}$, so about a factor 4 smaller than what the GRB rate delivers.

Lastly, the inferred merging BBH detection rate by the LIGO and Virgo Collaborations with a 90 per cent credible interval of $R = 23.9^{+14.3}_{-8.6} \ein{Gpc}^{-3}\ein{yr}^{-1}$ \citep{2021ApJ...913L...7A} results in $f^\nu_\text{BBH} = 1.72^{ + 9.96}_{ - 1.34} \cdot 10^{-4}$.
This is marked as the red point and error bars in \figref{fig:f_BBH}.

As is pointed out in \cite{2021arXiv211103634T}, the most recent BBH merger rates determined based on combined data until the most recent GW catalogue, GWTC-3, are consistent with the earlier results presented above.
They are marked in \figref{fig:f_BBH} as purple point and error bars. 
However, as only a range of $16 - 61 \ein{Gpc}^{-3}\ein{yr}^{-1}$ is given, we used the rate $R = 38.5^{+22.5}_{-22.5} \ein{Gpc}^{-3}\ein{yr}^{-1}$ for our calculations. 
This rate results in a neutrino to GW energy ratio of $f^\nu_\text{BBH} = 1.07^{ + 9.81}_{ - 0.83} \cdot 10^{-4}$.
\begin{table}
    \renewcommand{\arraystretch}{1.5}
	\centering
	\caption{BBH merger rates $R$ and corresponding values for $f^\nu_\text{BBH}$.}
	\label{tab:f_BBH}
	\begin{tabular}{ p{3.2cm}|c|c}
		& $R$ in $\text{Gpc}^{-3} \ein{yr}^{-1}$ 
		& $f^\nu_\text{BBH}$ 
		\\
		\hline
		GRB rate, Piran (2004)
		&  $ 10.22 \pm 3.41 $ 
		& $ 4.02^{ + 21.53}_{ - 2.96} \cdot 10^{-4}$ 
		\\
		LIGO/Virgo BBH merger rate based on data until GWTC-2 ($90 \%$ credible interval), Abbott et al.\ (2021c)
		& \multirow{3}{*}{$ 23.9^{ + 14.3}_{ - 8.6} $}
		& \multirow{3}{*}{$ 1.72^{ + 9.96}_{ - 1.34} \cdot 10^{-4}$}
		\\
		LIGO/Virgo BBH merger rate based on data until GWTC-3 ($90 \%$ credible interval), Abbott et al.\ (2021b)
		& \multirow{3}{*}{$ 38.5^{ + 22.5}_{ - 22.5} $}
		& \multirow{3}{*}{$ 1.07^{ + 9.81}_{ - 0.83} \cdot 10^{-4}$}
		\\
		Rate of events similar to 41.9+58, 
		Biermann et al.\ (2018)
		& \multirow{2}{*}{ $ 40^{ + 40}_{ - 20} $ }
		& \multirow{2}{*}{ $ 1.03^{ + 7.68}_{ - 0.85} \cdot 10^{-4}$ }
		\\
	\end{tabular}
\end{table}
\renewcommand{\arraystretch}{1}

It can be seen clearly in \figref{fig:f_BBH} that the GRB rate differs widely from the other two rates, as the GRB rate probably underestimates the rate of BBH mergers contributing to the production of neutrinos, which are detected as the diffuse neutrino flux.
On one hand, this rate involves all GRBs seen, short- and long-lived.
But only a part of short-lived GRBs could be associated with BBH mergers. 
Following this logic, this rate would overestimate the BBH merger rate.
However, as this rate is derived only from observed GRBs, the part of the BBH merger population, that is obscured, is not included in it because it is not detectable.
Neutrinos on the other hand are not absorbed where the electromagnetic spectrum is  \citep[see][]{2021ApJ...911L..18K} and can therefore be detected from obscured BBH mergers, creating obscured GRBs.
This means that more BBH mergers can add to the diffuse neutrino flux than can be detected in GRBs. This leads to an underestimation of the BBH merger rate with the GRB rate.

The BBH merger rates with a 90 per cent credible interval by the LIGO and Virgo Collaborations on the other hand are directly derived from observational data (GWTC-2 and GWTC-3 respectively) and thus reflect the presumed BBH merger rate in the local Universe.
These rates are therefore more accurate than the GRB rate.
However, only a fraction of the BBH merger rate contributes to the diffuse neutrino flux, as not all jets point towards Earth during the merger event.
Since this fraction cannot be determined without observing the jets, the corresponding value for the neutrino fraction $f^\nu_\text{BBH}$ represents a lower limit for the GW energy that goes into neutrinos during each BBH merger.
That is because the value for $f^\nu_\text{BBH}$ increases as the BBH merger rate decreases (see \figref{fig:f_BBH}).

The error bars of the estimated event rates similar to the compact source 41.9+58 (green) overlaps in great parts with the error bars of the BBH merger rates by the LIGO and Virgo Collaborations (red and purple).
The overlap with the BBH merger rate using data of all observed BBH mergers until GWTC-3 is even greater than only using data until GWTC-2.
This further supports the thesis that this event was caused by a BBH merger. 
However, this rate was estimated using simple assumptions (see \citealt{2018AdSpR..62.2773B} for detailed information), which could have led to a slight overestimation of the BBH merger rate.
In addition, the error bars were constructed using different values for the parameter $ t_{3.4}^{-1} $ (see \glref{eq:R_Biermann}), which could also result in an overestimation (and even a potential underestimation) of the BBH merger rate with this method.

\section{Results}
\label{sec:Results}

\subsection{Constraints on Gravitational Waves}
\label{subsec:constraints}
\begin{figure*}
	\centering
	\includegraphics[width=0.9\textwidth]{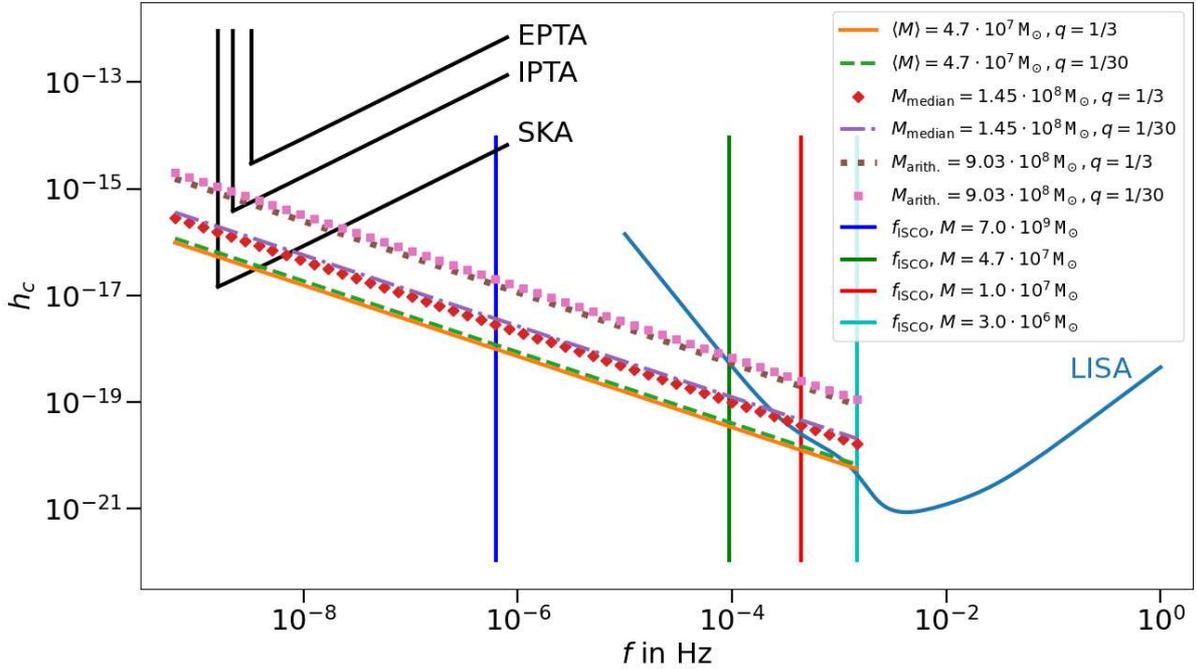}
    \caption{ Characteristic strain against frequency plot for the used SMBH distribution and several mean masses $M$. 
    The orange solid line, red diamonds and brown dotted line use a mass ratio of $q=1/3$. 
    The mass ratio $1/30$ is used for the green dashed line, purple dashed and dotted line and pink squares and is marked as such. 
    The vertical solid lines mark $f_{\rm fin}$ for corresponding total SMBBH masses.
    Sensitivity curves for the detectors SKA, IPTA and EPTA are constructed after \citet*{2015CQGra..32a5014M}. 
    LISA sensitivity curve are plotted after \citet*{2019CQGra..36j5011R}.     }
	\label{fig:strain_2}
\end{figure*}
Until now, the percentage of observed SMBHs that contribute to the diffuse astrophysical neutrino flux was estimated based on the time, the SMBBHs remain in the inspiral stage in comparison to the total time the binary needs to merge, including a preceding galaxy merger (see Sec.~\ref{sec:SMBBH_Merger_Rate}).
The question arises how accurate this estimate for the number of current SMBBH mergers is and what percentage of them does actually undergo a merger at the moment which is detectable in gravitational waves in the near future, for instance by the Laser Interferometer Space Antenna (LISA).
By comparing the expected GW signal, thus the characteristic strain, of the entire SMBH distribution with the detected characteristic strain, a constraint could be put on this distribution.
However, since GWs from SMBBH mergers are yet to be detected, the total characteristic strain is compared to current and future detection limits of PTAs.
The characteristic strain can be expressed as \citep{2001astro.ph..8028P}:
\begin{align}
	h_{c}(f) 
	&= 3 \cdot 10^{-24}\left(\frac{\mathcal{M}}{M_{{\sun}}}\right)^{5 / 6} \left(\frac{f}{10^{-3} \mathrm{Hz}}\right)^{-2 / 3}
	\notag \\ 
	& \qquad\qquad \cdot \left(\frac{ n_\text{SMBBH} }{\mathrm{Mpc}^{-3}}\right)^{1 / 2} \left(\frac{\left\langle(1+z)^{-1 / 3}\right\rangle}{0.74}\right)^{1 / 2} \,,
	\label{eq:chracteristic_strain}
\end{align}
with the the GW frequency $f$ and the current comoving number density of merging SMBBHs $n_\text{SMBBH}$.
The factor $ \left\langle(1+z)^{-1 / 3}\right\rangle $ is defined as:
\begin{equation}
	\left\langle(1+z)^{-1 / 3}\right\rangle=\frac{1}{n_\text{SMBBH}} \int_{z_{\min }}^{z_{\max}} \frac{N(z)}{(1+z)^{1 / 3}} dz \,,
	\label{eq:1_z_bracelets}
\end{equation}
with the number of SMBBH mergers in the comoving volume $N(z)$ between the redshift $z$ and $z + dz$.
The values $z_{\min }$ and $z_{\max }$ capture the minimal and maximal redshifts, in which SMBBH mergers occur, and are set to $0$ and $\infty$, respectively.
\glref{eq:1_z_bracelets} can be approximated with a value of $0.8$ in a flat $\Omega_\Lambda \approx 0.7 $ Universe (see \cite{2001astro.ph..8028P}).
As shown in \glref{eq:E_GW_1}, the chirp mass can be expressed in dependency of the mass ratio $q$ and the total mass of the merging SMBHs $M$.
Since the total BH mass before the merger can be expressed in form of the now detected mass as $M = \frac{M_{\rm post}}{(1 - k(q))}$ (\glref{eq:M_now}), the chirp mass yields:
\begin{equation}
	\mathcal{M} = \frac{10 \cdot k(q)}{1 - k(q)} \cdot M_{\rm post} \,.
	\label{eq:chirp_mass}
\end{equation}
The GW frequency is limited by the final frequency $f_{\rm fin}$ that the coalescing BHs can emit before the merger on one side and the initial frequency $f_{\rm in}$ on the other.
The latter describes the first gravitational wave that is emitted during the merger.
As, theoretically, it can get arbitrarily small and negligible, it depends mainly on the used detector and the detectable minimal frequency.
The final frequency on the other hand is characterized by the presence of the BHs in their innermost stable orbit.
Here, this orbit is taken as $a_{\rm ISCO} \approx 1.9 \, r_S$, which is the final separation of the GW event GW150914 and was used by \citet*{2005PhRvD..71h4025B} in their calculations as well.
In order to estimate this frequency, the third law of Kepler is used to obtain:
\begin{equation}
	f_{\rm fin} 
	= {6}^{-{3}/{2}}
	 \cdot \frac{c^3}{\pi \, G \, M} \,.
	\label{eq:f_fin}
\end{equation}
Taking the SMBBH density as $n_\text{SMBBH} = \bar{t}_{insp} / \bar{t}_{total} \cdot n_\text{SMBH}$, with $n_\text{SMBH} = 5.73 \cdot 10^{6\pm0.4} \ein{Gpc}^{-3}$ (see Sec.~\ref{sec:time_scales}), values for ${\bar{t}_{insp}}/{\bar{t}_{total}}$ from the Gergely and Biermann model in \tbref{tab:t_insp_t_total} and a fixed mass ratio, the characteristic strain of the SMBH distribution is estimated.
For that, the estimated masses from \figref{fig:number_mergers} are inserted for the observed mass $M_{\rm post}$.

Figure \ref{fig:strain_2} shows the characteristic strain to frequency plot.
The black solid lines represent the detector sensitivities of the PTAs  Square Kilometer Array (SKA), International Pulsar Timing Array (IPTA) and European Pulsar Timing Array (EPTA).
Only the SKA detector is currently not active.
For EPTA, 5 pulsars with 10 years of observations were taken to construct the detector sensitivity.
In comparison, 20 pulsars with 15 years of observational data and 100 pulsars with 20 observational years were taken for IPTA and SKA respectively.
However, the IPTA is not reaching the theoretical sensitivity with these number of pulsars yet. 
In its second data release, it sets a limit on the characteristic strain at $2.8^{+1.2}_{-0.8} \cdot 10^{-15}$ at a frequency of $1 \ein{yr}^{-1} \approx 3.2 \cdot 10^{-8} \ein{s}^{-1}$ in case of a signal with a spectral index of $-2/3$, as expected from a stochastic gravitational wave background (see \glref{eq:chracteristic_strain}) \citep{2022MNRAS.510.4873A}.
In comparison, with $24 \ein{years}$ of data, EPTA sets a limit of $2.95^{+0.89}_{-0.72} \cdot 10^{-15}$ for the characteristic strain at the same frequency of $1 \ein{yr}^{-1}$ with the same spectral index \citep{2021MNRAS.508.4970C}. 
Note that in both cases the spectral index had to be fixed to that value and has otherwise large error bars, so that a clear identification of a GW background cannot be made yet. 
Two other PTAs, the North American Nanohertz Observatory for Gravitational Waves (NANOGrav) and the Parkes Pulsar Timing Array (PPTA) constrain the characteristic strain at the same reference frequency and with the same spectral index with $1.92^{+0.75}_{-0.55} \cdot 10^{-15}$ and $2.2^{+0.4}_{-0.3} \cdot 10^{-15}$, respectively \citep{2020ApJ...905L..34A, 2021ApJ...917L..19G}.
Due to missing characteristics, in both cases, a GW background can not be identified either. However, with a longer observational time, all PTAs are getting more sensitive to a potential GW background at a smaller characteristic strain. 
The sensitivities of these two PTAs are not shown in \figref{fig:strain_2} due to the complexity of constructing them.

The light blue line in \figref{fig:strain_2} shows the sensitivity curve of LISA, as part of the SMBH distribution can be detected with it.
The orange line shows the characteristic strain with the mean mass $\langle M\rangle = 4.70 \cdot 10^7 \ein{M}_{\sun}$ for $M_{\rm post}$ and a mass ratio of $q=1/3$.
The green dashed line uses the same mean mass, but is made with $q = 1/30$.
Red diamonds use the mass ratio $1/3$ and the median mass $M_{\rm median} = 1.45 \cdot 10^8 \ein{M}_{\sun}$, which slightly over-weights the number of heavy SMBHs.
The mass ratio $1/30$ with the same mass is used for the purple dashed and dotted line.
The brown dotted line and pink squares are generated with the mass ratio $1/3$ and $1/30$ respectively and use the arithmetic mass of $M_{\rm arith.} = 9.03 \cdot 10^8 \ein{M}_{\sun}$, which represents a clear overestimate of the mean mass.
All curves were made using \glref{eq:chracteristic_strain}, with the chirp mass derived in \glref{eq:chirp_mass}.
The above mentioned masses were determined with the now detected SMBH distribution and were inserted for the now observed mass $M_{\rm post}$.

Vertical lines represent the cutoff of the frequencies $f_{\rm fin}$ for chosen SMBH masses.
A conversion from the total mass of the coalescing SMBBH to the SMBH mass after the merger is unnecessary, as they serve as examples of now detected masses. 
Only the green vertical line was created using the now detected SMBH mass $M_{\rm post}$ because it represents the final frequency of a SMBH with the mean observed mass $\langle M\rangle$. 
A binary SMBH with a total mass of $7 \cdot 10^9 \ein{M}_{\sun}$ has its final frequency in the $10^{-7} \ein{Hz}$ range and is therefore only detectable with the PTAs as a diffuse GW background,
whereas a total SMBBH mass of $3 \cdot 10^6 \ein{M}_{\sun}$ has its final frequency around $10^{-3} \ein{Hz}$ and is hence detectable with LISA as a point source. 
The same goes for a total observed SMBBH mass of $1 \cdot 10^7 \ein{M}_{\sun}$ as well as $4.70 \cdot 10^7 \ein{M}_{\sun}$, if the corresponding characteristic strain is high enough.
So only mergers of light SMBBHs are detectable with LISA.

Figure \ref{fig:strain_2} shows that the characteristic strain of the arithmetic mean mass $M_{\rm arith.}$ could be detectable by IPTA, if it reaches its maximal sensitivity.
But since this model clearly overestimates the mean SMBH mass of the distribution, it is taken as an overestimate of the emitted GW energy as well.
The other two models with the mean mass $\langle M\rangle$ and the median mass $M_{\rm median}$ are therefore more important, however, they will not be detectable with EPTA.
Therefore a constraint cannot be put on the SMBH distribution using this PTA.
With SKA, however, both models should be detectable, so that a constraint could be made in the future. 
However, if one of the above presented currents limits on the characteristic strain actually captures the gravitational wave background, it would put a constraint not only on the mean mass of SMBBH mergers, but also on the SMBBH density and therefore on the ratio $ \bar{t}_{insp} / \bar{t}_{total}$, which links the SMBH density to the SMBBH density.

Assuming all merging SMBBHs result in an observed mean mass of $\langle M\rangle = 4.70 \cdot 10^7 \ein{M}_{\sun}$, then they would be partly detectable in the lower frequency range with LISA, as the intersection of the the lightest part of the distribution, at $M = 3 \cdot 10^6 M_\odot$ (turquoise vertical line), with the orange solid and green dotted lines is slightly above the LISA sensitivity curve. 
Of course, noise could hide the signal. 
However, if the estimated percentage of observed SMBHs (see \tbref{tab:t_insp_t_total}) is smaller than the actual percentage, the characteristic stain increases proportional to the SMBBH merger density squared and could be detected better by LISA. 
A constraint on the distribution could thus be made, once LISA is active and started its observational run.

\subsection{Comparison}
\label{subsec:comparison}

\begin{figure*}
	\centering
	\includegraphics[width=0.95\textwidth]{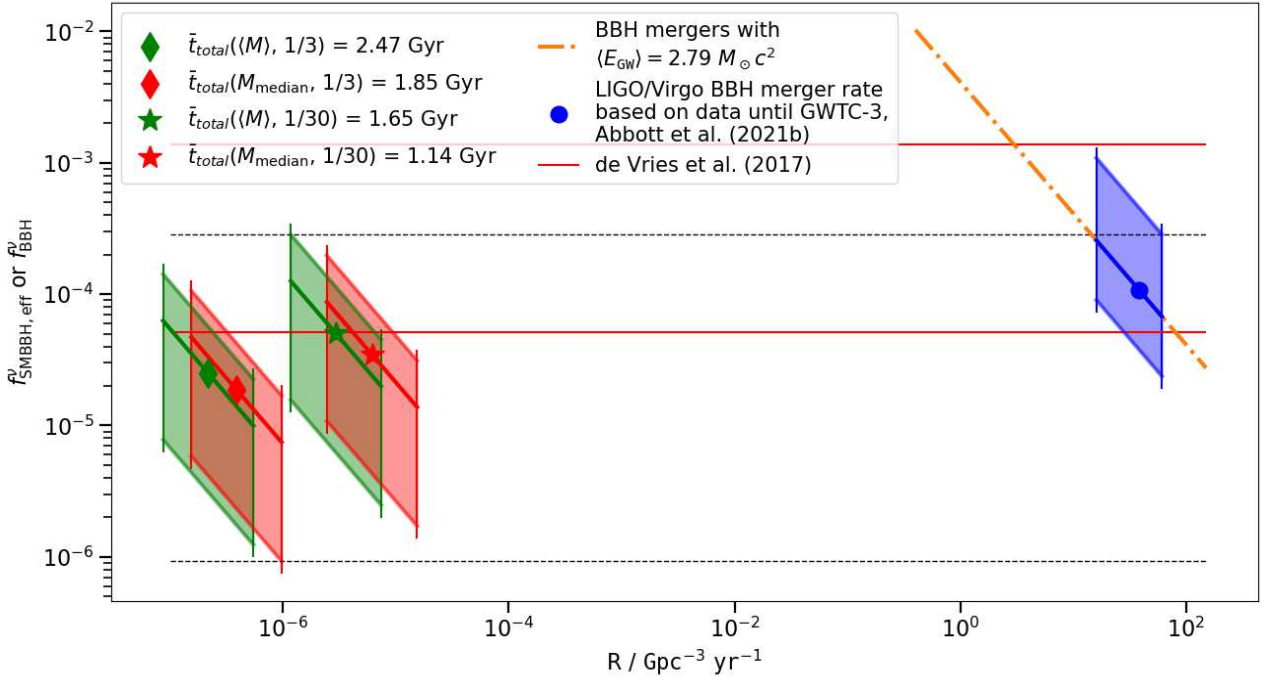}
	\caption{ Different SMBBH merger and BBH merger rates with the associated values for the fraction of gravitational wave energy that goes into neutrinos during each SMBBH merger $f^\nu_\text{SMBBH,eff}$ and stellar mass BBH merger $f^\nu_\text{BBH}$, respectively. 
	The x- and y-axis are in logarithmic scales. 
	BBH merger rates marked as dots, while SMBBH merger rates are marked with diamonds for a mass ratio of $q=1/3$ and stars for $q=1/30$.
	Red horizontal lines mark comparative values for $f^\nu_\text{BBH}$ \citep{2017PhRvD..96h3003D}.}
	\label{fig:beide_f}
\end{figure*}

The results found in this work are summarized in \figref{fig:beide_f}.
The blue point and its error bars represent the BBH merger rate using data of all observed BBH mergers until GWTC-3, as seen in \figref{fig:f_BBH}.
For the purpose of clarity, the other rates for stellar mass BBH mergers are omitted here.
The orange solid line describes the connection between the BBH merger rate and the fraction of GW energy that neutrinos receive during a BBH merger $f^\nu_\text{BBH}$, which is the fraction of neutrino energy from the gravitational potential and thus from the total emitted GW energy during the merger event (compare to \glref{eq:f_BBH}).
 
Diamond shaped points are selected for SMBBH mergers with the mass ratio $q=1/3$. 
Stars on the other hand stand for SMBBH mergers with the mass ratio $1/30$.
For the total mean SMBBH merger rates, the factor $\bar{t}_{insp} / \bar{t}_{total}$ from the Gergely and Biermann model in \tbref{tab:t_insp_t_total} was taken.
Other models result in lower rates, but the same value for $f^\nu_\text{SMBBH,eff}$ (see Sec.~\ref{subsec:neutrinos_SMMBH}).

The green diamond shaped point with its error bars stands for the SMBBH merger rate with the mean mass after the merger $\langle M\rangle$ and 6 mean mergers in the Hubble time.
These values are derived from a mass ratio $q=1/3$ and an accretion percentage between each SMBBH merger of $\eta_{\rm acc} = 33$ per cent.
The red diamond represents a median SMBH mass after the merger, $M_{\rm median}$, with the same parameters and 8 resulting mean SMBBH mergers.
The green star outlines 9 mean SMBBH mergers in the lifetime of the Universe, which is achieved with the same accretion percentage and the mean mass after the SMBBH merger $\langle M\rangle$, but with the mass ratio $1/30$.
The same applies for the red star, which was constructed using the median SMBH mass and the mass ratio $1/30$ with the same accretion percentage and results in 13 mean SMBBH mergers.
As the other rates in \figref{fig:f_SMBBH_gal} lie close to these and over-represent the heaviest SMBH masses, they are neglected at this point.

Horizontal grey dashed lines mark the highest and lowest values for $f^\nu_\text{SMBBH,eff}$, resulting from the mean total SMBBH merger rates considered, for better comparison. 

Red horizontal lines mark the values for $f^\nu_\text{BBH}$ by \cite{2017PhRvD..96h3003D}.
They were generated with the assumption that the whole seen diffuse neutrino flux is produced in BBH mergers in starburst galaxies with a rate between $R = 9 - 240 \ein{Gpc}^{-3}\ein{yr}^{-1}$ and a mean emitted GW energy of $3 \ein{M}_{\sun} \cdot c^2$.
This results in an upper range of the parameter $f^\nu_\text{BBH} \lesssim 5.15 \cdot 10^{-5} - 1.37 \cdot 10^{-3}$.

Figure \ref{fig:beide_f} shows that the fraction of gravitational wave energy that goes in neutrinos during stellar mass BBH mergers, $f^\nu_\text{BBH}$, determined with the stellar mass BBH merger rate from \cite{2021arXiv211103634T} overlaps quite well with this parameter range, making it in most parts consistent with the findings of \cite{2017PhRvD..96h3003D}. Moreover, we extended the allowed parameter range for $f^\nu_\text{BBH}$ by lower values. 

Comparing the values for the effective fraction of neutrino from GW energy during a SMBBH merger $f^\nu_\text{SMBBH, eff}$ with the mass ratios $q=1/3$ and $1/30$, it stands out that they overlap with the above determined parameter range for $f^\nu_\text{BBH}$, even though SMBBH mergers occur with a rate of about 7 to 9 orders of magnitude lower than BBH mergers.
At first glance this indicates that the same energy fraction from GW radiation goes into neutrinos during BBH mergers as well as SMBBH mergers.
That would mean that the efficiency of the neutrino production mechanism in both events is comparable for the assumption that 10 per cent of neutrinos is produced in BBH mergers and the rest in SMBBH mergers. 
However, as not all sources that are detectable via GWs contribute to the diffuse neutrino flux, but only those whose jets point towards Earth during the merger event, only an unknown fraction of the SMBBH and BBH merger rate can be responsible for the diffuse neutrino flux. 
Because of the anti-proportional relation, a smaller rate $R$ results in a larger value for $f^\nu_\text{SMBBH, eff}$ and $f^\nu_\text{BBH}$.
Therefore, as this fractions of the SMBBH and BBH merger rates cannot be determined, the specified values for $f^\nu_\text{BBH}$ with the BBH merger rates after \cite{2021arXiv211103634T} represent a lower limit: $f^\nu_\text{BBH} \gtrsim 2.4 \cdot 10^{-5} - 1.1 \cdot 10^{-3}$.

As for the parameter $f^\nu_\text{SMBBH,eff}$: the values presented here contain uncertainties, since they were estimated using an accretion percentage of $\eta_{\rm acc} = 33$ per cent, a mean or median SMBH mass and a mass ratio of $1/3$ and $1/30$ respectively.
However, 6 to 9 mean SMBBH mergers in the lifetime of a now observed SMBBH are evaluated as realistic values, achieved with the mean SMBH mass.
In contrast, 8 or 13 mergers are achieved with a slight over-representation of the highest SMBH masses in the distribution, so that they probably overestimate the number of SMBBH mergers as well.
Hence they are more suitable in defining a liberal lower limit for the parameter $f^\nu_\text{SMBBH, eff}$.
A mass ratio of $q=1/3$ with 8 mean SMBBH mergers results in lower values, while $q=1/30$ and 9 mean SMBBH mergers yield the highest values, leading to $f^\nu_\text{SMBBH, eff} \gtrsim 9.0 \cdot 10^{-7} - 2.8 \cdot 10^{-4}$ (see \tbref{tab:f_gal_1}).
Thus, the parameter ranges for $f^\nu_\text{SMBBH, eff}$ overlap with the ones for $f^\nu_\text{BBH}$, as can be seen with the horizontal grey dashed line.
At this point, it cannot be excluded that this happens by chance.

The figures \ref{fig:beide_f_muon}, \ref{fig:beide_f_HESE} and \ref{fig:beide_f_cascade} in the appendix show the same results with the diffuse astrophysical neutrino flux measured by IceCube with 9.5 years of muon tracks, HESE and cascade data respectively instead of the astrophysical starting tracks muon neutrino flux measured in 10.3 years. 
The relative position of the neutrino to GW energy fractions $f^\nu_\text{SMBBH,eff}$ for SMBBH mergers and $f^\nu_\text{BBH}$ for stellar mass BBH mergers remains the same, however, the absolute values for the fractions decrease by about a factor of $1/10$ compared to the muon tracks flux with 9.5 years of data and increase by about a factor $10$ for the HESE flux. For the cascade flux, the absolute values for the fractions are approximately the same as for the muon starting tracks flux detected in 10.3 years.

\section{Discussion of the results}
\label{sec:Discussion}

In this work, we show a possible connection between the diffuse astrophysical neutrino flux and the radiated gravitational wave energy from binary stellar mass black hole mergers in starburst galaxies as well as supermassive binary black hole mergers. 
For that, the assumption was made that neutrinos are produced during the inspiral stage of the merger events, due to a spin-flip of the leading jet of the black holes, which interacts with molecular clouds in its path.
Moreover, a relationship between the neutrino energy fraction of the emitted GW energy and the rate of the detected stellar mass BBH and SMBBH mergers was created.

The stellar mass BBH merger rate by the LIGO and Virgo Collaborations with a 90 per cent credible interval \citep{2021arXiv211103634T} leads to a lower limit for the fraction $f^\nu_\text{BBH}$ of GW energy that neutrinos receive during a BBH merger between {$\sim 2 \cdot 10^{-5}$ and $ \sim 1 \cdot 10^{-3}$}. 
The rate of stellar mass BBH mergers similar to the radio source 41.9+58 by \cite{2018AdSpR..62.2773B} is consistent with the BBH merger rate by the LIGO and Virgo Collaborations, thus could be a BBH merger and contribute to the above lower limit for $f^\nu_\text{BBH}$.
On the other hand, the GRB rate estimated by \cite{2004RvMP...76.1143P} is significantly lower than the other two and probably underestimates the stellar mass BBH merger rate due to not considering obscured BBH mergers \citep[see e.g.][]{2021ApJ...911L..18K}.

The neutrino energy fraction of emitted GW energy in SMBBH mergers was determined using 4 inspiral time models, leading to the rate of SMBBH mergers after the binary entered the inspiral stage.
This approach followed the SMBH mass distribution by \cite{2010A&A...521A..55C}, which captures $2.4 \cdot 10^{4 \pm 0.4}$ SMBHs in a radius of $100 \ein{Mpc}$ with a mass range between $3 \cdot 10^6 \ein{M}_{\sun}$ and $7 \cdot 10^9 \ein{M}_{\sun}$.
The SMBH mass density was determined as $2.69 \cdot 10^{5 \pm 0.4} \ein{M}_{\sun} \ein{Mpc}^{-3}$ and the SMBH density as $n_\text{SMBH} = 5.73 \cdot 10^{-3\pm0.4} \ein{Mpc}^{-3}$.
{We compare these values with other SMBH mass distributions.}
The Schechter function by \cite{2004MNRAS.354.1020S} delivers a SMBH mass density of about a factor 1.5 to 3 larger than the one determined in this work.
The same goes for \citet*{2003ApJ...582..559V} with an over 1.5 times larger SMBH density, whereas \cite{2003ApJ...583..616J} introduced a SMBH density of $h^3 \cdot 10^{-3} \ein{Mpc}^{-3} \approx 3 \cdot 10^{-4} \ein{Mpc}^{-3}$, which is about a factor $10$ smaller than the SMBH density in this work.
It can therefore not be excluded that the used SMBH mass distribution by \cite{2010A&A...521A..55C} over- or underestimates the distribution.
Though, it does not differ greatly from the other distributions.

Here, two mass ratios, $q=1/3$ and $q=1/30$, were considered, as the majority of SMBBH mergers occur in mass ratios between those two \citep{2009ApJ...697.1621G}.
An investigation delivered the shortest SMBBH merger rate after entering the inspiral stage of $R_{insp} = 5.1 \cdot 10^{-3 \pm 0.4} \ein{Gpc}^{-3} \ein{yr}^{-1}$ with the inspiral time model by \cite{2009ApJ...697.1621G} at the mass ratio $1/30$.
The largest SMBBH merger rate after entering the inspiral stage is achieved with the inspiral time model by \cite{PhysRev.136.B1224} with $q=1/3$.
It has a value of $R_{insp} = 24.61 \cdot 10^{-2\pm 0.4} \ein{Gpc}^{-3} \ein{yr}^{-1}$.
The rates following the other inspiral time models lie between these two.

Since these rates disregard the time scale of the preceding galaxy merger, the duration of the dynamical friction stage and the period of time to overcome the final parsec separation, the total SMBBH merger rate $R_{\rm tot}$ including all necessary time scales was introduced in Sec.~\ref{sec:SMBBH_Merger_Rate}. 
To estimate this merger rate, the total mean SMBBH merger time was determined using the simple assumption that each merger occurs with the same mass ratio $q$ and, moreover, that each SMBH increases its mass via accretion between each merger with the same accretion percentage $\eta_{\rm acc}$. 
The accretion percentage was chosen as $\eta_{\rm acc} = 33$ per cent, so that the mass accretion is as large as the mass increase via merger with the highest mass ratio of $q=1/3$.
It was discussed that a much lower accretion percentage value increases the number of SMBBH mergers until they become unrealistic (compare to \figref{fig:1_mal_10_9_M}), while a higher value lowers the number of SMBBH mergers, neglecting the influence of mass increase via mergers. 
An appropriate value for this parameter is therefore essential in order to estimate the number of SMBBH mergers in the Hubble time.
The value of $\eta_{\rm acc} = 33$ per cent was compared to the Eddington accretion and the accreted mass between SMBBH mergers determined by \citet*{2003ApJ...582..559V} with the conclusion that it probably overestimates the accretion between mergers of a mass ratio near $1/3$, but is sufficient in describing accretion between lower mass ratio mergers.
That is why the number of mergers at mass ratio $1/3$ was taken as a lower limit, as with lower accretion percentage between each merger, the number of mergers would increase.

The total mean merger time was compared to the total time duration for SMBBH mergers after the preceding galaxy merger.
The latter consists of the inspiral time model by \cite{2009ApJ...697.1621G}, which delivers the largest values among the 4 inspiral time models, the dynamical friction time scale by \cite{2002MNRAS.331..935Y} and an estimate for the final parsec time scale between $10^6 \ein{yr}$ and $10^9 \ein{yr}$ \citep{2016IAUFM..29B.285V}.
The time scale for overcoming the ISCO is several orders of magnitude lower than the other times and is therefore neglected in this comparison. 
In the most optimistic scenario with an accretion percentage of $\eta_{\rm acc} = 33$ per cent and the mass ratio $q=1/3$, an iteration yields $6$ mean SMBBH mergers between a redshift of $z=1$ and $z=3$, resulting in a total mean time for one merger with a slowing due to the expanding Universe of $2.47 \ein{Gyr}$ and a merger rate of $R_\text{tot} = 2.32 \cdot 10^{-3 \pm 0.4} \cdot {\bar{t}_{insp}}/{\bar{t}_{total}} \ein{Gpc}^{-3} \ein{yr}^{-1}$, based on the percentage of currently merging SMBHs in the distribution used, determined with the mean inspiral stage time (see Sec.~\ref{sec:SMBBH_Merger_Rate}).
A more pessimistic scenario delivers $8$ mean mergers with a total mean merger time of $1.85 \ein{Gyr}$ and the merger rate $R_\text{tot} = 3.09 \cdot 10^{-3 \pm 0.4} \cdot {\bar{t}_{insp}}/{\bar{t}_{total}} \ein{Gpc}^{-3} \ein{yr}^{-1}$.
In the first case, the total SMBBH merger time is larger than the duration for SMBBH mergers after the preceding galaxy merger.
This means that for the most optimistic scenario, even with an overestimation of the final parsec time scale with $10^9 \ein{yr}$, a SMBBH merging event including a galaxy merger is realizable.
For the second scenario, a final parsec time scale only up to $10^8 \ein{yr}$ is realistic for the most massive SMBHs.
As $8$ mean SMBBH mergers represent a more liberal model, its range for the effective neutrino energy fraction from the GW energy was taken as a lower limit, {while $q=1/30$ and $9$ mean SMBBH mergers yield the highest values, thus the upper limit: $f^\nu_\text{SMBBH, eff} \gtrsim 9 \cdot 10^{-7} - 3 \cdot 10^{-4}$.}
This range overlaps with the energy fraction that neutrinos receive from gravitational waves during a stellar mass BBH merger. 

The ratio ${\bar{t}_{insp}}/{\bar{t}_{total}}$ was introduced as the percentage of SMBHs that are currently undergoing a merger in \glref{eq:R_SMBH} and takes into account that all SMBBHs are equally distributed in their merging stages. If, however, more binaries are in their inspiral stage than in the galaxy merger or dynamical friction stage, this ratio will rise along with the SMBBH density. As a result, the SMBBH merger rate will rise accordingly. However, as $f^\nu_\text{SMBBH, eff}$ is independent of this ratio (see \glref{eq:f_SMBBH_4}), it will not change in such a case.

On the other hand, with the mass ratio $q=1/30$, all total mean SMBBH merger times for the highest and lowest SMBH masses in the distribution are smaller than the  time duration for SMBBH mergers after the preceding galaxy merger.
This led to the conclusion that these SMBHs require larger mass ratios to form.
As a closer review of \figref{fig:t_total_b} shows, only the merging times after the preceding galaxy merger of binaries with intermediate SMBBH masses ($\sim 10^7 \ein{M}_\odot - 10^9 \ein{M}_\odot$) are smaller than the determined total mean SMBBH merger times.
Lower masses require longer to merge, because of an increase of the dynamical friction time scale, while the duration of the inspiral stage increases with higher SMBH masses.
For the highest mass range, this also allows for the possibility that they became this heavy due to accretion after the last SMBBH merger or that another effect, like the influence of a third SMBH, accelerated the traversal through this stage.
For the lowest SMBH masses, the possibility exists that the radius of the central galaxy region and the velocity dispersion in that region are smaller which would decrease the dynamical friction time scale.
This is supported by the correlation of the SMBH mass and the velocity dispersion in the galaxy bulge \citep*{2003ApJ...582..559V}, though the velocity dispersion exerts minor influence on the dynamical friction time scale.

\section{Conclusions}
\label{sec:Conclusions}
To summarize the results, we derive the parameter space of $(f^\nu,\,R)$, which is allowed under the assumption that stellar mass BBH mergers contribute with 10 per cent and SMBBH mergers with 90 per cent to the diffuse neutrino flux. This ratio is taken under the assumption that the diffuse neutrino flux is connected to the flux of ultra-high energy cosmic rays and we therefore consider 10 per cent starburst galaxy contribution consistent with recent results by Auger. We further use that most mergers happened in the phase $z=1$ to $z=3$ as argued above. Under these assumptions, we find that the fraction of neutrino to gravitational energy, $f^\nu$, needs to be {$\sim 10^{-6} - 3 \cdot 10^{-4}$} for SMBBHs and is in a similar range for stellar mass BBHs, {$\sim 2\cdot 10^{-5} - 10^{-3}$}. In case only a fraction of the neutrino flux is made up by these sources, the fraction would go down correspondingly for both source types. 

The merger rates $R$ are expected to lie in the range $\sim 10^{-7}$ and $10^{-5}$~Gpc$^{-3}$~yr$^{-1}$ for SMBBHs and $\sim 10-100$~Gpc$^{-3}$~yr$^{-1}$ for stellar mass BBHs, consistent with the data \citep{2021arXiv211103634T, 2021ApJ...913L...7A}.

We also present a prediction of the gravitational wave signature, which lies in the sensitivity range of both SKA and LISA.

\section{Outlook}
\label{sec:Outlook}
There are some ways to further improve the model presented here.
First, the influence of the redshift dependent part $\xi_z$ on the evolution of SMBBHs was estimated with simple assumptions. 
This estimate can be improved by including the Madau plot with the galaxy formation evolution, as the growth of SMBHs is linked to galaxies (see e.g.~\citealt*{1998ApJ...498..106M}).
The eccentricity $e$ can be further investigated as well.
In this work, it has been assumed to be $e = 0$ during the inspiral stage.
However, as it only becomes negligible small at the time of the ISCO and the merger, it could be greater during the inspiral stage, especially at its beginning.
With a value of $e >> 0$, the binary can emit more orbital energy in the form of GWs and thus reduce its duration in the inspiral stage.
This could possibly explain mergers of the heaviest SMBHs, since their inspiral time exceeds their total merging time, as well as an accretion percentage of $\eta_{\rm acc}$ of mergers at $q = 1/3$, since with a smaller inspiral time, the time scale for accretion also decreases. 

As shown in Sec.~\ref{subsec:constraints}, a first constraint on the SMBH distribution can be made, once LISA and SKA are in operation make their first detections.
This constraint will affect the SMBBH merger rate, and thus its detection rate. 
A greater SMBBH detection rate will help to constrain the number of mergers in the SMBH distribution used and also the mean time between each merger.
If the latter gets smaller, the parameter $f^\nu_\text{SMBBH, eff}$ will decrease as well.
The BBH merger rate in starburst galaxies will also get more precise, the more binary black holes are detected with the LIGO and Virgo detectors.
This will also improve the accuracy for the parameter $f^\nu_\text{BBH}$.

Finally, the BH distribution of IMBHs between stellar mass BHs and SMBHs is not considered in this work.
The detection of a $\sim 5 \cdot 10^4 \ein{M}_{\sun}$ BH in the center of the dwarf galaxy RGG~118 showed, that BHs with masses below $10^6 \ein{M}_{\sun}$ exist at the center of dwarf galaxies \citep{2015ApJ...809L..14B}.
If these galaxies also merge, the central BHs merge as well.
As some of them are active and produce a possible jet, neutrinos could then be produced in the jet path and during the merger. 
Even if the jets are faint or obscured, the resulting neutrinos could also contribute to the diffuse astrophysical neutrino flux, which is detected at IceCube.
An investigation, whether these galaxies could have merged at least once by resolving the jet structures, could help expanding the model presented here with the merger rates of these BHs {\citep[see e.g.][for a merger rate investigation of IMBHs until $200 \ein{M}_{\sun}$]{2022A&A...659A..84A}}.

An interesting follow-up investigation could also be whether CRs, or UHECRs, and the gamma-ray background could also be explainable with SMBBH and stellar mass BBH mergers by introducing a parameter which describes the fraction of gravitational wave energy that is converted into UHECRs or gamma-rays during each merger.

As discussed in \cite{2020ApJ...905L..13D}, TXS~0506+056 could be an ongoing supermassive binary black hole merger, which currently merges and thus be detectable during the first LISA run, depending on its mass ratio.
The latest high-energy neutrino detected in September 2022 from this blazar by IceCube \citep{IC220918A_gcn1} was predicted in \cite{2020ApJ...905L..13D} based on a model describing the precession of its jet due to the spin-flip of the jet, updated in \cite{txs_spinflip2022}. 
The possible detection of yet another high-energy neutrino from the same direction by the Baikal-GVD neutrino telescope in April 2021 \citep{2022arXiv221001650B} is also in accordance with the jet precession model, strengthening the possibility of TXS~0506+056 being a SMBBH merger close to its final merger.
Since its mass is estimated around $3 \cdot 10^8 \ein{M}_{\sun}$, a successful detection could help putting a constraint on the merger rates as well as the SMBH density and thus the stochastic GW background.

\section*{Acknowledgements}
We acknowledge support from the Deutsche Forschungsgemeinschaft via the following projects: DFG-ARN Project MICRO (Project Number TJ 62/8) and the DFG funded Collaborative Research Center SFB~1491.

\section*{Data Availability}
The data underlying this article will be shared on reasonable request to the corresponding author.

%%%%%%%%%%%%%%%%%%%%%%%%%%%%%%%%%%%%%%%%%%%%%%%%%%



\bibliographystyle{mnras}
\bibliography{literature} % if your bibtex file is called example.bib

%%%%%%%%%%%%%%%%%%%%%%%%%%%%%%%%%%%%%%%%%%%%%%%%%%

\appendix
\section{Cosmological density parameters}
\label{app:A}
\clearpage

\begin{table}
	\centering
	\caption{Current values for the cosmological density parameters and the Hubble constant according to Planck data \citep{2020A&A...641A...6P}. Here, a $\Lambda$-CDM Universe is assumed.}
	\label{tab:Planck}
	\begin{tabular}{ c|c}
	    \hline
		parameter & value\\
		\hline
		$H_0$ & $67.66 \pm 0.42 \ein{km}\ein{s}^{-1}\ein{Mpc}^{-1}$\\
		$h$ & $0.6766 \pm 0.0042 $\\
		$t_H$ & $(1.446 \pm 0.0090) \cdot 10^{10} \ein{yr}$\\
		$\Omega_\text{M}$ & $0.3111 \pm 0.0056$\\
		$\Omega_\Lambda$ & $0.6889 \pm 0.0056$\\
		$\Omega_\text{k}$ & $0.0007 \pm 0.0037$\\
		\hline
	\end{tabular}
\end{table}

\section{Results for different IceCube diffuse astrophysical neutrino fluxes}
\label{app:B}

\begin{figure*}
	\centering
	\includegraphics[width=0.95\textwidth]{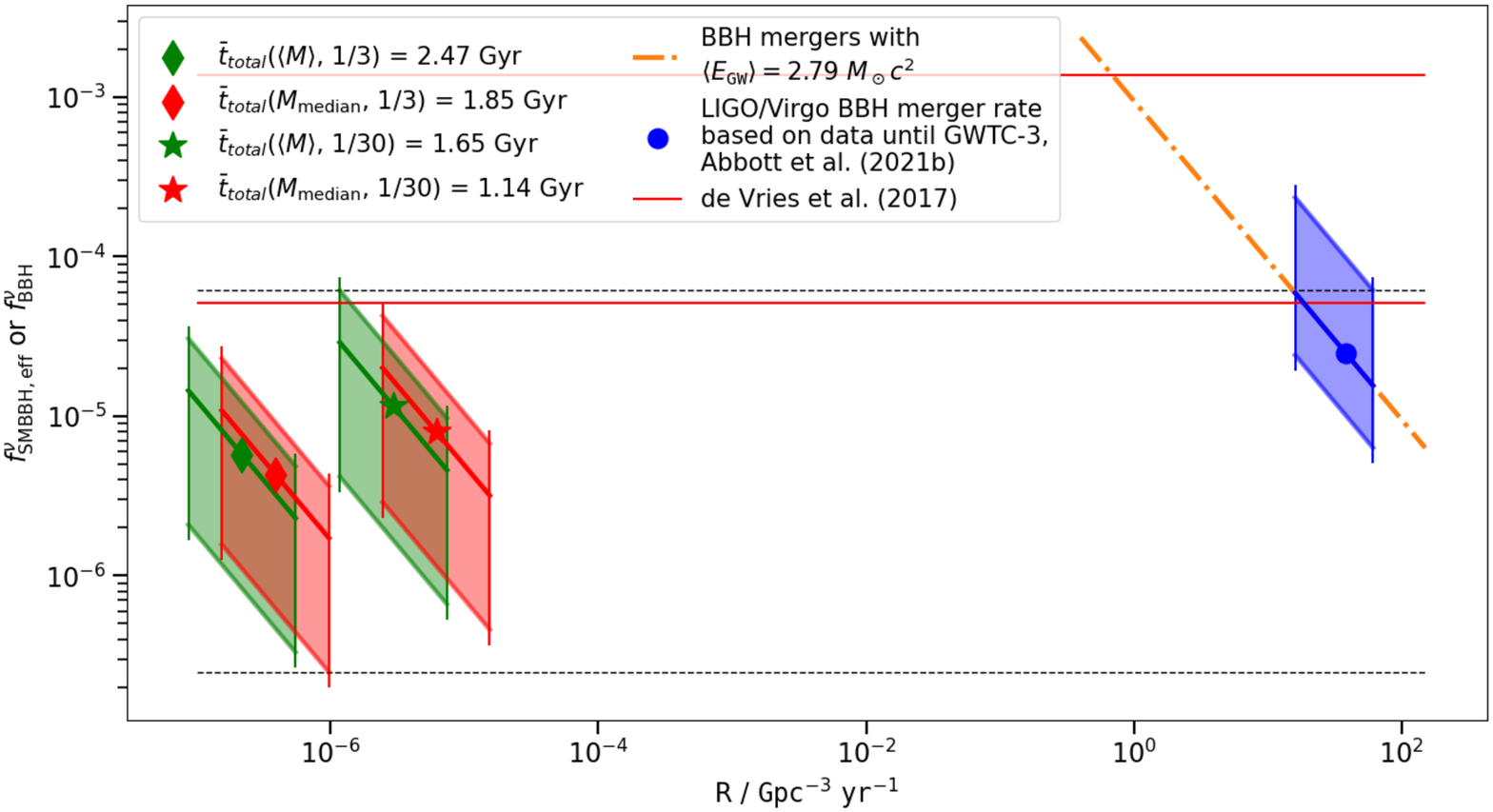}
	\caption{Results for the IceCube diffuse astrophysical neutrino flux measured in 9.5 years of muon tracks data \citep{2019ICRC...36.1017S} instead of the starting tracks data measured in 10.3 years as described in Sec.~\ref{subsec:Compasison_diffuse_flux}. 
	Different SMBBH merger and BBH merger rates with the associated values for the fraction of gravitational wave energy that goes into neutrinos during each SMBBH merger $f^\nu_\text{SMBBH,eff}$ and stellar mass BBH merger $f^\nu_\text{BBH}$, respectively. 
	The x- and y-axis are in logarithmic scales. 
	BBH merger rates marked as dots, while SMBBH merger rates are marked with diamonds for a mass ratio of $q=1/3$ and stars for $q=1/30$.
	Red horizontal lines mark comparative values for $f^\nu_\text{BBH}$ \citep{2017PhRvD..96h3003D}.}
	\label{fig:beide_f_muon}
\end{figure*}

\begin{figure*}
	\centering
	\includegraphics[width=0.95\textwidth]{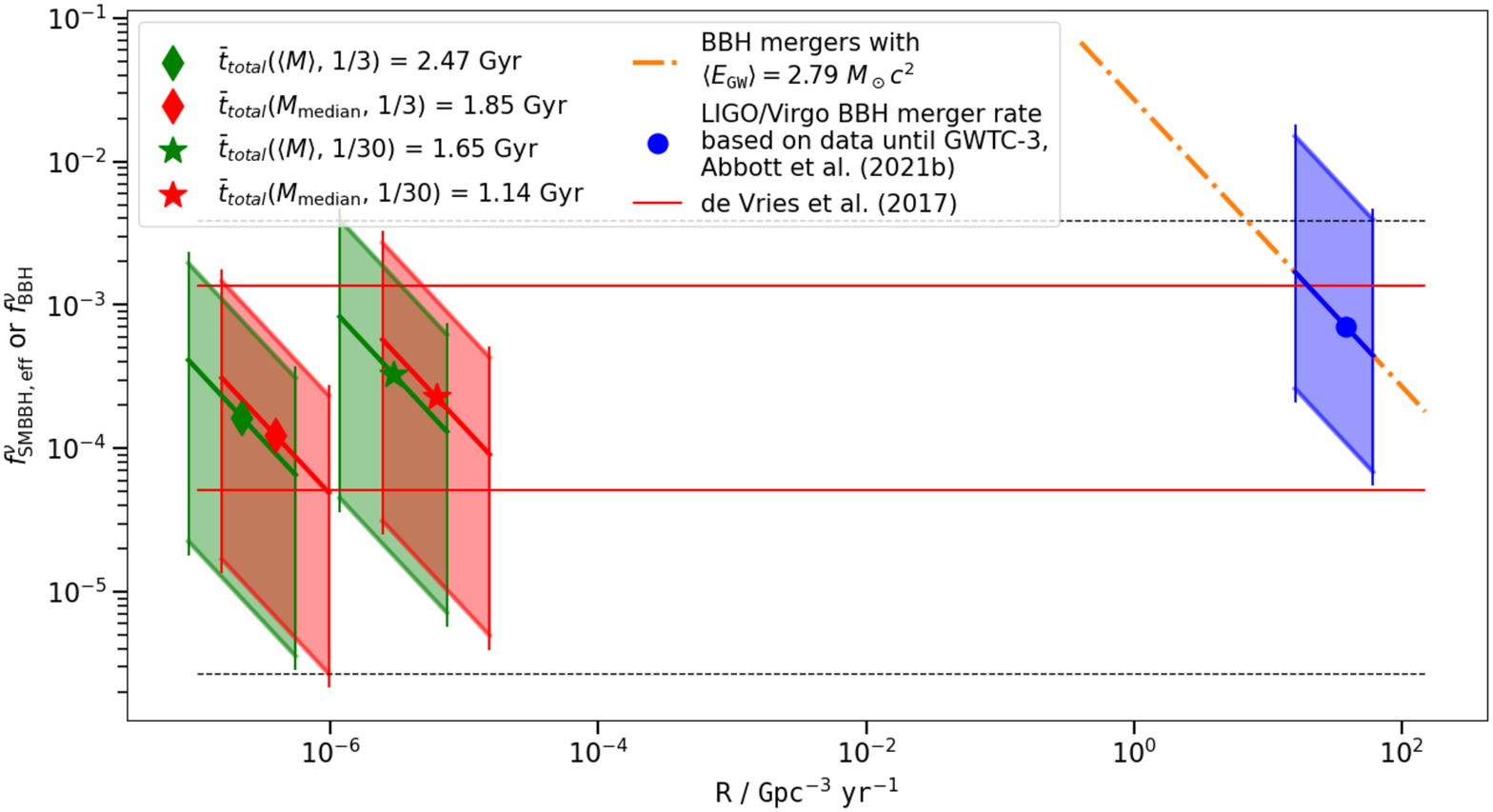}
	\caption{Results for the IceCube diffuse astrophysical neutrino flux measured using high-energy starting event, HESE, data \citep{2021PhRvD.104b2002A} instead of the starting tracks muon neutrino data measured in 10.3 years as described in Sec.~\ref{subsec:Compasison_diffuse_flux}. 
	Different SMBBH merger and BBH merger rates with the associated values for the fraction of gravitational wave energy that goes into neutrinos during each SMBBH merger $f^\nu_\text{SMBBH,eff}$ and stellar mass BBH merger $f^\nu_\text{BBH}$, respectively. 
	The x- and y-axis are in logarithmic scales. 
	BBH merger rates are marked as dots, while SMBBH merger rates are marked with diamonds for a mass ratio of $q=1/3$ and stars for $q=1/30$.
	Red horizontal lines mark comparative values for $f^\nu_\text{BBH}$ \citep{2017PhRvD..96h3003D}.
	}
	\label{fig:beide_f_HESE}
\end{figure*}

\begin{figure*}
	\centering
	\includegraphics[width=0.95\textwidth]{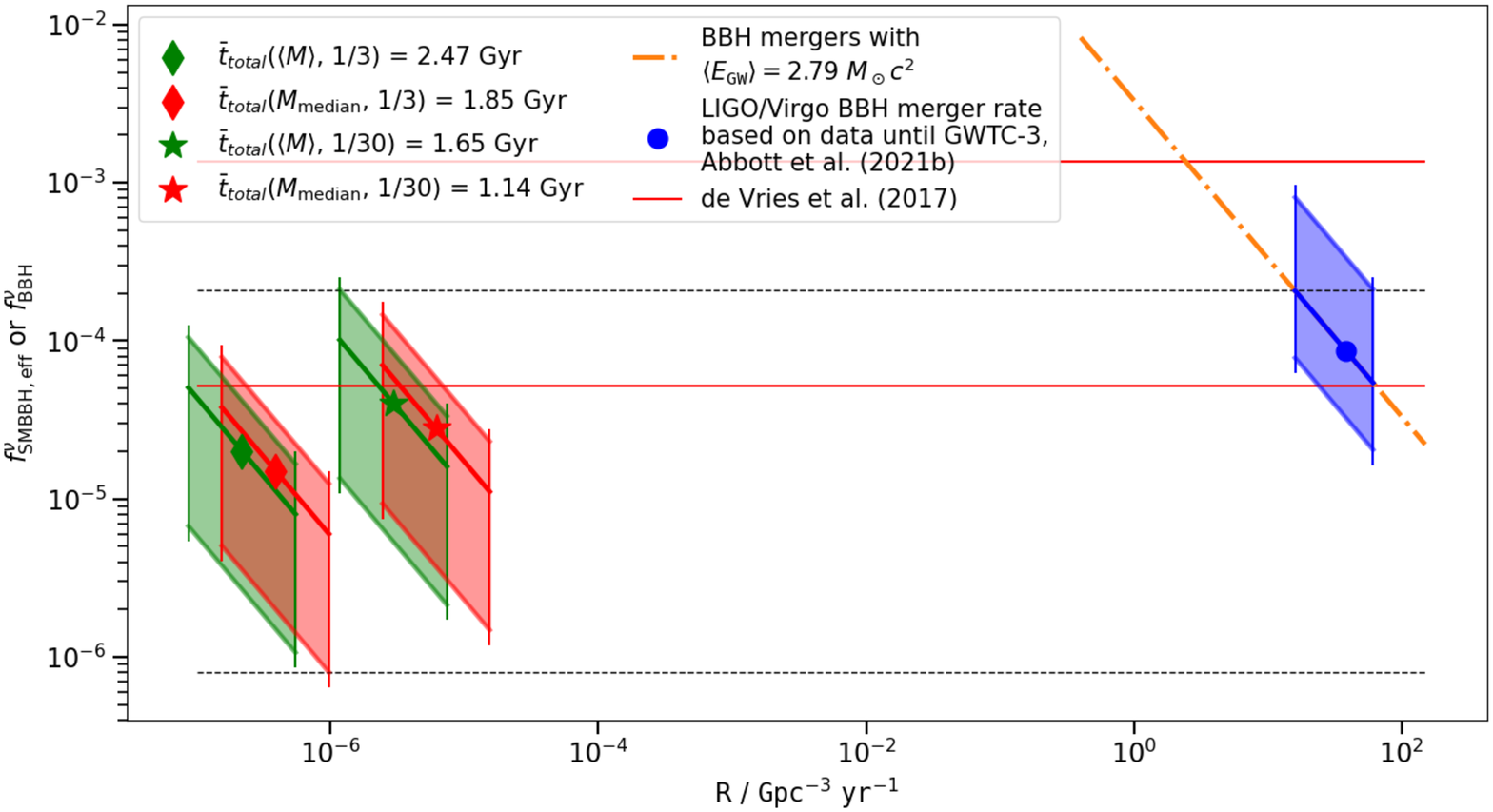}
	\caption{Results for the IceCube diffuse astrophysical neutrino flux measured using cascade data \citep{2020PhRvL.125l1104A} instead of the starting tracks muon neutrino data measured in 10.3 years as described in Sec.~\ref{subsec:Compasison_diffuse_flux}. 
	Different SMBBH merger and BBH merger rates with the associated values for the fraction of gravitational wave energy that goes into neutrinos during each SMBBH merger $f^\nu_\text{SMBBH,eff}$ and stellar mass BBH merger $f^\nu_\text{BBH}$, respectively. 
	The x- and y-axis are in logarithmic scales. 
	BBH merger rates are marked as dots, while SMBBH merger rates are marked with diamonds for a mass ratio of $q=1/3$ and stars for $q=1/30$.
	Red horizontal lines mark comparative values for $f^\nu_\text{BBH}$ \citep{2017PhRvD..96h3003D}.}
	\label{fig:beide_f_cascade}
\end{figure*}

% Don't change these lines
\bsp	% typesetting comment
\label{lastpage}
\end{document}